\documentclass[12pt, a4paper]{article}
\pdfoutput=1

\usepackage[utf8]{inputenc}
\usepackage[T1]{fontenc}

\usepackage{amsmath}
\usepackage{amssymb}
\usepackage{amsfonts}
\usepackage{bbm}
\usepackage{verbatim}
\usepackage{dsfont}
\usepackage{booktabs}
\usepackage{slashed}
\usepackage{bm}
\usepackage{cancel}

\usepackage{mathtools}

\usepackage{epsfig}
\usepackage{color}
\usepackage[table]{xcolor}
\usepackage{graphicx}
\usepackage[]{caption}
\usepackage{float}
\usepackage{overpic}

\usepackage{enumerate}
\usepackage{hhline}
\usepackage{multirow}
\usepackage{xspace}
\usepackage{setspace}
\usepackage[title]{appendix}

\usepackage{soul}
\renewcommand{\theequation}{\arabic{section}.\arabic{equation}}

\renewcommand{\Re}{{\rm Re}\,}
\renewcommand{\Im}{{\rm Im}\,}

\def\beq{\begin{equation}}
\def\eeq{\end{equation}}
\def\bea{\begin{eqnarray}}
\def\eea{\end{eqnarray}}
\def\bes{\begin{subequations}}
\def\ees{\end{subequations}}

\renewcommand{\bm}{\boldmath}

\usepackage[square,numbers,compress]{natbib}
\usepackage{bibentry}

\counterwithin*{equation}{section}

\usepackage[hidelinks, linkcolor=blue, citecolor=blue, urlcolor=blue]{hyperref}

\renewcommand\vec[1]{\ensuremath\boldsymbol{#1}}

\usepackage[caption=false]{subfig}

\usepackage{xcolor}

\usepackage[normalem]{ulem}
\begin{document}

\thispagestyle{empty}

\begin{center}
  {\Large {\bf\bm Brane Nucleation in Supersymmetric Models}}\\[12pt]

%\bigskip
%\bigskip
\bigskip
{\bf Igor Bandos,}\footnote{igor.bandos@ehu.eus}\textsuperscript{,*,$\dagger$,$\ddagger$}
{\bf Jose J. Blanco-Pillado,}\footnote{josejuan.blanco@ehu.eus}\textsuperscript{,*,$\dagger$,$\ddagger$}
{\bf Kepa Sousa}\footnote{kepa.sousa@uah.es}\textsuperscript{,$\mathsection$}\\
{\bf and Mikel A. Urkiola}\footnote{mikel.alvarezu@ehu.eus}\textsuperscript{,$\dagger$,$\mathparagraph$}

\setcounter{footnote}{0}
%To avoid starting the main text with a footnote number of 5.

%\bigskip
\bigskip
\vspace{0.23cm}
\textsuperscript{*}{\it Department of Physics, University of the Basque Country, Bilbao, Spain}\\[5pt]
\textsuperscript{$\dagger$}{\it EHU Quantum Center, University of the Basque Country UPV/EHU, Bilbao, Spain} \\[5pt]
\textsuperscript{$\ddagger$}{\it IKERBASQUE, Basque Foundation for Science, 48011, Bilbao, Spain} \\[5pt]
\textsuperscript{$\mathsection$}{\it University of Alcal\'a, Department of Physics and Mathematics, 28805 Alcal\'a de Henares (Madrid), Spain} \\[5pt]
\textsuperscript{$\mathparagraph$}{\it Department of Applied Mathematics, University of the Basque Country UPV/EHU, Plaza Ingeniero Torres Quevedo 1, 48013 Bilbao, Spain}\\[20pt]
%\bigskip

\end{center}

\begin{abstract}
\noindent

\vspace{-10pt}

This paper explores the process of vacuum decay in supersymmetric models related to flux compactifications.
In particular, we describe these instabilities within supersymmetric Lagrangians for a single three-form multiplet. This
multiplet combines scalar fields, representing the moduli fields in four dimensions, with 3-form fields that influence the potential
for these moduli via the integer flux of their associated 4-form field strength. Furthermore, using supersymmetry as a guide we
obtain the form of the couplings of these fields to the membranes that act as sources to the 3-form potentials. Adding small
supersymmetry breaking terms to these Lagrangians one can obtain instanton solutions describing the decay of the vacua
in these models by the formation of a membrane bubble. These instantons combine the usual Coleman-de Luccia and the
Brown-Teitelboim formalisms in a single unified model.  We study simple numerical examples of theories with and without
gravity in this new framework and generalize known Euclidean methods to accomodate the simulataneous inclusion
of scalar fields and charged membranes to these instanton solutions. Moreover, we show explicitly in these examples how
one recovers the static supersymmetric solutions in the limiting case where the supersymmetry breaking terms vanish. In this
limit, the bubble becomes infinite and flat and represents a hybrid between the usual supersymmetric domain walls of field
theory models and the brane solutions interpolating between the supersymmetric vacua; a sort of dressed supermembrane
BPS solution. Finally, we briefly comment on the implications of these solutions in cosmological models based on the String
Theory Landscape where these type of 4d effective theories could be relevant in inflationary scenarios.

\end{abstract}

\newpage
\setcounter{page}{2}

%%%%%%%%%%%%%%%%%%%%%%%%%%%%%%%%%%%%%%%%%%%%%%%%%%%

\tableofcontents

\section{Introduction}

Many high energy extensions of the Standard Model make use in some way or another of supersymmetry.
Furthermore, the presence of multiple vacua is quite generic in these models. Some of these vacua
preserve part of the original supersymmetry while others break it completely. Understanding the
stability of these vacua and their possible decay processes is, therefore, an essential aspect of the low
energy description of these theories.

Some of the most interesting examples of this type of scenarios are the effective four dimensional models
obtained from string theory compactifications. Compactifying $10d$ string theory to four dimensions leaves
us with a low energy effective theory with a collection of scalar fields (the moduli fields) that parametrize the
possible deformations of the internal compact manifold. One can further impose that the compactification
mechanism preserves some supersymmetry so that we end up with a low energy theory
that can be classified as a supersymmetric scalar field theory.\footnote{More rigorously, one should speak about supermultiplets
involving scalar fields; in the case of $\text{N=1}$ supersymmetry, these include not only the scalar supermultiplet but also the so-called three form supermultiplets (see below).
 } This led the authors in \cite{Cvetic:1992st} to consider the non-perturbative stability of a $\text{N=1}$ model of supergravity.
They demonstrated that supersymmetric vacua are stable with respect to the usual
process of bubble nucleation. Furthermore, they also showed that this
stability is due to the restriction imposed by supersymmetry on the tension
of the wall that interpolates between the two vacua. In fact, this quenching
phenomenon is nothing more than the Coleman-deLuccia \cite{Coleman:1980aw} suppression.
In this limit the bubble radius would be infinite and the solution would be a planar
domain wall that preserves part of the supersymmetry \cite{Cvetic:1992bf}.

On the other hand, recent developments in models of string compactification based
on the use of fluxes along the internal dimensions has led us to the idea of
an extremely rich Landscape of possible $4d$ Effective Field Theories (EFTs), also referred to as \emph{flux vacua} \cite{Bousso:2000xa}. Each of these vacua
is characterized by the presence of a set of integer fluxes that thread some
cycles in the internal manifold. The stabilizing potential for the moduli in each of these
sectors of the theory is also quite complicated and could easily have many
local minima itself. One could therefore use the conclusions discussed earlier
for the scalar field potential in each of these sectors to study their stability
and find the bounce solutions using the techniques derived by Coleman
and collaborators \cite{Coleman:1977py,Callan:1977pt}.

One can then ask whether there is an analogous process that
would take us from one sector with some set of fluxes to another one where one
or several of those fluxes have been changed. Naively
this would seem impossible since the flux is quantized and therefore
cannot be continuously changed. However, similarly to what happens
in the Schwinger process \cite{Schwinger:1951nm}, one could reduce the flux by the
creation of sources charged with respect to the same field that
produces it.\footnote{See \cite{Blanco-Pillado:2009lan} for a simple description of
several field theory models in different dimensions of spacetime
that exhibit a similar behaviour to the one described in this paper.} In fact, this type of process had
been already discussed in a purely four dimensional context a number
of years ago by Brown and Teitelboim (BT) \cite{Brown:1987dd,Brown:1988kg}. In this model the presence of a 4-form field strength
in four dimensions induces an effective cosmological constant that can
only be changed by the nucleation of membranes charged with respect
to its 3-form potential. The instanton solutions describing this type
of instability of the model were studied in detail in connection to the
possible self-tuning mechanism of the cosmological constant.

In models of string theory compactification the situation is quite similar,
and one can assume that each of the forms present in the 4d theory
would have an associated brane charged with respect to that specific form. Indeed
one can identify $4d$ membrane objects by wrapping higher dimensional
branes along some internal cycles. Similarly, one can also find the effective
4-form field that couples to the 4-dimensional membrane and
understand its higher dimensional origin. Taking this point of view,
these models of flux compactification seem to lead to a model
closely related to the Brown-Teitelboim idea \cite{Bousso:2000xa}.

There is however an important difference between these two models. In the original
Brown-Teitelboim's model the nucleation of the membranes would
lead to a jump in the value of the $4d$ cosmological constant. In models
with higher dimensions this change in the flux would lead us to a
different sector with a different moduli potential. This is
interesting since it allows us to avoid the so called "empty universe problem"
in the purely $4d$ BT model by postulating a period of inflation after the
bubble nucleation driven by the compactification potential (See for example \cite{Freivogel:2005vv}).

The previous arguments suggest the idea of combining both models
into a unifying picture where we can describe within the same theory
the presence of the scalar field moduli and the 4-form fields.
Moreover, following what was done before in the purely scalar field model,
we will look for guidance in supersymmetric models that include both
types of degrees of freedom, the 4-form field as well as the moduli fields.
The inclusion of form fields in supersymmetric and supergravity multiplets
has been done before in  \cite{Gates:1980ay,Binetruy:1996xw,Ovrut:1997ur,Kuzenko:2010am,Kuzenko:2010ni,Bandos:2011fw,Farakos:2017jme}. Furthermore, their   interaction with supersymmetric membranes (supermembranes) was studied in
\cite{Ovrut:1997ur,Bandos:2010yy,Bandos:2012gz,Bandos:2018gjp}
 and their role in
the low energy description of flux compactifications has been recently discussed in  \cite{Farakos:2017jme,Bandos:2018gjp,Lanza:2019xxg,Lanza:2019nfa}. In this paper we will study the non-perturbative
stability of these models and their modified versions once we introduce
small supersymmetry-breaking terms. More concretely, we will deform the supersymmetric field-theoretic part of the action by including soft supersymmetry breaking terms \cite{Chung:2003fi}, while keeping the (super)membrane part of the action untouched.

This paper is organized as follows. In section \ref{sec:global_susy} we will
review the results of  global supersymmetric scalar field theories  (more concretely, self-interacting scalar supermultiplet models which can be described in terms of a generic chiral superfield), and the existence
of supersymmetric domain wall solutions in these models. We will show
with explicit examples how these solutions appear as limiting cases
of a bubble decay process from a non-supersymmetric vacua. In sec.~\ref{sec:susy_membranes}
we will explain how to introduce 3-form gauge fields in our supersymmetric theory, as
well as membranes that naturally couple to these forms. This will lead us to discuss the so-called 3-form supermultiplets described by a special type of chiral superfields. Furthermore,
we will obtain the instanton solutions describing the decay processes in
these theories, all the while explicitly showing how they connect in a proper limit with the supersymmetric solutions found in the previous section. Finally, in secs.~\ref{sec:sugra_membranes} and \ref{sec:sugra_tunnel} we will describe
a similar situation in the case of supergravity and find explicit solutions
of the equations of motion describing the bubble nucleation. We end
with some conclusions in section \ref{sec:conclusions}.

\section{Rigid supersymmetric scalar field theory models}
\label{sec:global_susy}

As a warm up exercise,  in this section, we will consider the stability of vacua in a scalar field Lagrangian
of an $\text{N=1}$ four dimensional globally supersymmetric field theory of a single
chiral superfield, which describes the so-called scalar supermultiplet.\footnote{In the main part of this paper we will only deal with the bosonic components of the supermultiplets. See, e.g., \cite{Wess:1992cp} for a more complete treatment of the simplest  cases.}. Many of the results presented in this section can be
found in the literature, in particular in \cite{Cvetic:1992st,Cvetic:1992bf}.
In the following sections we will discuss more complicated models
following a similar reasoning to the one presented here.

\subsection{Field theory supersymmetric domain walls}

The model we will be considering here is given by the most general Lagrangian
for a complex scalar field $(\phi)$ which
is given by the leading component of a generic chiral superfield (which describes the so-called scalar multiplet). This Lagrangian reads
\begin{equation}
{\cal L} = - K_{\phi \bar \phi} \ \partial_{\mu} \phi \partial^{\mu} \bar \phi - K^{\phi \bar \phi} |W_{\phi}|^2
\label{eq:scalar_theory}
\end{equation}
where we have introduced the two functions that define the model: the real Kahler
potential $K(\phi, \bar \phi)$ and the complex holomorphic superpotential $W(\phi)$. We
will denote their derivatives as
$\partial_{\phi}\partial_{\bar \phi} K = K_{\phi \bar \phi} = 1/ K^{\phi \bar \phi} $ and $W_{\phi} = \partial_{\phi} W(\phi)$, thus
following the conventions of \cite{Wess:1992cp}.\footnote{In this paper we will use the $(- + + + )$ signature convention.}

The equations of motion for this theory are
\begin{equation}
K_{\phi \bar \phi} \partial_{\mu} \partial^{\mu} \phi -  K_{\phi \phi \bar\phi} \partial_{\mu} \phi  \partial^{\mu} \bar{\phi} + K_{\phi \bar \phi \bar \phi} (K^{\phi \bar \phi})^2   |W_{\phi}|^2 - K^{\phi \bar \phi} W_{\phi} \bar{W}_{\bar \phi \bar \phi} = 0
\label{eq:2nd_order_eqs}
\end{equation}
and its complex conjugate. These reduce to the usual Klein-Gordon equation
for a complex scalar field with a scalar potential given by $V(\phi,\bar\phi) = |W_{\phi}|^2$
if the kinetic term is canonical (which particularly requires $K=\phi \bar\phi$).

We are looking for a domain wall solution in this model that interpolates between two supersymmetric minima, in other
words, between two points whose superpotential satisfies $W_{\phi}( \phi_{\pm}) = 0$. Recall that all supersymmetric minima have a vanishing potential, and therefore degenerate in energy. For that matter, let us consider a flat domain wall whose transverse direction is given by the coordinate $z$. One can then
show \cite{Cvetic:1992bf} that the static solution preserving half of the $\text{N=1}$ supersymmetry solves the first-order equation
\begin{equation}
\partial_z \phi(z) = e^{i \theta} K^{\phi \bar \phi} ~\bar W_{\bar \phi}(\bar \phi(z)),
\label{eq:BPS}
\end{equation}
known as the BPS equation, where the phase $\theta$ is given by
\begin{equation}
e^{i\theta} = \frac{\Delta W}{|\Delta W|},  \qquad \text{with} \qquad \Delta W \equiv W(\phi(z=\infty))-W(\phi(z=-\infty)).
\end{equation}

Of course, given appropriate boundary conditions, both the first-order and second-order equations should yield the
same static solution for the supersymmetric  domain wall. The tension of the domain wall in this model can be computed writing the
energy per unit area as follows:
%&&%
\begin{equation}
\sigma  =  \left[\int_{-\infty}^{\infty} dz \ K_{\phi \bar \phi} ~\left |\partial_z \phi(z) - e^{i \theta} K^{\phi \bar \phi} \bar W_{\bar \phi}(\phi(z))\right |^2 \right]+ 2 ~\text{Re}[e^{-i\theta} \Delta W]
\end{equation}
One can readily see that in the case of supersymmetric bosonic solutions, where \eqref{eq:BPS} holds, the tension becomes
\begin{equation}
\sigma_{BPS} = 2 | \Delta W| .
\end{equation}

\subsubsection{Example. Double well potential}
	
Let us illustrate all of the above by considering the model defined by:
\begin{equation}
K(\phi, \bar \phi)  = \phi \bar \phi, \qquad W (\phi) = \left( \frac{1}{3} \phi^3 - a^2 \phi \right)
\label{eq:cubic_susy_model}
\end{equation}
where we are assuming that $a>0$. For this particular model,   the Lagrangian
\eqref{eq:scalar_theory} simplifies  to
\begin{equation}
{\cal L} = - \partial_{\mu} \phi \partial^{\mu} \bar \phi - |\phi^2 - a^2|^2.
\end{equation}
The potential of the theory, restricted to the real part of $\phi$, has been plotted in figure \ref{fig:BPS_DW}(a) (dashed line) where
we see the double well potential form with the two supersymmetric minima located at $\phi_{\pm} = \pm a$ . The second-order equations of motion \eqref{eq:2nd_order_eqs} read, in this case,
\begin{equation}
\partial^2_z \phi(z) - 2 \bar \phi (\phi^2 - a^2) = 0.
\end{equation}
On the other hand, the  first-order BPS  equation \eqref{eq:BPS} reads
\begin{equation}
\partial_z \phi(z) = -  (\bar \phi^2 - a^2)~,
\label{eq:first-order-eq-ex}
\end{equation}
where we have chosen  $W(\phi(z=\infty))<W(\phi(z=-\infty)$, which implies\footnote{On the other hand, $e^{i \theta} = +1$ would yield the mirrored profile, often denoted as the anti-domain wall. This simply corresponds to flipping the boundary conditions imposed at $\pm \infty$.} $e^{i \theta} = -1$. The solution of this equation is given by the real field configuration,
\begin{equation}
\phi (z) = a \tanh (a z),
\end{equation}
while the tension of this domain wall is given by
\begin{equation}
\sigma_{BPS} = 2 | \Delta W| =  \frac{8}{3} a^3 ~.
\end{equation}
	
As we mentioned earlier, any solution to the BPS equation preserves some supersymmetry by construction, so it is clear that
it cannot represent the decay of the vacuum. We can also see this noting that both vacua
are supersymmetric, degenerate in energy, and the wall is flat and infinite, so there is no
way these vacua can decay.
	
On the other hand, the solution we found here is purely real. This is consistent with the
potential we have constructed since its form is such that perturbations around the solution in the imaginary
field directions are stabilized. We can check this by expanding the potential in the real and imaginary parts of the field, namely
\begin{equation}
\phi (z) = \psi (z) + i \ s(z) \ ,
\end{equation}
so the potential reads
\begin{equation}
V(\psi,s)= (\psi^2 - a^2)^2 + 2(\psi^2+a^2) s^2 + s^4 \ .
\end{equation}
This is why we can concentrate on the solution along the $s=0$ line. In all of the examples
we show here, we have checked that this is indeed the case; therefore, in all of our illustrations
we will simply draw the results concerning the real part of the fields.

\subsection{Stability and Vacuum decay}
	
In the previous section we showed how one can find supersymmetric domain wall configurations
that interpolate between supersymmetric vacua in our model. The fact that these
vacua are stable is not surprising since supersymmetry imposes them to be degenerate global minima of the potential.
Let us now consider the case where there is a small supersymmetry breaking term in our
potential and study the stability of the resultant vacua.

\begin{figure}[t]
\centering
\subfloat[]{
\includegraphics[width=0.47\textwidth]{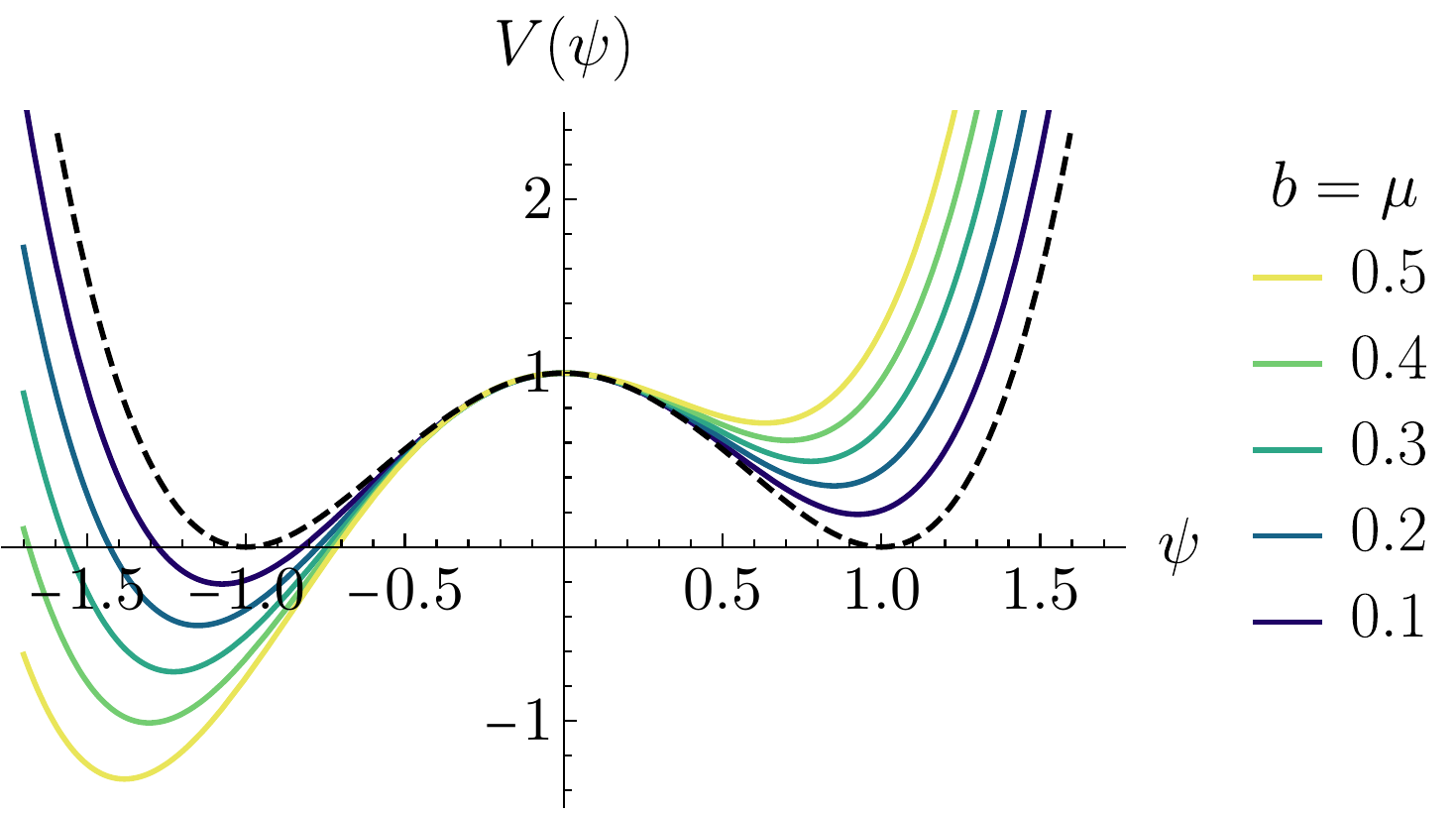}
}
\hfill
\subfloat[]{
\includegraphics[width=0.47\textwidth]{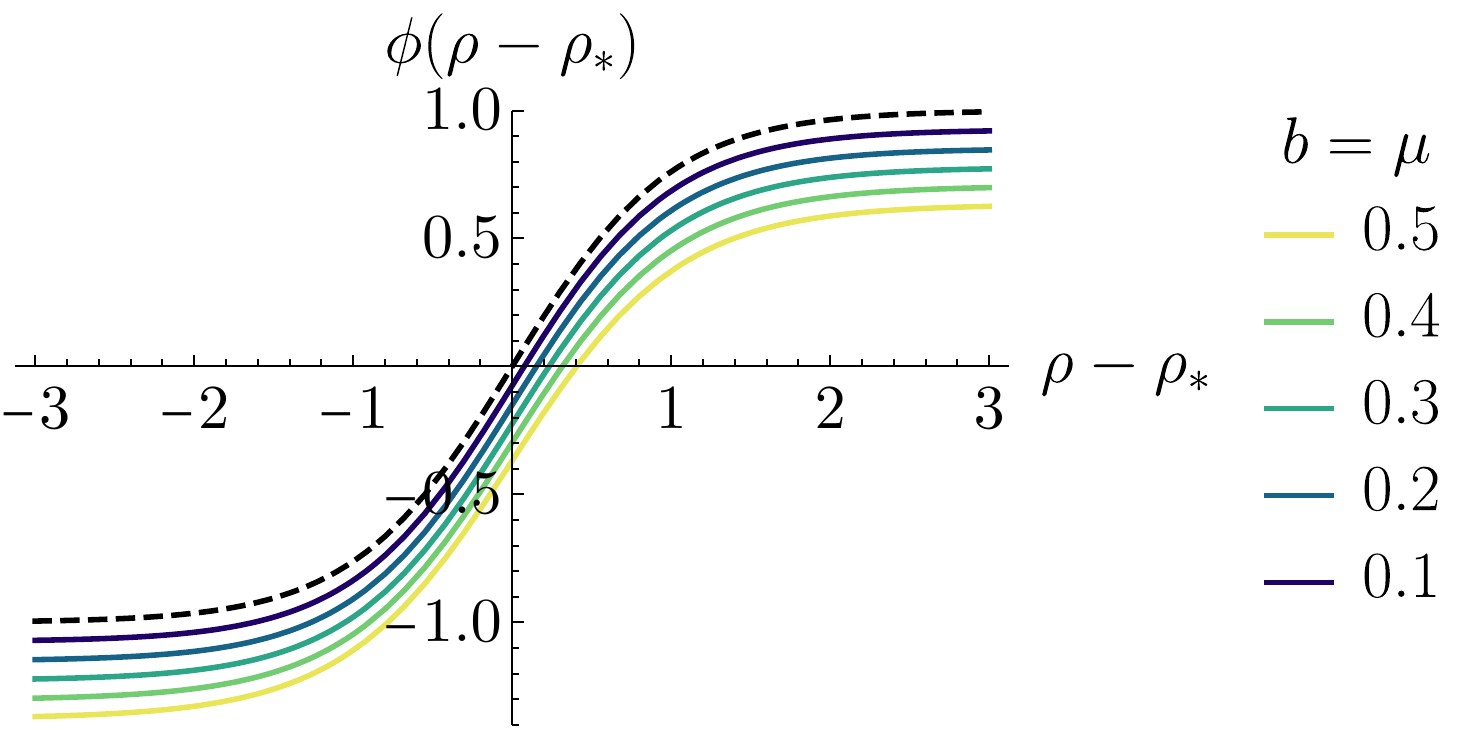}
}
\caption{(a) Potential \eqref{eq:V_soft} with $a=1$, for several supersymmetry-breaking parameters. (b) Solutions to \eqref{eq:bounce}, centered around the inflection point corresponding to each profile, labeled as $\rho_*$. The BPS limit is shown with a dashed line, representing the domain wall solution arising from the dashed potential in (a). }
\label{fig:BPS_DW}
\end{figure}

For simplicity let us assume that we introduce in the theory a couple of {\it soft supersymmetry
breaking terms} \cite{Chung:2003fi} of the form
\begin{equation}
S_{\text{soft}} = - \int{d^4 x \sqrt{-g} \left[ \mu^2 \phi \bar \phi + b \left( \phi^3 + \bar \phi^3\right)\right]}~,
\label{eq:susy_break}
\end{equation}
which break supersymmetry explicitly. In the following, we will consider the coefficients to be small enough so that many of the properties
of the solution found earlier will still hold. This means we will consider the case where $\mu^2 \ll a^2$
as well as $b\ll a$, see the coloured curves of \ref{fig:BPS_DW}(a), where for simplicity we have plotted some curves for the cases of $b=\mu$.
	
In this regime we can see that the theory still has two minima given by
\begin{equation}
\phi_{\pm} = \pm a + \delta_{\pm} (\mu^2,b) \ ,
\end{equation}
where the solutions have only shifted slightly, i.e. $|\delta_{\pm}/a| \ll 1$.
The interesting point now is that both of these minima break supersymmetry. If one takes $b>0$, the potential at $\phi_{+}$ becomes slightly
higher than the other minimum; this means that this vacuum will be unstable
with respect to the nucleation of bubbles of the true vacuum at $\phi_{-}$. Furthermore,
the form of the supersymmetry-breaking terms allows for the tunneling to happen along the real
direction of the field, see fig.~\ref{fig:BPS_DW}(a).

%%%%%%%%%%%%%%%%%%%%%%%%%%%%%%%%%%%%%%%%
%%%%%%%%%%%%%%%%%%%%%%%%%%%%%%%%%%%%%%%%

In order to compute the probability of the decay and its bubble profile in terms of the scalar field, we will resort
to the usual methods developed by Coleman and collaborators \cite{Coleman:1977py,Callan:1977pt} in the
context of False Vacuum Decay. Thus, we only have to extremize the Euclidean action associated to the
Lagrangian \eqref{eq:scalar_theory} completed by the supersymmetry-breaking terms.
Assuming a standard kinetic term ($K=\phi\bar{\phi}$) and an $O(4)$ symmetry in Euclidean space, the equations of motion for the field are given by
\begin{equation}
\phi'' + \frac{3}{\rho} \phi' = \frac{\partial V}{\partial \bar{\phi}}
\label{eq:bounce}
\end{equation}
where $\rho = \sqrt{\tau^2 + \vec{x}^2}$, $\tau$ being  the Euclidean time,  and primes denote derivatives
with respect to $\rho$. This equation is to be solved considering the boundary conditions
\begin{align}
\lim_{\rho \to \infty} \phi (\rho) = \phi_+ \ , \qquad \phi'(0) = 0.
\label{eq:bounce_boundary_conds}
\end{align}
Note that the Euclidean $O(4)$ symmetry will turn to $O(1,3)$ when transporting the solution back to
Lorentzian spacetime. Among other things, this will mean that the profile $\phi(\rho)$ obtained via \eqref{eq:bounce}
will correspond to the emergent scalar profile for the bubble at formation. Furthermore, this symmetry implies
that the radius of the bubble so formed will expand outward following a constant acceleration trajectory.
	
Following the example presented in the previous subsection, we will work with the following potential
\begin{equation}
V(\phi, \bar{\phi}) = (\phi^2 - a^2)  (\bar\phi^2 - a^2) + \mu^2 \phi \bar{\phi} + b (\phi^3 + \bar{\phi}^3)
\label{eq:V_soft}
\end{equation}
which already includes a contribution from supersymmetry-breaking terms within its definition.
	
In this one-dimensional setup, the profile of the scalar field can be easily found in Euclidean radial
coordinates using an \emph{undershoot/overshoot} algorithm \cite{Coleman:1977py}. Essentially, since
the boundary conditions \eqref{eq:bounce_boundary_conds} do not specify the initial starting point
$\phi (0)$, we can first obtain a couple of points where the field either undershoots or overshoots the
 false vacuum at $\phi_{+}$. Iteratively reducing this field range, we will eventually
 find a starting point $\phi (0)$ which stays sufficiently close to the true vacuum at $\phi_{-}$ for large values of $\rho$. The
 solutions found this way have been contrasted with ones obtained using the software \verb|AnyBubble| \cite{Masoumi:2016wot}, which  applies an alternative multiple shooting method,
 and have been shown to be essentially identical.
	
Using this numerical algorithm, we can check that indeed the form of the
domain wall forming the bubble does not change qualitatively
in comparison to the supersymmetric case given above. In order to visually
make this comparison we have plotted in figure \ref{fig:BPS_DW}(b) the resultant
domain wall profiles for the bubble nucleation centered on the same point.
It is clear that dialling back to zero the supersymmetry breaking coefficients $b, \mu$
one recovers the profiles obtained with the first-order BPS equation \eqref{eq:first-order-eq-ex}, i.e., the supersymmetric domain wall profile. Of
course, this is consistent since in this case the bubble radius is infinite
and the tunnelling rate would be zero, as this limit takes us back to the
stable supersymmetric vacua presented in the previous section.

\section{Rigid supersymmetric models with 3-form potentials}
\label{sec:susy_membranes}

As we described in the introduction, we are interested in studying supersymmetric
models with some new degrees of freedom beyond the complex scalar
field presented in the previous section. In particular, we would like to
introduce a 3-form potential into our model. Actually, using supersymmetry as a guiding principle, we can describe both the scalar field and the 3-form gauge field into a single multiplet described in terms of a special type of chiral superfield.
We show in Appendix \ref{ch:3f_multiplets} how one can construct such a  special chiral superfield
describing the so-called {\it single 3-form supermultiplet} from unconstrained real scalar superfields. In this case, the bosonic content of the multiplet is simply
given by complex scalar fields and auxiliary fields, the real part of which will be related
to a field strength of the 3-form potential. Following this prescription we find that
the simplest bulk action for such fields is given by
\begin{align}
\label{bulk-action}
	S_{\text{bulk}} &= \int d^4 x \left[ - K_{\phi \bar \phi} ~\partial_{\mu} \phi \partial^{\mu} \bar \phi - \frac{1}{4\cdot 4!} K_{\phi \bar \phi} ~F^{\mu \nu \sigma \rho} F_{\mu \nu \sigma \rho} + \frac{1}{2\cdot 4!} \left(  W_{\phi} + \bar{W}_{\bar \phi}\right) ~\epsilon^{\mu \nu \sigma \rho} F_{\mu \nu \sigma \rho} \right. \nonumber\\
	&\hspace{50pt} \left. + \frac{1}{4}  K^{\phi \bar{\phi}} \left(  W_{\phi} - \bar{W}_{\bar \phi}\right)^2 \right] \ ,
\end{align}
where  $F_{\mu \nu \rho \sigma} = 4 \partial_{[\mu}A_{\nu \rho \sigma]}$ and as before $K$ and $W$ denote the Kähler potential and superpotential
which define the model for the complex scalar field $\phi(x)$.\footnote{For the sake of simplicity, will study models consisting of a single scalar field here. See  \cite{Farakos:2017jme} for a study of similar systems with an arbitrary number of scalar fields .}

As is usually the case in models involving gauge fields, this action must be supplemented with some boundary terms that are required
to make the variation of the action well  posed \cite{Brown:1988kg,Dyer:2008hb}. In our present case, the boundary term is given by
\beq\label{Sbd=}
S_{\text{bd}} = \frac{1}{2 \cdot 3!}  \int d^4 x \, \partial_\mu \left[ A_{\nu \rho \sigma} \left( K_{\phi \bar{\phi}} F^{\mu \nu \rho \sigma} - \epsilon^{\mu \nu \rho \sigma} \left( W_{\phi} + \bar{ W}_{\bar{ \phi}} \right) \right) \right] \,,
\eeq
see appendix \ref{ch:3f_multiplets} for a derivation of this result.

The equation of motion for the 4-form field strength can be written as
\beq
\partial_{\mu} \left[K_{\phi \bar{\phi}} F^{\mu \nu \rho \sigma} - \epsilon^{\mu \nu \rho \sigma} \left( W_{\phi} + \bar{ W}_{\bar{ \phi}} \right)   \right] = 0 ~,
\eeq
which can be integrated to give,
\beq\label{F4=}
F^{\mu \nu \rho \sigma} = K^{\phi \bar{\phi}} \epsilon^{\mu \nu \rho \sigma} \left( W_{\phi} + \bar{ W}_{\bar{ \phi}} - 2 n \right) ~,
\eeq
where $n\in \mathbb{R}$ is a real integration constant. This expression allows us to integrate out the 3-form potential from the original action
to obtain a new effective theory written in terms of the scalar field alone, namely,
\beq\label{S=W-n}
S =  \int d^4 x \left[  - K_{\phi \bar \phi} \ \partial_{\mu} \phi \partial^{\mu} \bar \phi - K^{\phi \bar \phi} \left(W_{\phi} - n\right) \left(\bar W_{\bar \phi} - n\right)\right]\; .
\eeq
It is important to stress that to arrive at this result the contribution coming from
the boundary term \eqref{Sbd=} on the surface of 3-form equations \eqref{F4=} have to be taken into account.

Thus, quite interestingly, when the 3-form is set on-shell, the action reduces to the usual theory for a complex
scalar field, albeit with a potential modified by a constant of motion that parametrizes the value of the 4-form
flux. As we mention in the introduction, we will assume that our $4d$ theory is in fact obtained from
a higher dimensional theory so we will consider these fluxes to be quantized.

\subsection{Supersymmetric membranes coupled to 3-form potentials}

Incorporating a 3-form field in our model immediately suggests the presence of sources
of this field in the theory. For 3-form fields, these should be fundamental membranes (2-branes), i.e., objects of codimension one in our 4d setup.
It is therefore clear that one should add new terms to the
action in order to describe the dynamics of these objects. In  \cite{Brown:1988kg}
the proposed terms were
\beq\label{S=t+A}
S_{\text{memb (BT)}} = - T_m \int_{\mathcal{M}} d^3 \xi \sqrt{-h} ~ + ~\frac{q}{3!} \int_{\mathcal{M}} d^3 \xi A_{\mu \nu \rho}
\frac{\partial x^{\mu}}{\partial \xi^{a}} \frac{\partial x^{\nu}}{\partial \xi^{b}} \frac{\partial x^{\rho}}{\partial \xi^{c}} \epsilon^{abc} ~,
\eeq
where $\epsilon^{abc}$ denotes the fully antisymmetric unit tensor. The physical origin of each these terms in the previous action is pretty clear. The first one describes a
Nambu-Goto-like contribution to the action, where $T_m$ represents the tension of the membrane and $h$  is the determinant
of the induced metric on the brane whose worldvolume is parametrized  by the coordinates $\xi^a=(\xi^0,\xi^1,\xi^2)$, that is,  $h={\rm det} \left( g_{\mu\nu} \frac{\partial x^{\mu}}{\partial \xi^{b}} \frac{\partial x^{\nu}}{\partial \xi^{b}}\right) $. The second term shows that, as indicated earlier, the branes are sources for the 3-form potential,
in other words, they are coupled (minimally or electrically) to $A_{\mu \nu \rho}$ and carry a charge $q$ with respect to this field.
In the case of supersymmetric membranes (i.e., supermembranes) this second contribution to the action \eqref{S=t+A} is also known as the Wess-Zumino term.

However, our model is a little bit different from the one in  \cite{Brown:1988kg} since
we are also interested in studying the dynamics of the complex scalar field and its influence on the supermembrane. Therefore, one needs to consider
the possibility of having some new couplings of these branes to the scalar fields.
Here we will look for supersymmetry as a guiding principle and consider the most
generic supersymmetric action that describes our bulk model given by eq.~\eqref{bulk-action}
coupled to a membrane. This problem was already addressed in \cite{Bandos:2012gz}
where the authors concluded that the appropriate form of the brane terms
in a global supersymmetric action (after setting all fermions to zero) should be
\beq
S_{\text{memb.}} = - \int_{\mathcal{M}} d^3 \xi  \sqrt{-h} ~ 2|q \phi| ~ + ~\frac{q}{3!} \int_{\mathcal{M}} d^3 \xi A_{\mu \nu \rho}
\frac{\partial x^{\mu}}{\partial \xi^{a}} \frac{\partial x^{\nu}}{\partial \xi^{b}} \frac{\partial x^{\rho}}{\partial \xi^{c}} \epsilon^{abc} ~.
\label{eq:s_memb}
\eeq
Note that the constant parameter in the previous action, $T_m$, has been replaced by a scalar field dependent tension proportional to the  modulus of the 3-form charge $q$. This means that our branes will be coupled to the scalar field as well. As we
will see in the following, this will have important consequences for the scalar field profile in the
presence of these branes. Moreover, since the tension of the brane is also linked to the value of $q$, this is the only independent parameter of the membrane action.

In the following we will consider the bosonic part of the  full action given by the combination of all the terms
given above, namely,
\beq\label{ST=}
S_T = S_{\text{bulk}} + S_{\text{bd}} + S_{\text{memb.}}~.
\eeq

The equation of motion for the 3-form field in the presence of these sources gets now modified to be,
\beq
\partial_{\mu} \left[K_{\phi \bar{\phi}} F^{\mu \nu \rho \sigma} - \epsilon^{\mu \nu \rho \sigma} \left( W_{\phi} + \bar{ W}_{\bar{ \phi}} \right)   \right] =   \, -2q\,  \int d^3 \xi~ \frac{\partial x^{\nu}}{\partial \xi^{a}} \frac{\partial x^{\rho}}{\partial \xi^{b}} \frac{\partial x^{\sigma}}{\partial \xi^{c}} \epsilon^{abc}  \delta^4 (x - x(\xi))~.
\eeq
Adding this source, one can show that the solution of the field strength will
change as we cross the membrane by a quantity that is proportional to the
charge, $q$. In other words, the solution is given by
\beq
F^{\mu \nu \rho \sigma} = K^{\phi \bar{\phi}} \epsilon^{\mu \nu \rho \sigma} \left( W_{\phi} + \bar{ W}_{\bar{ \phi}} - 2 (n + q H(x))\right) ~,
\eeq
where the function $H(x)$ is defined implicitly as
\beq
\label{H-definition}
\partial_{\mu} H(x) = J_{\mu}=  \frac{1}{3!}\epsilon_{\mu\nu\rho\sigma} \int d^3 \xi~ \frac{\partial x^{\nu}}{\partial \xi^{a}} \frac{\partial x^{\rho}}{\partial \xi^{b}} \frac{\partial x^{\sigma}}{\partial \xi^{c}} \epsilon^{abc}  \delta^4 (x - x(\xi)) \, .
\eeq
Thus, for a flat membrane perpendicular to the $z$ axes and placed at $z=0$ this function reduces to the usual Heaviside step function that changes its value at the surface of the brane, namely,
\begin{equation}
H(z)=\Theta(z)=\begin{cases}
1, \quad z>0\; , \cr
0, \quad z<0 \end{cases}.
\end{equation}
Using this solution we can
rewrite the effective action in terms of the scalar field alone to give
\begin{equation}
S = \int d^4x \sqrt{-g} \left[-K_{\phi \bar \phi} ~\partial_{\mu} \phi \partial^{\mu} \bar \phi - K^{\phi \bar \phi} |\hat{W}_{\phi}|^2  \right] - \int_{\mathcal{M}} d^3 \xi  \sqrt{-h} ~ 2|q \phi| \,,
\label{eq:on_shell_L}
\end{equation}
where $\hat{W} (\phi)$ is the following effective superpotential:
\begin{equation}\label{hW=}
\hat{W}(\phi) \equiv W(\phi) - (n + q H (x))\phi .
\end{equation}
Just as in the case with no membrane source (see eq.~\eqref{S=W-n}), the boundary term contributes essentially
to the effective action. Furthermore, in this case, the Wess-Zumino term of the membrane action can be shown to be absorbed in the above effective superpotential.

Let us now consider a flat brane located at $z=0$. The form of the effective superpotential
indicates that the membrane separates two regions of space where the potential is
different. However, one can still find a solution of the second order equations
that satisfies the first order BPS equations, namely a solution of the
form,
\begin{equation}
\partial_z \phi(z) = e^{i \eta} K^{\phi \bar \phi} ~\hat{\bar{W}}_{\bar{\phi}} (\bar \phi(z)),
\label{eq:BPS-2}
\end{equation}
with the condition that
\begin{equation}
\label{eq:BPS_susy}
e^{i \eta} = \frac{\Delta \hat W}{|\Delta \hat W|} = - \left. {\frac{q \phi}{|q \phi|}}\right|_{z=0}
\end{equation}
As it was shown in \cite{Bandos:2018gjp}, these configurations preserve part of the supersymmetry, see Appendix \ref{sec:BPS_eqs} for a discussion of this point.
Furthermore, equation \eqref{eq:BPS-2} requires the derivative of the field to have a discontinuity at the position of the
brane given by,
\begin{equation}
\left[ \partial_z \phi(z) \right ]_{|z=0} = - q e^{i \eta} K^{\phi \bar \phi}
\end{equation}
where we have introduced the notation for a jump of any quantity at some point as $\left[ A \right]_{|z=0} = A|_{z\rightarrow 0^+} - A|_{z\rightarrow 0^-}$. 
One can easily check that this is indeed consistent with the second order equations of motion
obtained directly from the action given in eq. (\ref{eq:on_shell_L}). In the following we will present a simple example for such a membrane.

\subsection{Example with quadratic superpotential}

\begin{figure}[t]
\centering
\subfloat[]{
\includegraphics[width=0.45\textwidth]{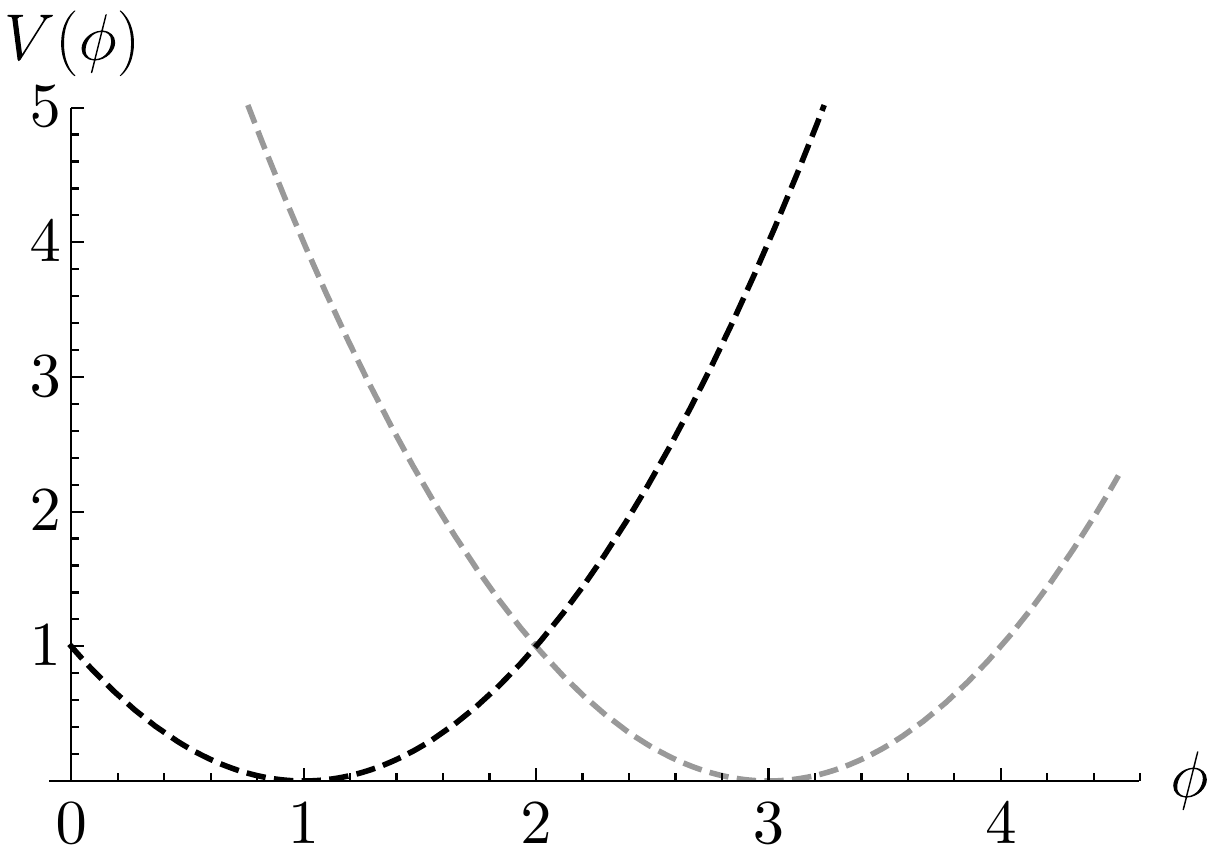}
}
\hfill
\subfloat[]{
\includegraphics[width=0.45\textwidth]{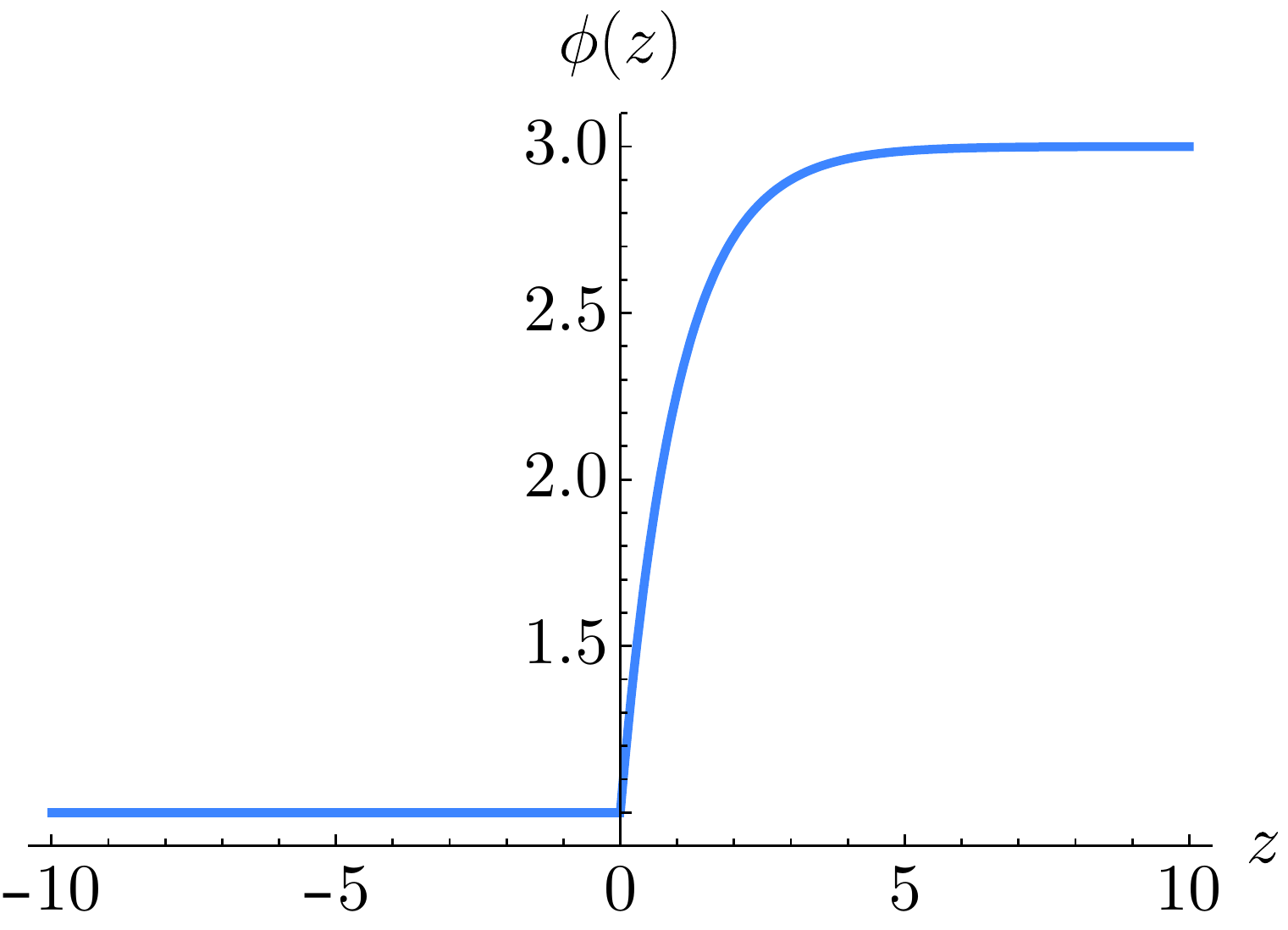}
}
\caption{(a) Scalar potential, corresponding to eq.~\eqref{eq:BPS_pot_flux}, with $n=1$ and $q=2$. The darker line represents the potential for $z<0$, while the lighter one is for $z>0$. (b) Scalar field profile interpolating between the minima of the previous potential, with a membrane sitting at $z=0$, see eq. \eqref{eq:BPS_sol}.}
\label{fig:BPS_pot_flux}
\end{figure}

Let us consider the simplest model where we can study the type of solutions discussed earlier. In particular, let us investigate the following Kähler potential and superpotential\footnote{See \cite{Lanza:2019nfa} for a complete treatment of this particular example.}
\begin{equation}
K(\phi, \bar \phi) = \phi  \bar \phi,  \qquad W(\phi) = \frac{1}{2} a \phi^2,
\label{eq:basic_susy_model}
\end{equation}
where $a$ is a constant with dimensions of energy which we will take for simplicity to be $a=1$. Then, following the description of the previous section, we integrate out the 3-form gauge field in the action \eqref{ST=}. We thus arrive at the effective action \eqref{eq:on_shell_L} where the effective superpotential is given by
\begin{equation}
\hat{W}(\phi) = \frac{1}{2} \phi^2 - (n + q ~\Theta (z))\phi \,
\end{equation}
where $q\in\mathbb{R}$ is the membrane charge and  $n\in \mathbb{R}$ is the constat 3-form flux.
This means that the scalar field potential at each side of the membrane is given by
\begin{equation}
V_{-} (\phi) = |\phi -n|^2, \qquad \qquad V_{+} (\phi) = |\phi - (n+q)|^2
\label{eq:BPS_pot_flux}
\end{equation}
for $z<0$ and $z>0$, respectively. We show in figure \ref{fig:BPS_pot_flux}(a) the
potential around their respective supersymmetric minima at $\phi_{-}= n$ and $\phi_{+} = n+q$. As explained
above, the perturbations of $\phi$ around its imaginary part are completely stabilized, and thus we will only
consider the physics of its real part. This implies in our case that $e^{i \eta} = \pm1$. We will take the
solution with $e^{i \eta} = -1$.
	
The BPS equation \eqref{eq:BPS_susy} can be now split into
	\begin{equation}\label{dphi=n+q}
	\left\lbrace\begin{array}{lr}
	\partial_z \phi(z) = - \bar{\phi} (z) + n & \quad z<0 , \\[5pt]
	\partial_z \phi(z) = - \bar{\phi} (z) + n +q  & \quad z>0 .
	\end{array}\right.
	\end{equation}
	This system is easily solved by considering the field to be real, imposing continuity of $\phi$ at $z=0$, and taking the field to asymptotically approach the supersymmetric minimum of the potential at each side of the membrane. This yields
	\begin{equation}
	\phi (z) = n + \Theta(z) q \left( 1 - e^{-z}\right)
	\label{eq:BPS_sol}
	\end{equation}
	which we have plotted in figure \ref{fig:BPS_pot_flux}(b). Indeed, the profile interpolates between the two different minima $\phi_{\pm}$ of their
	respective quadratic potentials on both sides of the brane. Notice the jump in the first derivative of the scalar field across the brane; as we will see below, this is entirely due to the tension of the membrane involving the scalar field $\phi$.
	
One can also compute the tension of the membrane dressed with the scalar field, by integrating the energy of the whole system across the membrane.
%&&%
In the case at hand this becomes
\begin{equation}
\sigma_{\text{DW+memb}} =  2 \left|\hat{ W}_{z=\infty} - \hat{ W}_{z=-\infty}\right|=
2 \left|\frac{1}{2} (n+q)^2 -   \frac{1}{2} n^2 -2nq-q^2\right| = 2nq + q^2
\label{eq:BPS_memb_T}
\end{equation}
where we have noted that the superpotential is actually different at each side of the
membrane. Note that this value is different from the one inferred from the Nambu-Goto
contribution of the membrane which is $T_{\text{NG}} = 2nq$. If we consider the
fact that the original flux ($n$) should be quantized in units of $q$ and assume
the situation where $n\gg q$, we notice that the correction to the tension is
therefore small compared to the NG one.

\subsection{Membrane nucleation}

\label{sec:flat_tunnel}
	
Now that we have studied the interplay of scalar fields and fluxes in the presence of a static membrane, we are ready to take another step forward. In the following, we will analyze how the Coleman-de Luccia \cite{Coleman:1977py,Coleman:1980aw} and Brown-Teitelboim \cite{Brown:1987dd,Brown:1988kg} schemes can be combined to yield a very interesting perspective on membrane nucleation in the presence of scalar fields.
	
In order to have a potential with non-degenerate vacua, we will add soft su\-per\-sym\-me\-try-breaking terms to the Lagrangian to get some
shifted new minima which will explicitly break su\-per\-symmetry. This will allow the false vacuum solution (the highest of the two minima) to decay to the
other vacuum by the formation of a membrane bubble that interpolates between
them. This membrane bubble will have a structure locally similar to the
flat membrane solution found earlier and should approach the supersymmetric profile as the supersymmetry-breaking terms are dialed down.

In order to study the tunnelling from a field configuration in the false vacuum everywhere in space to a state with a spherical membrane coupled to the field, we will apply Euclidean methods and assume $O(4)$ symmetry as in the usual false vacuum decay process. In this case, the Euclidean
action for our problem becomes,
\begin{align}
S_E = 2 \pi^2 \int d\rho \ \rho^3 \left[ K_{\phi \bar{\phi}} \left| \frac{d\phi}{d\rho} \right|^2 + V(\phi,\bar{\phi}) + 2 \left| q \phi \right| \delta(\rho - R) \right]~,
\label{eq:eu_action_memb}
\end{align}
where $\rho$ denotes the radial coordinate in 4-dimensional Euclidean space, and $V(\phi,\bar{\phi})$ is the potential for the scalar field which includes the contribution from the supersymmetry-breaking terms in \eqref{eq:susy_break}, namely
\begin{align}
V= K^{\phi \bar{ \phi}} \left| \hat{W}_{\phi} \right|^2 + \mu^2 \phi \bar{ \phi} + b (\phi^3 + \bar{\phi}^3), \qquad \hat{W} \equiv W - (n + q \Theta(r-R)) \phi~.
\end{align}
Finally, the last term of the action describes the Euclidean contribution due to the Nambu-Goto term of a spherical bubble of radius $R$.

The equation of motion for the complex scalar field is then
\begin{align}
\frac{1}{\rho^3} \partial_{\rho} \left( \rho^3 K_{\phi \bar{ \phi}} \frac{d\phi}{d\rho} \right) = K_{\phi \bar{ \phi} \bar{ \phi}} \left| \frac{d\phi}{d\rho} \right|^2 + \frac{\partial V}{\partial \bar{\phi}} - q e^{i \eta}  \delta (\rho - R)
\end{align}
where $e^{i\eta} = - \frac{q\phi}{|q\phi|}$. This equation is greatly simplified if the scalar field has a canonical kinetic term, namely if $K(\phi,\bar{\phi})=\phi \bar{ \phi}$ so that $K_{\phi \bar{ \phi}}=1$. In that particular case, the equation of motion for the scalar field reads
\begin{align}
\frac{d^2 \phi}{d\rho^2} + \frac{3}{\rho} \frac{d\phi}{d\rho} = \frac{\partial V}{\partial \bar{\phi}} - q e^{i \eta}  \delta (\rho - R) \ ,
\label{eq:bounce_membrane}
\end{align}
which should be supplemented with appropriate boundary conditions, namely the ones described in eq.~\eqref{eq:bounce_boundary_conds}. This last equation is quite similar to the one which is generally used for false vacuum decay, see eq.~\eqref{eq:bounce}. The only difference is that here we have derived it for a complex scalar field and that it has a contribution proportional to a Dirac delta function, due to the presence of a membrane of radius $R$ which appears to be charged with respect to this scalar.

To find the instanton solution in this case we should proceed with a little
bit of care since the potential for the scalar field will be different on both
sides of the wall. Most notably, the true and false vacuum of the theory
now belong to two different potentials, as opposed to the usual case where
they are both local minima of the same function.

We should integrate this equation starting at the true vacuum state at the center
of the bubble at $\rho=0$ up to some distance $R$. At that point we should
change the potential and integrate the new equations of motion with the new
form of the potential $V_{+}(\phi)$ ending up in the false vacuum state.
There is, however, another important effect coming from the Nambu-Goto
contribution to the action and its dependence on the field $\phi$. As we
can see in the flat membrane case, the scalar field undergoes a jump on its
first derivative at the brane (see eq.~\eqref{dphi=n+q}). In the case of the bubble profile we should
have the same situation. In fact, integrating \eqref{eq:bounce_membrane}
between $R-\epsilon$ and $R+\epsilon$, where $R$ is the radius of the Euclidean
membrane, and making $\epsilon \to 0$, assuming $\phi$ is everywhere continuous, we find
\begin{align}
\left[\partial_{\rho} \phi\right]_{|\rho=R} = - q e^{i \eta}
\label{eq:jump}
\end{align}
across the membrane. Taking this jump in the first derivative we can now integrate
the profile for the field outside of the bubble all the way to infinity.
	
This procedure for finding a solution of the Euclidean equation of motion with spherical $O(4)$ symmetry
has a free parameter: the radius of the membrane bubble, $R$. There are actually several
different methods one can use to fix it in this non-gravitational case. Here we will use one
that is easy to do numerically and use some of the other methods to check for the validity
of the solution we find.

The method we use is based on the numerical evaluation of
the Euclidean action \eqref{eq:eu_action_memb} at the solution as a function
of this parameter, $R$. Indeed, one may compute several profiles (and their corresponding Euclidean action) for several
values of $R$. Having done that, we only need to extremize this action with respect to $R$. Recall that the Euclidean
action describing the instanton of bubble nucleation must be a maximum in terms of the radius of the bubble, in the $R$ direction \cite{Coleman:1977py}.
\subsubsection{Numerical Example}

\begin{figure}[t]
\centering
\includegraphics[width=0.7\textwidth]{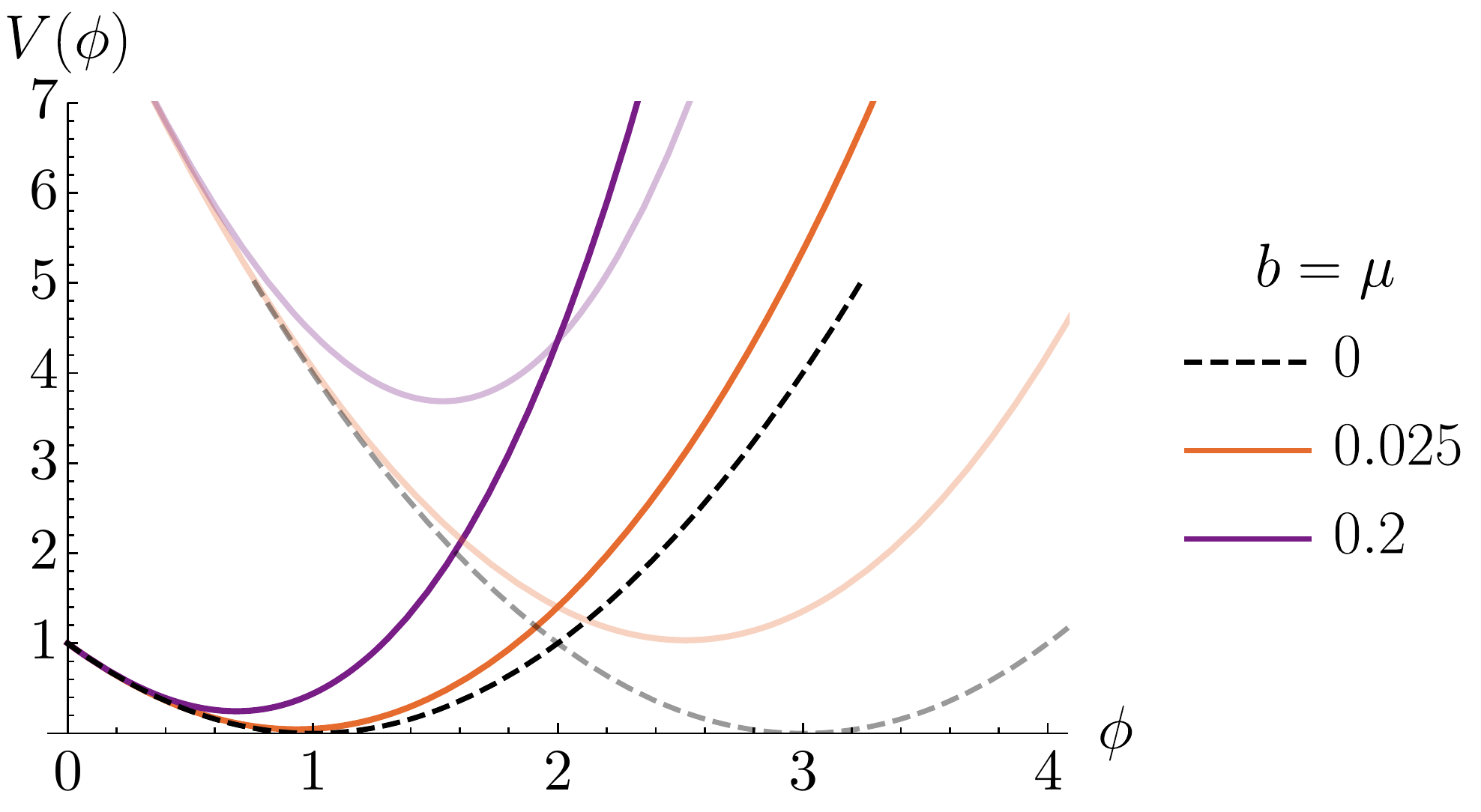}
\caption{Scalar potential \eqref{eq:V_break} for $n=1$, $q=2$ and some values of the supersymmetry-breaking
parameters. The darker curve shows the potential inside the membrane ($\rho < R$), while the lighter one represents the potential outside it ($\rho > R$).}
\label{fig:susy_potentials}
\end{figure}
	
In order to visualize the procedure described above we have implemented the numerical
integration of the equation of motion in the model defined by \eqref{eq:basic_susy_model}, albeit
with the inclusion of soft supersymmetry-breaking terms. In this case, the effective potential of the theory is
\begin{equation}
V(\phi, \bar{\phi}) = \left| \phi - \left(n + q \Theta (r-R) \right) \right|^2  + \mu^2 \phi \bar{\phi} + b (\phi^3 + \bar{\phi}^3)~,
\label{eq:V_break}
\end{equation}
which is shown in figure \ref{fig:susy_potentials} along the real part of $\phi$. Once again, we will not consider the imaginary
part of the scalar field, since the system is perturbatively stable along that direction. We are therefore interested in finding the
correct instanton that interpolates between the false vacuum at higher energy density generated by the supersymmetry
breaking terms and the true vacuum.
	
\begin{figure}[t]
\centering
\centering
\subfloat[]{
	\includegraphics[width=0.51\textwidth]{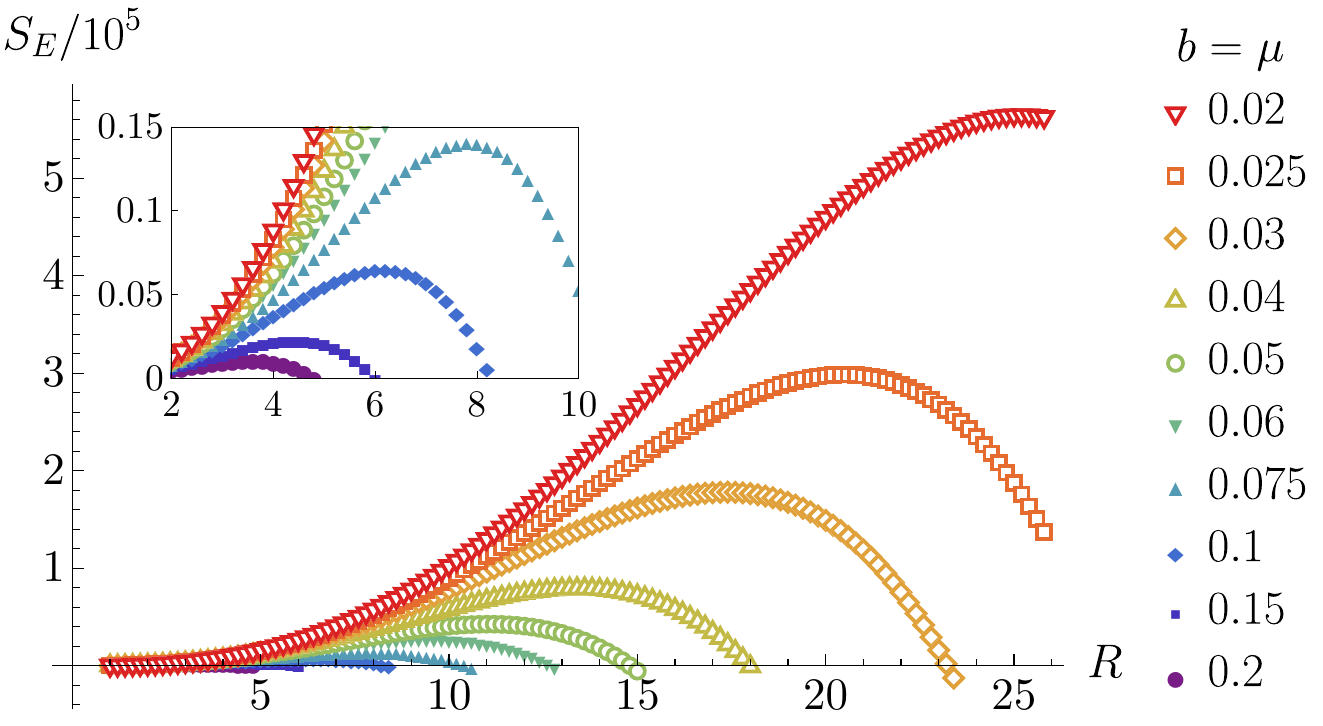}
}
\hfill
\subfloat[]{
	\includegraphics[width=0.44\textwidth]{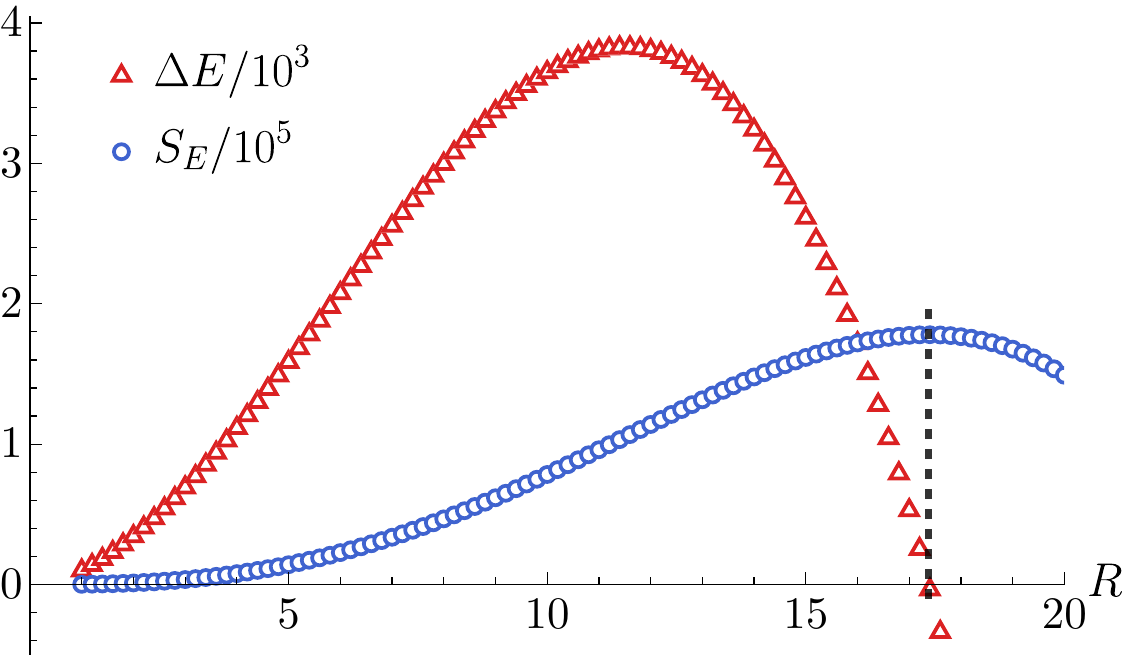}
}
\caption{(a) Euclidean action for solutions of the equation of motion \eqref{eq:bounce_membrane}, with $n=1$, $q=2$ and several values for $R$, $b$ and $\mu$. (b) Energy and Euclidean action difference with respect to the false vacuum state for several radii, in the case where $\mu=b=0.03$. The maximum of $B$ coincides with the root of $\Delta E$, as expected.}
\label{fig:actions_and_energy}
\end{figure}

As we explained in the previous section, the radius of the membrane corresponding to the instanton solution is a priori unknown.
Therefore, we have computed the scalar field profiles for the bubbles with different values of $R$. Using these profiles we can
now calculate their action as a function of $R$. We show the results of this procedure in fig.~\ref{fig:actions_and_energy}(a)  for particular
values of the supersymmetry breaking parameters $b$ and $\mu$. We note in these figures that the bounce action for
each model reaches a maximum for a particular value of the radius $R$. We take this to be the correct value
of the radius of the instanton that mediates the vacuum decay of interest to us.

We also checked the variation in the total energy of the profile with respect to the background at the time of nucleation. Indeed, there should be no energy loss or gain for the instanton solution at the time of its emergence. Thus, the correct bubble profile should correspond to the roots of the energy density difference with respect to the false vacuum background. This quantity is defined by
\begin{align}
	\Delta E =  4 \pi \int_0^\infty dr \ r^2 \left[ K_{\phi \bar{\phi}} \left| \frac{d\phi}{dr} \right|^2 + V(\phi,\bar{\phi}) + 2 \left| q \phi \right| \delta(r - R) - V_{\text{fv}} \right]_{t=0} \, .
\end{align}
This last expression can be easily evaluated from the Euclidean solution, since the profile obtained in Euclidean space corresponds to the profile of the emerging bubble in Lorentzian space at $t=0$. In all of the profiles that we have computed, the value of R corresponding to no energy loss or gain with respect to the background has been found to correspond with the maximum of the Euclidean action, see figure \ref{fig:actions_and_energy}(b) for an explicit example.
			
\begin{figure}[t]
\centering
\subfloat[]{
\includegraphics[width=0.48\textwidth]{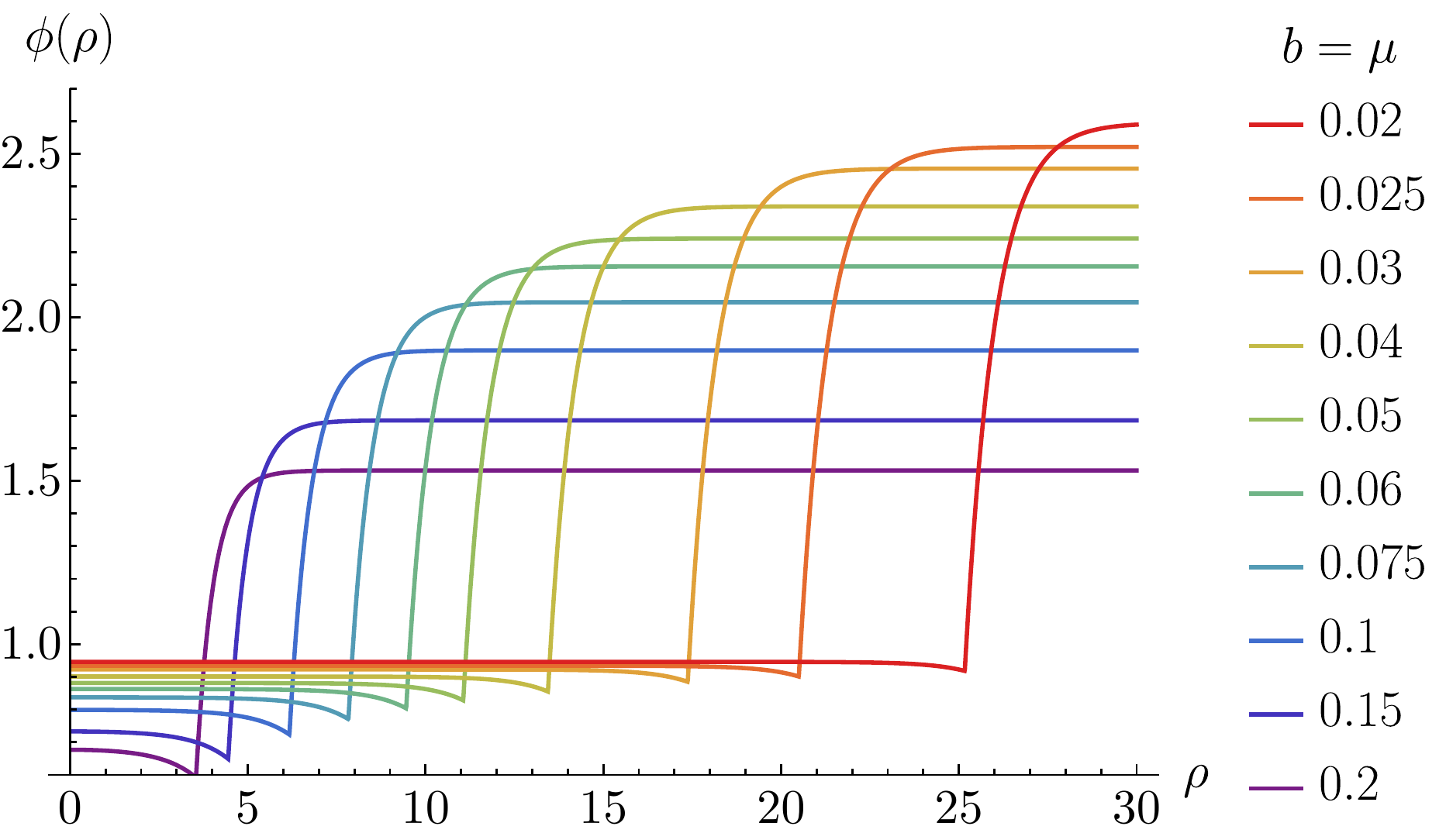}
}
\hfill
\subfloat[]{
\includegraphics[width=0.48\textwidth]{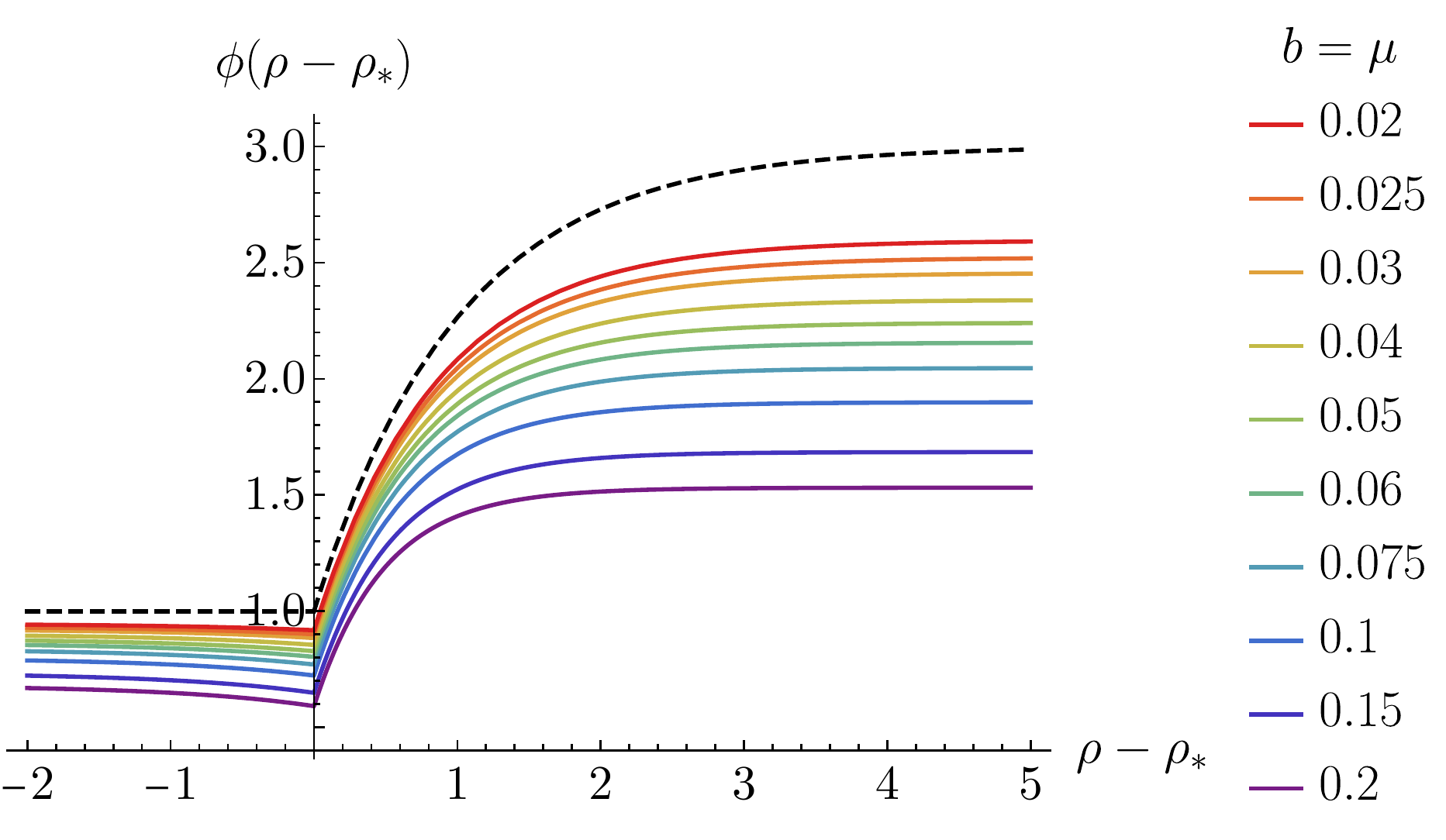}
}
\caption{(a) Scalar field profiles corresponding to each maximum Euclidean action for some supersymmetry-breaking parameters. (b) The same profiles as before, centered around the membrane radius corresponding to each profile, denoted by $\rho_{\ast}$. The dashed line represents the BPS solution from eq.~\eqref{eq:BPS_sol}, with $n=1$ and $q=2$, which is clearly the asymptotic behaviour of the profiles as the supersymmetry-breaking parameters $b$ and $\mu$ tend to 0. }
\label{fig:profiles}
\end{figure}

The profiles corresponding to the solutions which extremize $S_E$  have been plotted in figure \ref{fig:profiles}(a). As expected, as the supersymmetry-breaking parameters are made smaller, the radius of the emerging membrane increases and the scalar field profiles progressively tend towards the BPS solution derived above, as shown in fig.~\ref{fig:profiles}(b).
	
A curious feature of these profiles is that, when considered as a particle in the inverted potential $-V$, they first tend to get away from the false vacuum, only to then be projected in the $r>R$ potential with enough velocity to asymptotically reach the false vacuum.

\section{Supergravity models}
\label{sec:sugra_membranes}

	We now turn to study the same setup as in the previous section, with gravity taken into consideration. We will work in the context of $\text{N=1}$, $D=4$ Supergravity coupled to chiral matter. We will be interested in generalizing the false vacuum decay discussed in the previous section including gravity. However, before analysing more generic situations, we will first study the supersymmetric limit of flat membrane solutions in supergravity. Thus, in this section, we will start by analyzing the action of the system composed by scalar fields, real three-forms and flat membranes in a spacetime of Lorentzian signature.
	
	The action we will be considering is once again given by the sum of the following terms,
\begin{equation}\label{Sint=}
S = S_{\text{bulk}} + S_{\text{membrane}}+S_{\text{boundary terms}}.
\end{equation}
	
	As we review in appendix \ref{ch:3f_multiplets}, the bosonic part of the bulk action of the system, which includes gravity, scalar fields and 3-forms, is given by the following supergravity action
\begin{align}
S_{\text{bulk}}= \int d^4 x \sqrt{-g} &\left[ \frac 1 2 \, \mathcal{R} -  K_{\phi \bar{\phi}}\partial_\mu \phi \partial^\mu\bar{\phi} - \frac 1 3 e^{-K} (M +  K_{\bar{\phi}}\bar{\mathcal{F}})\, (\bar{M}+   K_{\phi} \mathcal{F} ) -  M\bar{W} - \bar{M} W \right. \nonumber \\[6pt]
&+ \left. e^{-K} K_{\phi\bar{\phi}}\mathcal{F}\bar{\mathcal{F}} + \mathcal{F} W_\phi +\bar{\mathcal{F}}\bar{W}_{\bar{\phi}} \right]
\label{eq:sugra_L}
\end{align}
where, as before, $K(\phi,\bar{\phi})$ and $W(\phi)$ describe the real K\"ahler potential and the holomorphic superpotential functions of the scalar field and
\begin{align}
\mathcal{F} = \frac 1 2 \left(D_\mu A^{\mu} + i{\rm d} \right)+\frac 2 3 \bar{\phi}{} M
+\frac 1 3 \phi \bar{M}.
\end{align}
 In the above equations we use  $M_{\text{Pl}}=1$ and denote the Ricci scalar by $\mathcal{R}$. Furthermore, $M$ is a complex scalar auxiliary field of the minimal supergravity multiplet, $\text{d}$ is a real scalar auxiliary field and $A^\mu$ is the Hodge dual of the three-form.  In our setting this latter field belongs to the matter supermultiplet, which also includes the scalar field\footnote{We note that a 3-form fields may also be included as an auxiliary field of the supergravity supermultiplet
\cite{Stelle:1978ye,Ogievetsky:1980qp,Gates:1980az,Grisaru:1981xm,Ovrut:1997ur,Kuzenko:2005wh,Bandos:2012da,Kuzenko:2017vil}. Even though this provides a more straightforward supersymmetric generalization of the
 Brown-Teitelboim construction \cite{Brown:1987dd,Brown:1988kg}, we will not consider it here since the advantage of
 inclusion of the 3-form in the matter supermultiplet is that this construction has a smooth and nontrivial flat space limit.
 }.

 Just as in the non-gravitational case, boundary terms must be included in the full action to ensure the variation of the action with respect to the form field is well posed. We leave their derivation for appendix \ref{sec:appendix_sugra}. Note that these boundary terms will also include the Gibbons-Hawking term \cite{Gibbons:1976ue} in order for the variational problem to be well defined also in the gravitational sector.
	
The action of the membrane can be fixed by requiring the interacting system \eqref{Sint=}, with fermionic terms taken into account in the first two terms, to remain invariant under a half of local supersymmetry after the bosonic membrane is included \cite{Bandos:2001jx,Bandos:2005ww}.
It reads
\begin{align}
S_{\text{memb.}} &= - \int_{\mathcal{M}}{d^3 \xi \sqrt{-h} ~ 2 e^{K/2} |q \phi|}~ + ~\frac{q}{3!} \int_{\mathcal{M}}{d^3 \xi A_{\mu \nu \rho}
\frac{\partial x^{\mu}}{\partial \xi^{a}} \frac{\partial x^{\nu}}{\partial \xi^{b}} \frac{\partial x^{\rho}}{\partial \xi^{c}} \epsilon^{abc}}
\end{align}
which as before describes the coupling between the membrane and the 3-form potential as well as the scalar field. Note however that in supergravity the Nambu-Goto term receives a correction from the exponential of the Kähler potential.\footnote{The exponential factor $e^{K/2}$ arises due to the super-Weyl rescaling and field redefinitions required to bring the action to Einstein frame, see appendix \ref{sec:appendix_sugra} for more detail.}
	
It is easy to check that setting the form field and auxiliary fields on-shell and taking into account the contribution of the boundary terms, the action reads \cite{Bandos:2018gjp}
\begin{align}
S = \int d^4x \sqrt{-g} \left[ \frac{\mathcal{R}}{2} - K_{\phi\bar{\phi}} \partial_\mu \phi \partial^\mu \bar{\phi} - V (\phi,\bar{\phi}) \right] + S_{\text{GH}} - \int_{\mathcal{M}} d^3 \xi  \sqrt{-h} \, 2 e^{K/2} |q \phi|
\label{eq:S_os_sugra}
\end{align}
where $S_{\text{GH}}$ represents the Gibbons-Hawking boundary term and $V (\phi,\bar{\phi})$ is the $\text{N=1}$, $D=4$ matter-coupled supergravity scalar potential of the form
\begin{align}
V (\phi,\bar{\phi}) = e^K \left[ D_{\phi} \hat{W} K^{\phi \bar{\phi}}  D_{\bar{\phi}} \hat{\bar{ W}} - 3 |\hat{W}|^2 \right]
\end{align}
with the usual Kähler-covariant derivative denoted by $D_\phi = \partial_\phi + K_\phi$. As in the global supersymmetric models
the information about the 3-form flux is encapsulated in the form of the  effective superpotential:
\begin{align}
\hat{W} = W - (n + q H(x))\phi~,
\end{align}
where, just as in the non-gravitational case, $H(x)$ is given by eq.~\eqref{H-definition}.
This modified superpotential will allow us to find a profile for the dressed membrane which interpolates
between supersymmetric minima of \emph{different} potentials.

\subsection{Flat membrane solutions in supergravity models}

Similarly to what we did in the flat space case, one can find smooth domain wall solutions in the supergravity
context. This has been extensively studied in the literature starting in  \cite{Cvetic:1992bf} (see also a more recent discussion
about this topic in \cite{Ceresole:2006iq} and references therein). Here we show how one can generalize these solutions to include
the presence of a supermembrane as it was described in \cite{Bandos:2018gjp}.
	
	Let us begin with the case of a flat membrane interpolating between two supersymmetric vacua. Let us assume this static membrane sits at $z=0$. In order to study the profile across such a membrane, we will assume the following ansatz for the metric\footnote{Note that with this choice of metric and in the so-called static gauge $x^a(\xi)=\xi^a$, for flat membrane, $z(\xi)=0$, we will have $\sqrt{-h} = \sqrt{-g}$, where the r.h.s. is calculated at $z=0$.} \cite{Ceresole:2006iq,Bandos:2018gjp}:
\begin{align}
ds^2 = e^{2D(z)} (-dt^2 + dx^2 + dy^2) + dz^2
\label{eq:flat_sugra}
\end{align}
so that $\sqrt{-g} = e^{3D(z)}$. Let us turn to study the equation of motion for the scalar field, which is
\begin{align}
\frac{1}{\sqrt{-g}} \partial_{\mu} \left( \sqrt{-g} K_{\phi \bar{ \phi}} g^{\mu \nu} \partial_\nu \phi \right) = K_{\phi \bar{ \phi} \bar{ \phi}} \partial_\mu \phi \partial^\mu \bar{\phi} + \frac{\partial V}{\partial \bar{ \phi}} + e^{K/2} \left[ K_{\bar{ \phi}} |q\phi| - q e^{i\eta} \right]\delta(z),
\end{align}
where $e^{i\eta}$ is defined as
\begin{align}
	e^{i\eta} = - \left. {\frac{q \phi}{|q \phi|}}\right|_{z=0} .
\end{align}
It is natural to assume that scalar field only depends on the transverse coordinate to the membrane, i.e., $\phi = \phi (z)$. In that particular case, the field obeys
\begin{align}
\partial_z (K_{\phi \bar{ \phi}} \partial_z \phi) + 3 K_{\phi \bar{\phi}} \ \partial_z D \ \partial_z \phi = K_{\phi \bar{ \phi} \bar{ \phi}} \left| \partial_z \phi  \right|^2 + \frac{\partial V}{\partial \bar{\phi}} + e^{K/2} \left[ K_{\bar{\phi}} \left| q \phi \right| - q e^{i\eta} \right] \delta (z).
\label{eq:2nd_order_field}
\end{align}

On the other hand, the Einstein equations for the metric can be combined to give
\begin{align}
\partial_z^2 D + 3(\partial_z D)^2 &= -V - e^{K/2} \left| q \phi \right| \delta (z)
\label{eq:2nd_order_scale}
\end{align}
		
Note that the deltas at $z=0$ will yield jumps in the first derivative of both $\phi$, as in the non-gravitational case,  and in the scale factor $D$.
	
A supersymmetric and static domain wall may interpolate between non-degenerate minima, since essentially the gravitational contribution may compensate the difference in scalar potential between both vacua \cite{Cvetic:1992bf}. If supersymmetry is partly conserved across the profile of the domain wall, then the minima are bound to be either Minkowski or AdS vacua (note, however, that no supersymmetric domain wall may interpolate between two Minkowski vacua when gravity is included \cite{Cvetic:1992bf}).
	
With these remarks at hand, the BPS equations acquire the form \cite{Cvetic:1992bf,Ceresole:2006iq,Bandos:2018gjp}
\begin{align}
\phi'(z) &= \mp e^{K/2} e^{i \arg (\hat{W})} K^{\bar{\phi} \phi} D_{\bar{\phi}} \hat{\overline{W}} \label{dzphi}\\ \label{dzD}
D'(z) &= \pm e^{K/2} |\hat{W}|
\end{align}
where primes denote derivatives with respect to $z$,
The second order equations are obeyed by the solutions of the  first-order BPS equations when suitable boundary conditions are imposed. In particular, taking the derivative of eq.~\eqref{dzD} and using \eqref{dzphi}, we find that the second order equation  \eqref{eq:2nd_order_scale} is satisfied provided\footnote{Actually this identification can be obtained straightforwardly as it implies that on the worldvolume of the membrane the supersymmetry preserved by the solution coincides with $\kappa$--symmetry of the supermembrane action \cite{Bergshoeff:1997kr}.}
 \begin{equation}\label{argW=eta}
\left. e^{i \arg (\hat{W})}\right|_{z=0} = \mp e^{i\eta}\; .
\end{equation}

	It is convenient, following \cite{Ceresole:2006iq}, to write these equations in terms of
	\begin{align}
	\mathcal{Z} \equiv e^{K/2} \hat{W}.
	\end{align}
	Note that the value of the scalar potential at supersymmetric critical points is then given by $V_{\text{susy}} = -3|\mathcal{Z}|^2$. In terms of $\mathcal{Z}$, the BPS equations simplify to
	\begin{align}
	\phi ' (z) &= \mp 2 K^{\bar{\phi} \phi} \partial_{\bar{\phi}} |\mathcal{Z}| \label{eq:phi'}\\[5pt]
	D'(z) &= \pm |\mathcal{Z}|\; . \label{eq:D'}
	\end{align}
The sign to use in the integration of the BPS equations is determined by the value of $|\mathcal{Z}|$ at $z\to \pm\infty$. In order to see this, let us firstly note that
\begin{align}
	\frac {d |{\mathcal Z}|}{dz}= (\phi^\prime \partial_\phi |{\mathcal Z}|+\bar{\phi}{}^\prime \partial_{\bar{\phi}}|{\mathcal Z}|) \mp e^{K/2} |q\phi| \delta (z) = \mp \left[ 4 K^{\phi\bar\phi}\partial_\phi |{\mathcal Z}| \partial_{\bar{\phi}}|{\mathcal Z}|+ e^{K/2}|q\phi| \delta (z) \right]
\end{align}
where, in the second step, we have used  \eqref{eq:phi'}. Then it is easy to observe that, if for example we choose the lower sign in eqs.~\eqref{eq:phi'} and \eqref{eq:D'}, which then become
\begin{align}
	\phi ' (z) &= 2 K^{\bar{\phi} \phi} \partial_{\bar{\phi}} |\mathcal{Z}| \label{eq:phi'=}\\[5pt]
	D'(z) &= - |\mathcal{Z}|\; , \label{eq:D'=}
	\end{align}
we find that the derivative of $|{\mathcal Z}|$ is positive
\begin{align}
   \frac  {d|{\mathcal Z}|}{dz}=  4 K^{\phi\bar\phi}\partial_\phi |{\mathcal Z}| \partial_{\bar{\phi}}|{\mathcal Z}|+ e^{K/2} |q\phi| \delta (z) > 0.
\end{align}
Therefore, $|{\mathcal Z}|$ must increase monotonically, which implies $|\mathcal{Z}|_{-\infty} < |\mathcal{Z}|_{+\infty}$. On the other hand, if $|\mathcal{Z}|_{-\infty} > |\mathcal{Z}|_{+\infty}$, the lower sign of the BPS equations eqs.~\eqref{eq:phi'}-\eqref{eq:D'} will apply.

Finally, it should be noted that in case $|\mathcal{Z}|$ has some root along $z$, the situation becomes a bit more complicated. Indeed, the signs must be swapped after crossing a root of the superpotential in order to have a positive overall tension of the domain wall \cite{Ceresole:2006iq,Bandos:2018gjp}. However, as noted in \cite{Cvetic:1992bf}, such cases do not correspond to the limiting case of a bubble instability. In fact the spacetime induced by these solutions is asymptotically quite different and resembles the one in the Randall-Sundrum scenario \cite{Randall:1999vf} with a single positive tension brane\footnote{We will defer the exploration of these types of solutions in our context for a future publication.}.

\subsection{Example: quartic superpotential}
	
\begin{figure}[t]
\centering
\includegraphics[width=0.7\textwidth]{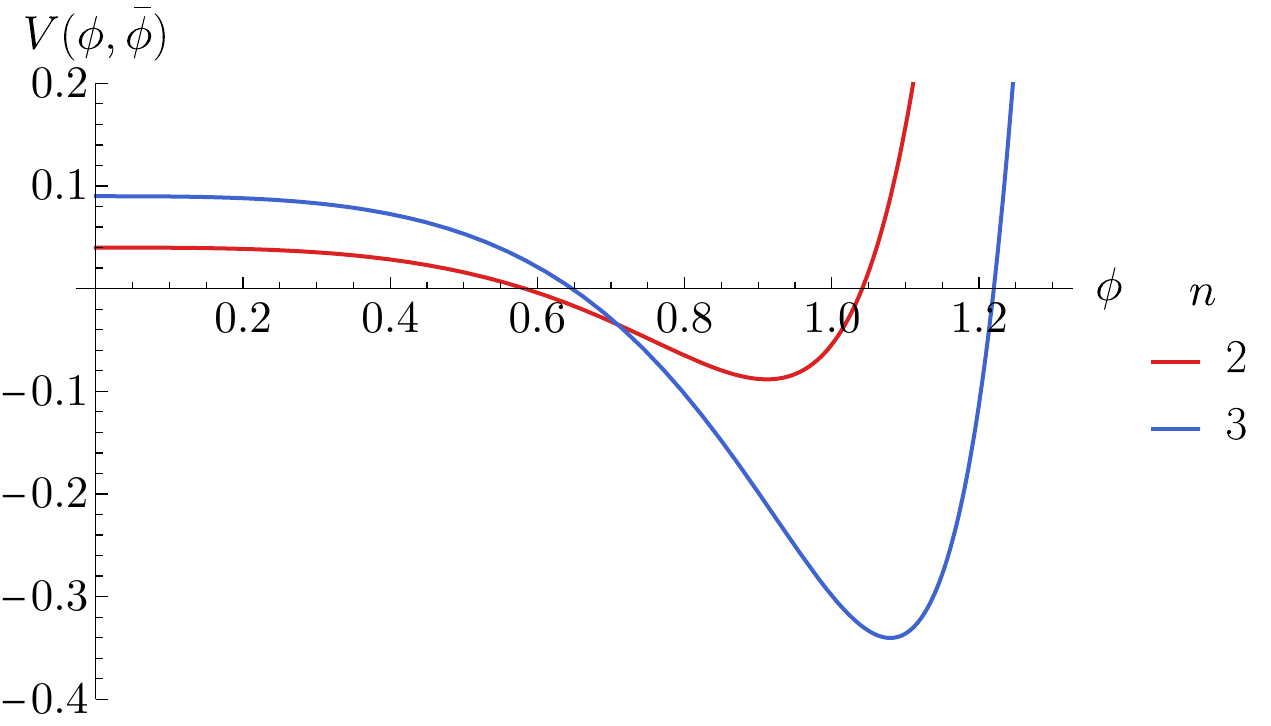}
\caption{Scalar potential for the model described in this section, for two different values of the flux value $n$. }
\label{fig:BPS_sugra}
\end{figure}
\begin{figure}[t]
\centering
\subfloat[]{
\includegraphics[width=0.48\textwidth]{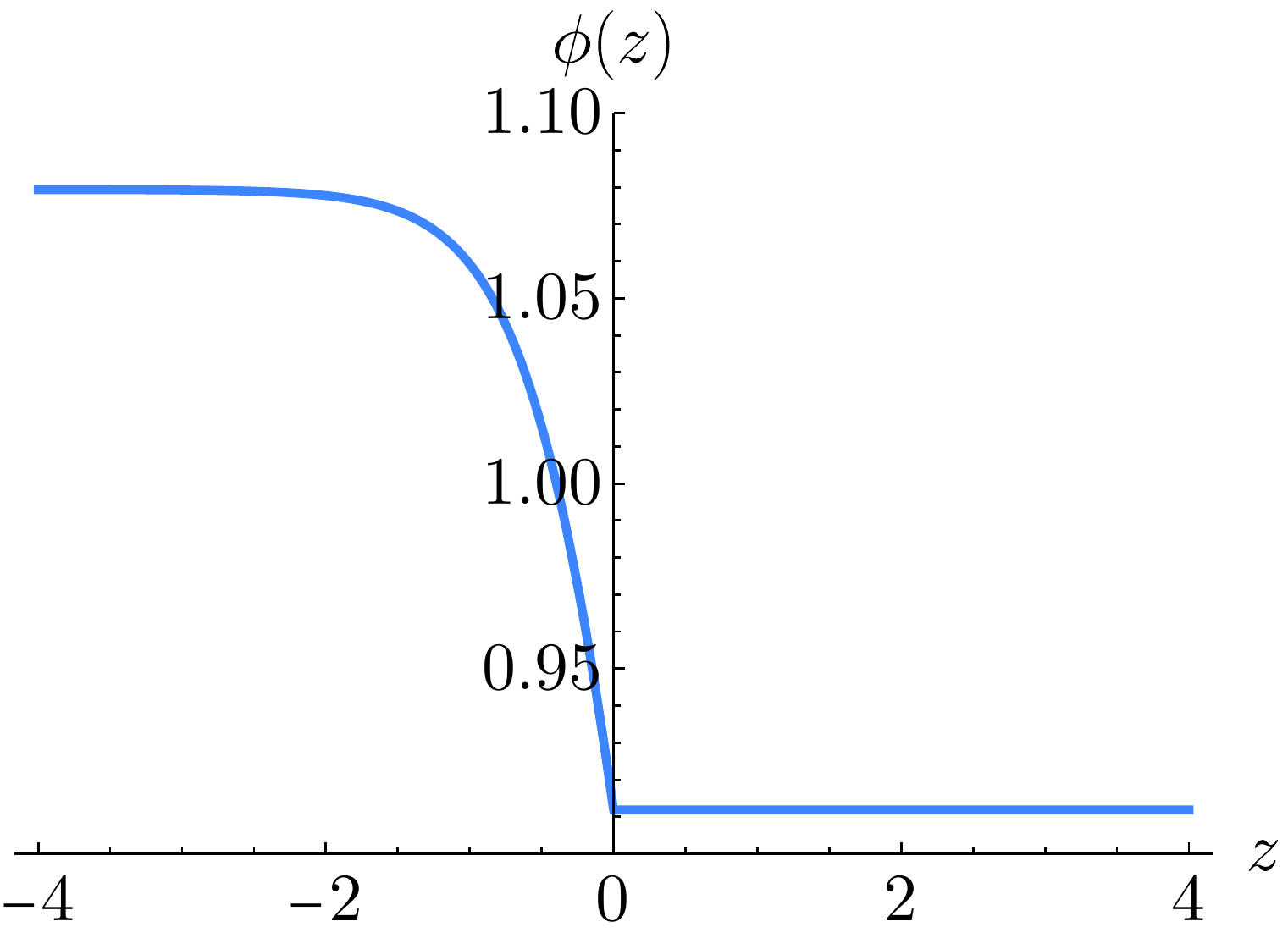}
}
\hfill
\subfloat[]{
\includegraphics[width=0.48\textwidth]{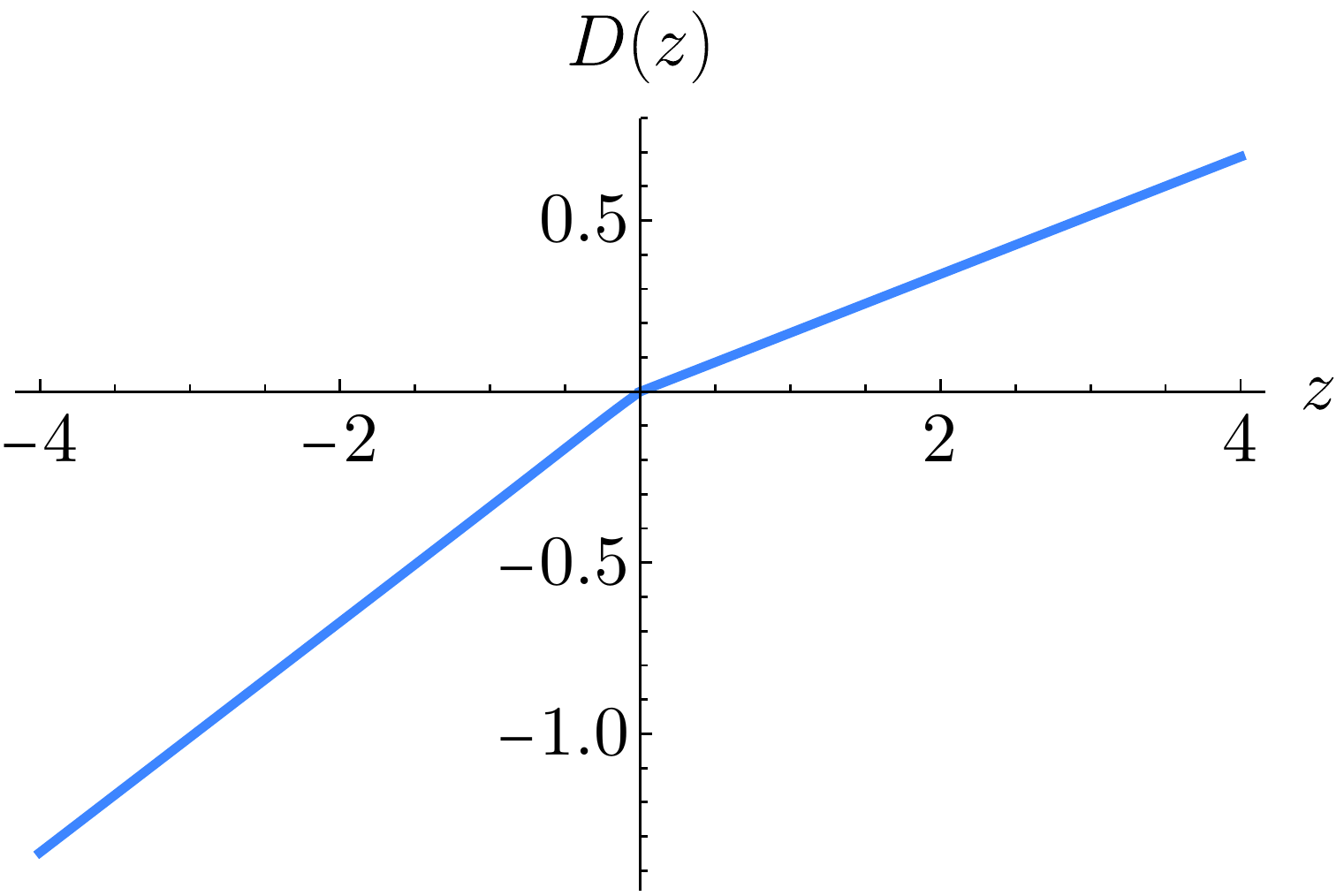}
}
\caption{Solution of the BPS equations to the model defined in \eqref{eq:sugra_model_BPS}, for (a) the scalar field (b) the scale factor, for the metric defined in \eqref{eq:flat_sugra}. The potential at $z<0$ corresponds to the case $n=3$, while $z>0$ corresponds to $n=2$; thus, the membrane is charged with $q=-1$. }
\label{fig:BPS_sugra_profs}
\end{figure}
	
The above equations can be used to find the profiles of a scalar field interpolating between different minima of the model defined by
\begin{align}
K(\phi,\bar{\phi}) = \phi \bar{\phi}, \qquad W(\phi) = (10 M_{\text{Pl}})^{-1} \phi^4.
\label{eq:sugra_model_BPS}
\end{align}
where we have restored the Planck mass momentarily in order to show explicitly the energy scales involved in this example. The scalar potential defined by this Kähler potential and superpotential is shown in figure \ref{fig:BPS_sugra}, once the 3-form has been integrated out. The minimum featured by each branch can be shown to be supersymmetric, i.e., it satisfies $D_\phi \hat{W} = 0$.
	
Going back to units where $M_{\text{Pl}} = 1$, the numerical BPS profile arising from this potential for both the scalar field and the scale factor D is shown in figure \ref{fig:BPS_sugra_profs}, where we have placed the lower minimum to the left (for easier comparison later on). We have also checked that the second-order equations \eqref{eq:2nd_order_field} and \eqref{eq:2nd_order_scale}, which explicitly incorporate the first-derivative jumps as Dirac deltas, yield exactly the same profiles. Note that for this particular solution the supersymmetric minima on both sides
of the wall describe an anti-deSitter vacua with different values of their cosmological constant.

\section{Membrane nucleation in Supergravity}
\label{sec:sugra_tunnel}

As we discussed in the introduction membrane nucleation in a model with a 4-form flux
has been studied in the literature in the context of models similar to Brown-Teitelboim \cite{Brown:1987dd,Brown:1988kg}.
Our supergravity model includes a 3-form field and a brane charged with respect to
it, however, as we have shown in the previous section it can also be casted exclusively in terms of a scalar field by integrating
out the 3-form potential. We can then ask whether there will be bounce solutions similar to
the flat space ones where the field interpolates between two minima of the flux-dependent
potential for the scalar field as one crosses the wall.

In order to illustrate these vacuum decay processes in our model, we will follow a similar
procedure to the one we presented in the flat space limit in section (\ref{sec:flat_tunnel})
where we introduced small corrections to the supersymmetric Lagrangian.
These terms will allow the possibility of having supersymmetry breaking vacua that could be
susceptible to decay. In particular, we will study the Euclidean action given by\footnote{We have omitted the
supersymmetry-breaking cubic term for simplicity, as its effect (in this model, at least) was identical to turning on the quadratic term.}
\begin{align}
S_E = \int d^4 x \sqrt{g} \left[ - \frac{R}{2} + K_{\phi \bar{ \phi}} g^{ab} \partial_a \phi \partial_b \bar{\phi} + \tilde{V} (\phi,\bar{ \phi}) \right] + 2 \int_{\mathcal{M}} d^3 \xi \sqrt{h} e^{K/2} |q\phi| + S_{\text{GH}}
\label{eq:SE_sph_memb}
\end{align}
where,
\begin{align}
\tilde{V}(\phi,\bar{ \phi}) = e^K \left( K^{\phi \bar{\phi}} \left| D_{\phi} \hat{W} \right|^2  - 3 \left|\hat{W} \right|^2 \right) + \mu^2 \phi \bar\phi , \qquad \hat{W} \equiv W - (n + q H(x)) \phi .
\end{align}
As we will see afterwards, the Gibbons-Hawking boundary term will become important once we evaluate the actual value of the Euclidean action.
	
Furthermore, assuming an $O(4)$ symmetric Euclidean solution for the instanton, we introduce the following ansatz for the metric
\begin{align}
ds^2 = d\chi^2 + \rho (\chi)^2 d \Omega_3^2 ~.
\label{eq:metric_sugra}
\end{align}
Using these coordinates we search for a solution featuring a spherical membrane sitting at a fixed value
of the radial coordinate, which we call $\chi = R$. With this setting, one arrives to the following equations of motion for the metric function and the scalar field,
\begin{align}
\partial_{\chi} \left( K_{\phi \bar{\phi}} \partial_\chi \phi \right) + \frac{3 \ \partial_\chi \rho}{\rho}  K_{\phi \bar{\phi}} \partial_{\chi} \phi &= K_{\phi \bar{ \phi} \bar{ \phi}} \left| \partial_\chi \phi  \right|^2 + \frac{\partial \tilde{V}}{\partial \bar{\phi}} + e^{K/2} \left[ K_{\bar{ \phi}} |q\phi| - q e^{i\eta} \right] \delta (\chi - R) \\[6pt]
\rho'' &= - \frac{1}{3} \rho \left( 2 K_{\phi \bar{\phi}} |\phi'|^2 + \tilde{V} \right) - \rho e^{K/2} |q\phi| \delta (\chi - R) ,
\end{align}
which in the case of a canonical kinetic term for the field $\phi$, that is,
$K(\phi,\bar{\phi}) = \phi \bar{ \phi}$, reduce to
\begin{align}
\phi'' + \frac{3 \rho'}{\rho} \phi' &= \frac{\partial \tilde{V}}{\partial \bar{\phi}} + e^{|\phi|^2/2} \left[ \phi |q\phi| - q e^{i\eta} \right] \delta (\chi - R)\\
\rho'' &= - \frac{1}{3} \rho \left( 2  |\phi'|^2 + \tilde{V} \right) - \rho e^{|\phi|^2/2} |q\phi| \delta (\chi - R) ,
\end{align}
which are very similar to the set of equations to be solved in the case of a purely
scalar field bounce solution in curved space first found by Coleman and de Luccia in \cite{Coleman:1980aw}
with the important difference that now both first derivatives of the functions $\rho$ and $\phi$
present a jump at the position of the membrane. This is to be expected since locally the
behaviour of the functions should be the same as the flat membrane case in these models.

Finally, note that the radius of the membrane, $R$, is still a free parameter of our system. Therefore, in order to fix it, we can take a similar approach as the one we took in the non-gravitational case. Namely, we can try solving the equations for several choices of $R$ and find which one extremizes the Euclidean action \eqref{eq:SE_sph_memb}.

\subsection{Example: quartic superpotential}
	
	In this subsection, we will apply the machinery described above to the simple model of eq.~\eqref{eq:sugra_model_BPS}.
Taking into account the different values of the supersymmetry breaking parameter $\mu$ one can encounter several
situations depending on the nature of the false vacuum. In particular,  the geometry of the Euclidean space will depend on the
sign of the potential at the false vacuum, i.e., while for $V_{\text{fv}}\le 0$, the background will be a non-compact
space, the case with $V_{\text{fv}}>0$ will yield a compact de Sitter space for the background. This different
properties of the background will become important for the way we handle our numerical solutions.
	
All the examples below were solved numerically using a simple o\-ver\-shoot-un\-der\-shoot algorithm to find the correct
initial condition for the scalar field. Note that, as far as the boundary conditions are concerned, requiring the spacetime to be
regular at the center of the bubble implies that
%&&%
$\rho(\chi) = \chi + \mathcal{O}(\chi^3)$
%&&%
for $\chi \to 0$, see \cite{Coleman:1980aw} for further details.
	
\subsubsection{AdS/Minkowski to AdS transitions}
	
	As we mentioned above, this tunnelling event occurs within a non-compact space. Therefore, the integral of the Euclidean action
corresponding to the instanton with a membrane fixed at a particular value of the coordinate radius $R$ is given by,
\begin{align}
S_E &= 2\pi^2 \int_{0}^\infty d\chi \left[ \rho^3 \left( |\phi'|^2 + \tilde{V}(\phi,\bar{\phi}) \right) + 3 \left( \rho^2 \rho'' + \rho (\rho')^2 - \rho \right) \right] +  4\pi^2 \left[ \rho^3 e^{|\phi|^2/2}  |q\phi| \right]_{\chi = R} + S_{\text{GH}} \nonumber \\
&= 2\pi^2 \int_{0}^\infty d\chi \left[ \rho^3 \left( |\phi'|^2 + \tilde{V}(\phi,\bar{\phi}) \right) - 3 \left( \rho (\rho')^2 + \rho \right) \right] + 4\pi^2 \left[ \rho^3 e^{|\phi|^2/2}  |q\phi| \right]_{\chi = R}
\label{eq:SE_grav}
\end{align}
where, in the last step, the term we have integrated out cancels the contribution from the GH term (see \cite{Weinberg:2012pjx,Masoumi:2016pqb}).

On the other hand, the integral corresponding to the background is given by,
\begin{align}
S_{E,\text{bg}} = 2\pi^2 \int_{0}^{\infty} d\chi \left[ \rho_{\text{fv}}^3 V_{\text{fv}} - 3 \left( \rho_{\text{fv}} (\rho_{\text{fv}}')^2 + \rho_{\text{fv}} \right)  \right],
\end{align}
where
\begin{align}
\rho_{\text{fv}}(\chi) = H^{-1} \sinh \left( H \chi \right)
\end{align}
is the expression for the scale factor of Euclidean anti-deSitter space in the particular
slicing given by Eq. (\ref{eq:metric_sugra}) and where we have introduced the Hubble parameter $H=  \sqrt{\frac{|V_{\text{fv}}|}{3}}$. We can take
the Minkowski limit of this expression to find that in the case of $V_{\text{fv}}=0$, the scale factor becomes,
$\rho_{\text{fv}} (\chi)= \chi$.

It is easy to see that these background Euclidean actions as well as the ones obtained from the instanton solutions are, in fact, divergent. However, the
physically relevant quantity is the difference between the instanton action and the background one.
This action is finite.
	
The key to computing this difference correctly lies in performing the integrations up to a certain $\rho_{\text{max}}$, such that its corresponding
radial coordinate $\chi_{\text{max}}$ satisfies $\chi_{\text{max}} \gg R$. See \cite{Masoumi:2016pqb} for more detail and an explicit proof of the
convergence of this difference.
	
\begin{figure}[t]
\centering
\subfloat[]{
\includegraphics[width=0.48\textwidth]{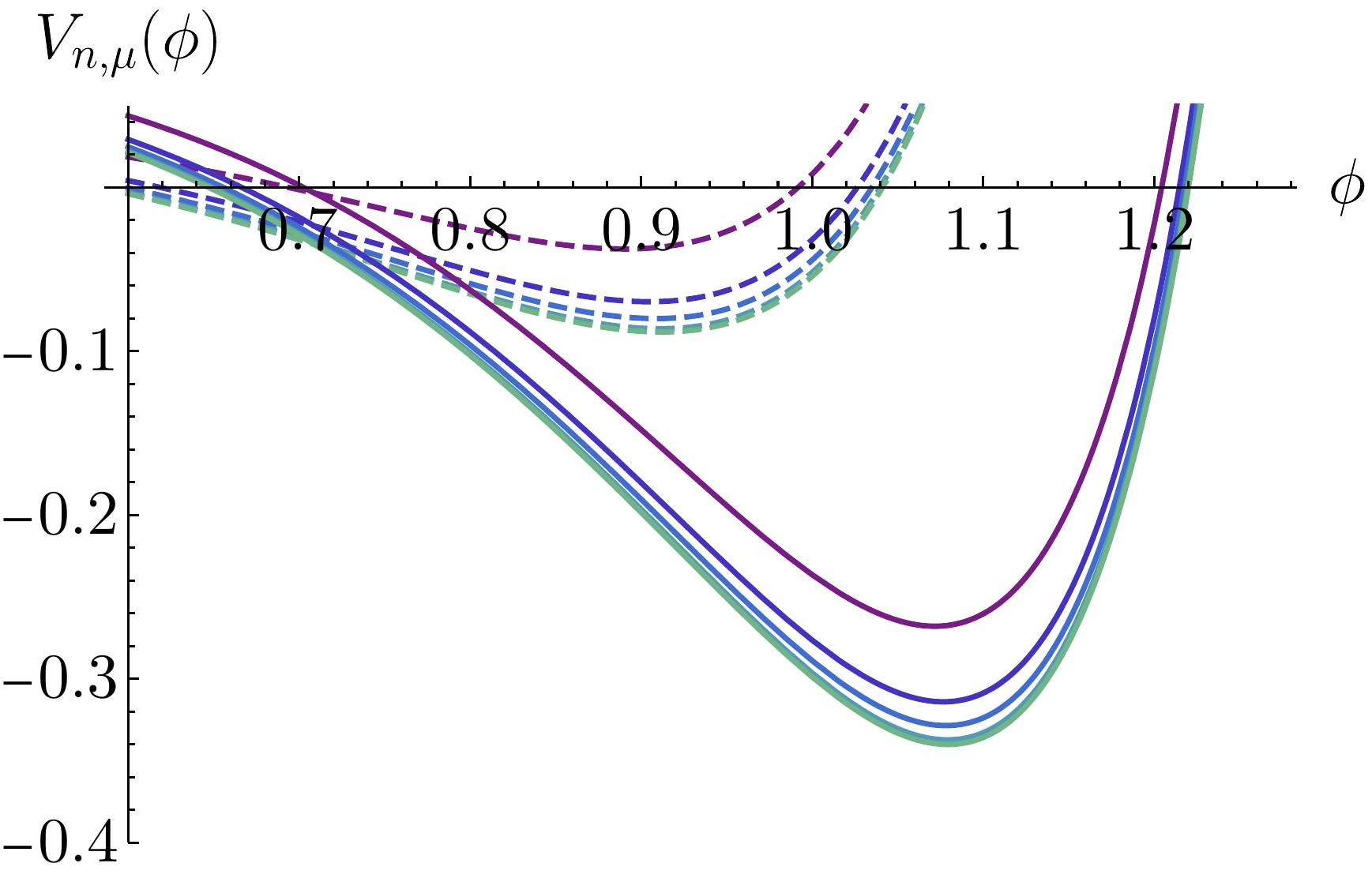}
}
\hfill
\subfloat[]{
\includegraphics[width=0.48\textwidth]{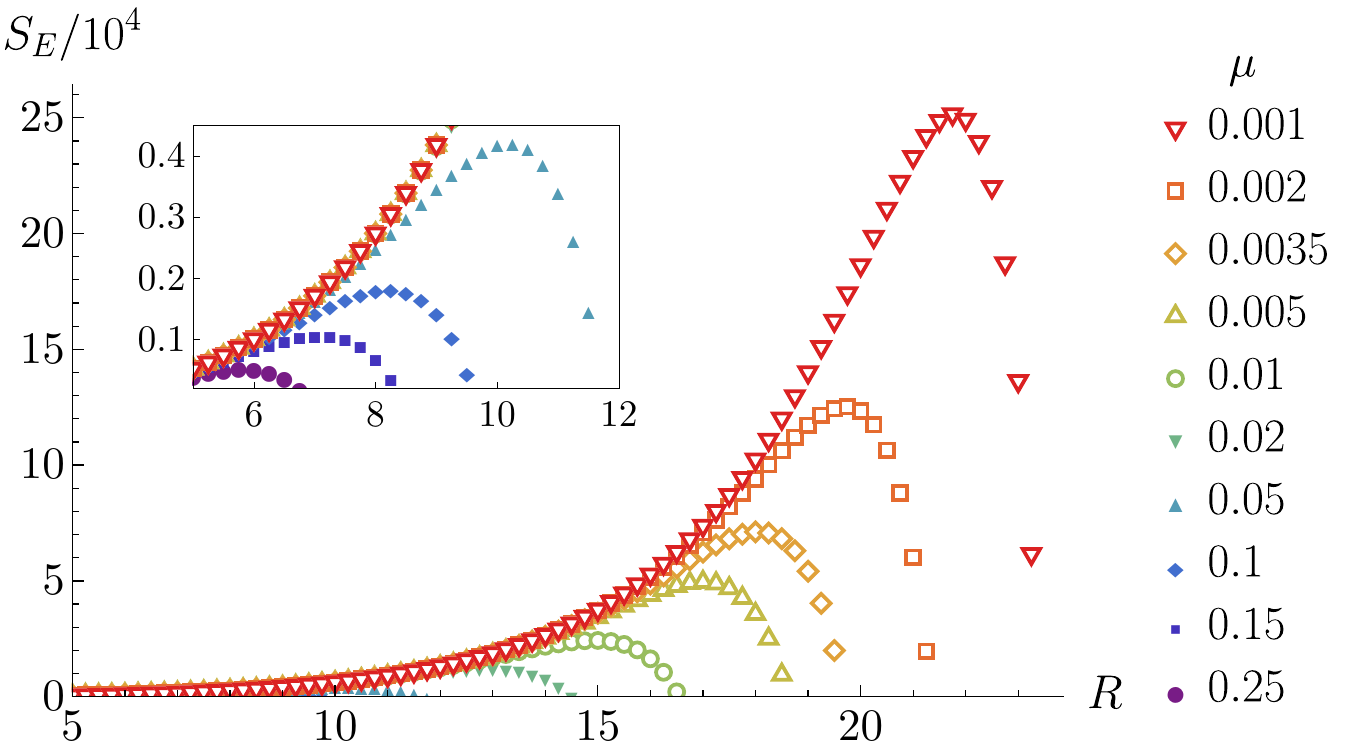}
}
\caption{(a) Scalar potential for $n=2$ (dashed) and $n=3$ (solid). (b) Euclidean action for different fixed membrane radii $R$ for a list of supersymmetry breaking parameters $\mu$. The color corresponding to each $\mu$ is the same for both figures. Note that the radius $R$ which extremizes the action increases as we diminish the supersymmetry-breaking parameter $\mu$.}
\label{fig:sugra_break_pot}
\end{figure}

	In figure \ref{fig:sugra_break_pot} we show the result of computing this difference for several membrane radii and supersymmetry-breaking parameter values. We can clearly see that the difference between actions is finite and reaches a maximum at a certain R, depending on $\mu$. Furthermore, as the potential tends towards its original and supersymmetric form, the radius of the membrane interpolating between both branches of the scalar potential gets bigger, which is consistent with the fact that at the supersymmetric limit the tunneling transitions becomes completely suppressed.
	%&&%

\begin{figure}[t]
\centering
\subfloat[]{
\includegraphics[width=0.48\textwidth]{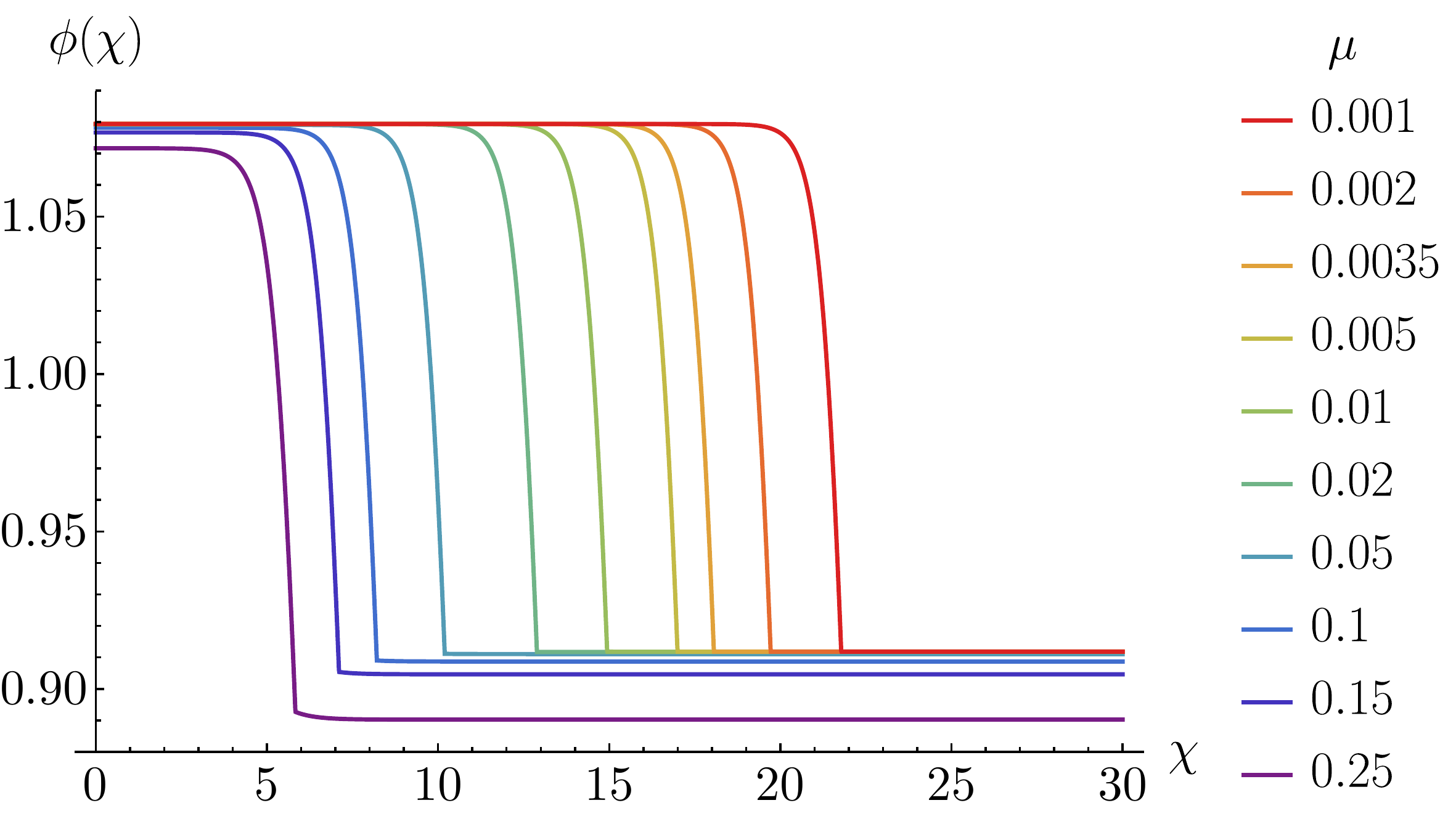}
}
\hfill
\subfloat[]{
\includegraphics[width=0.48\textwidth]{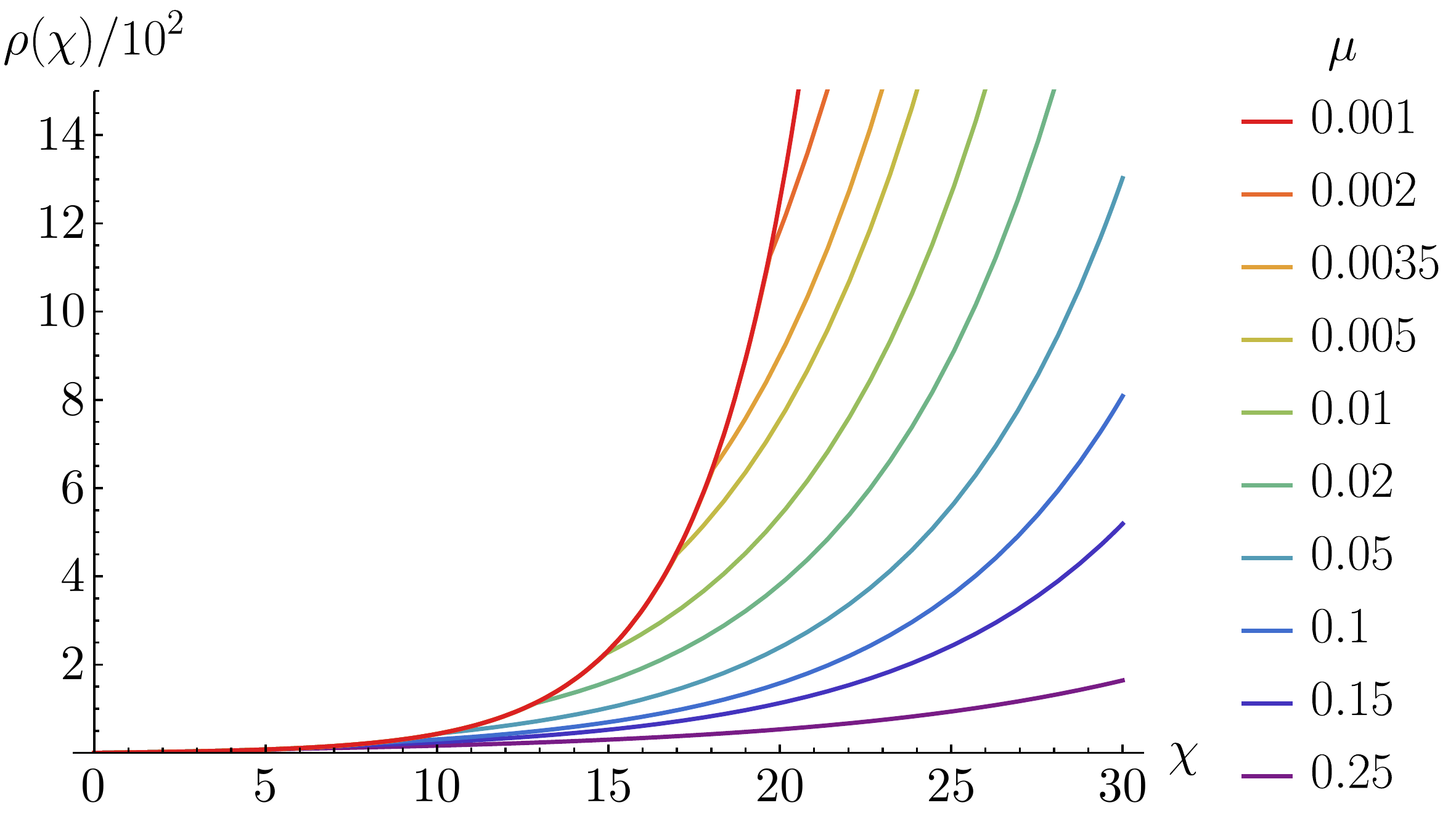}
}
\caption{Evolution of (a) scalar field and (b) scale factor, for different supersymmetry parameters $\mu$. Each case corresponds to the radius of maximum Euclidean action obtained in figure \ref{fig:sugra_break_pot}.}
\label{fig:sugra_profiles}
\end{figure}

	The profiles for both the scalar field and the scale factor corresponding to the radii with maximum Euclidean action difference for each $\mu$ are shown in figure \ref{fig:sugra_profiles}. The scalar field profiles are all quite similar in shape, with the only differences resting on the positions of the true and false vacuum, and in the radius where the jump happens. On the other hand, the scale factor shows very clearly where the jump happens as well, and the exponential behaviour seems to pick up quite fast once the membrane has been crossed.
	%&&%
	
\begin{figure}[t]
\centering
\includegraphics[width=0.6\textwidth]{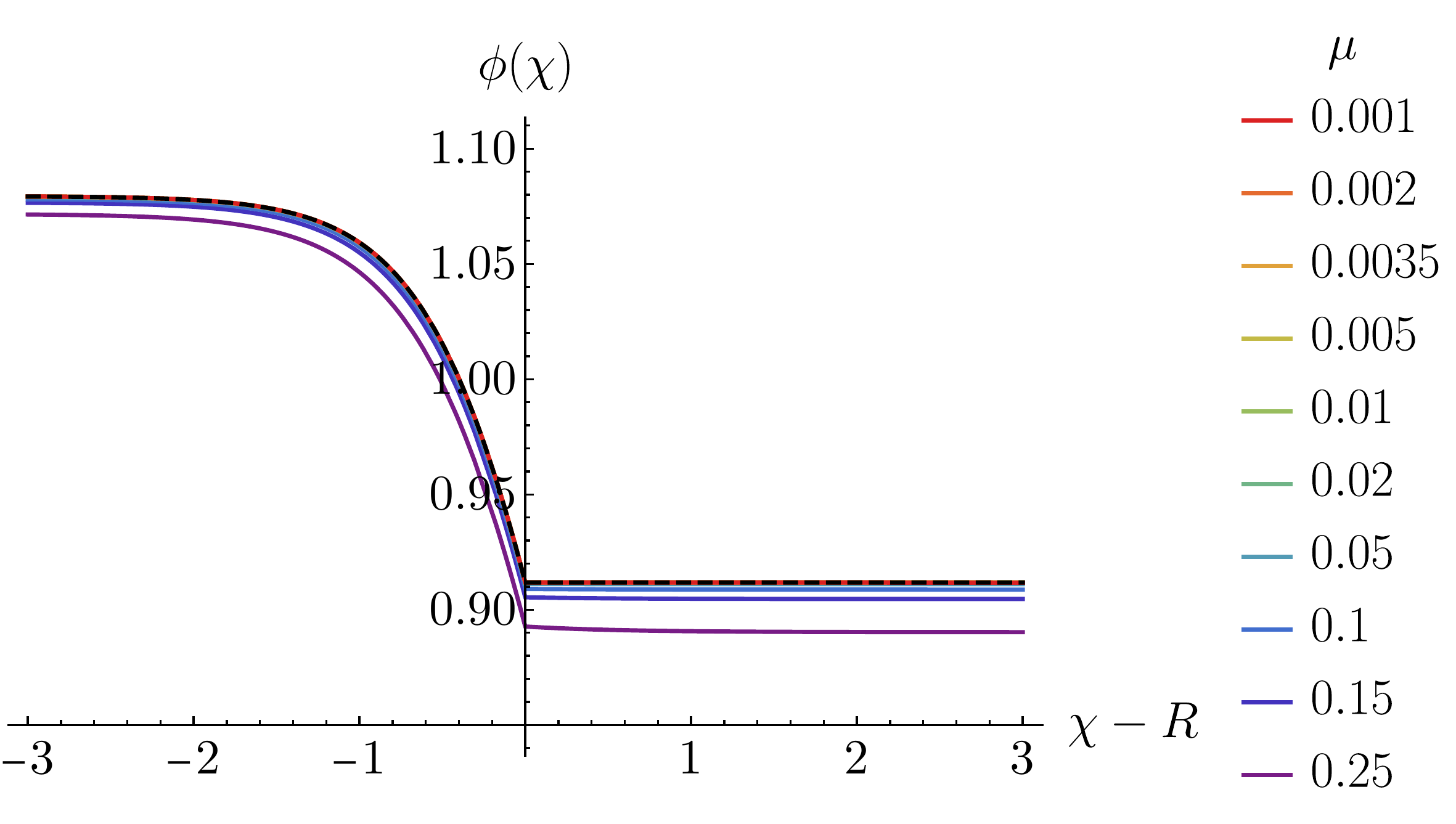}
\caption{Evolution of scalar field for different supersymmetry parameters $\mu$ and around the membrane (i.e., $R$ corresponds to the membrane's radial coordinate). The BPS case has been singled out with a dashed black line.}
\label{fig:sugra_profiles_comp}
\end{figure}

In order to compare all these profiles with the limiting BPS case, we show in figure \ref{fig:sugra_profiles_comp} a close-up plot around the membrane for all of them. Essentially, we see that the BPS profile is actually a limiting case for the scalar field, which concurs with our results in the non-gravitational case.
	
\begin{figure}[t]
\centering
\subfloat[]{
\includegraphics[width=0.48\textwidth]{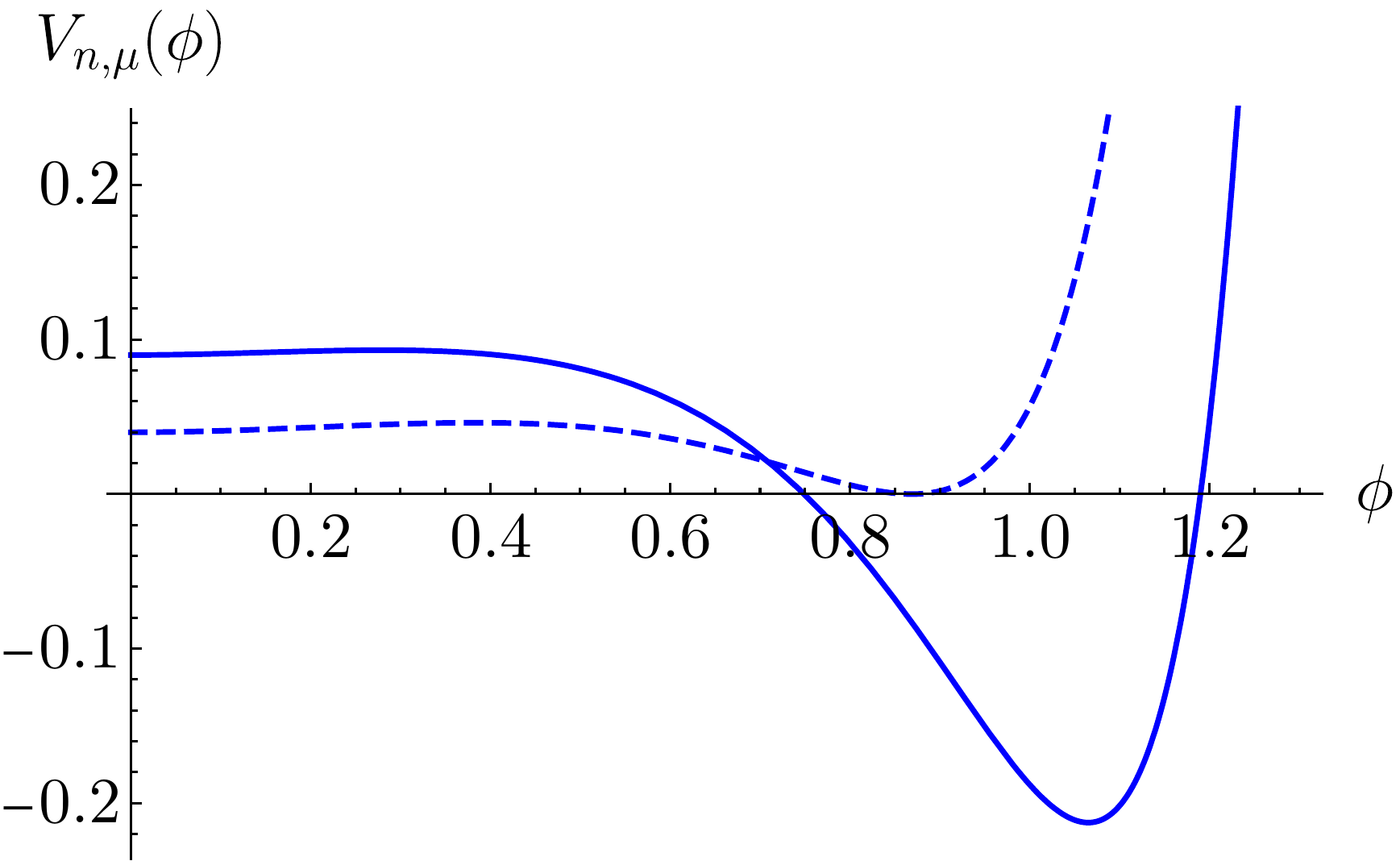}
}
\hfill
\subfloat[]{
\includegraphics[width=0.48\textwidth]{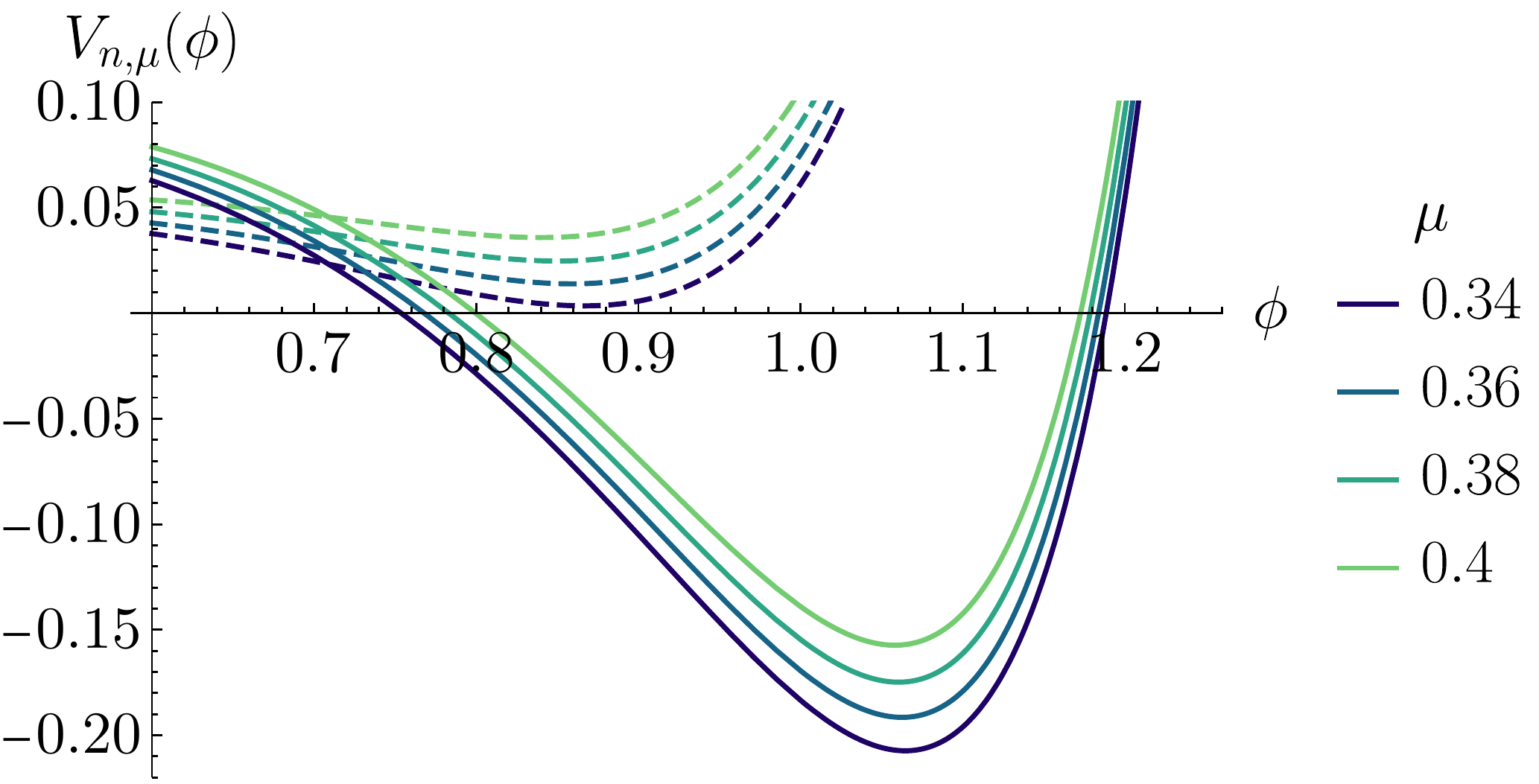}
}
\caption{Scalar potential featuring (a) a Minkowskian false vacuum ($\mu=0.33$) (b) Scalar potential with dS false vacua, for several supersymmetry breaking parameter values $\mu$.}
\label{fig:pots}
\end{figure}

\begin{figure}[t]
\centering
\subfloat[]{
\includegraphics[width=0.48\textwidth]{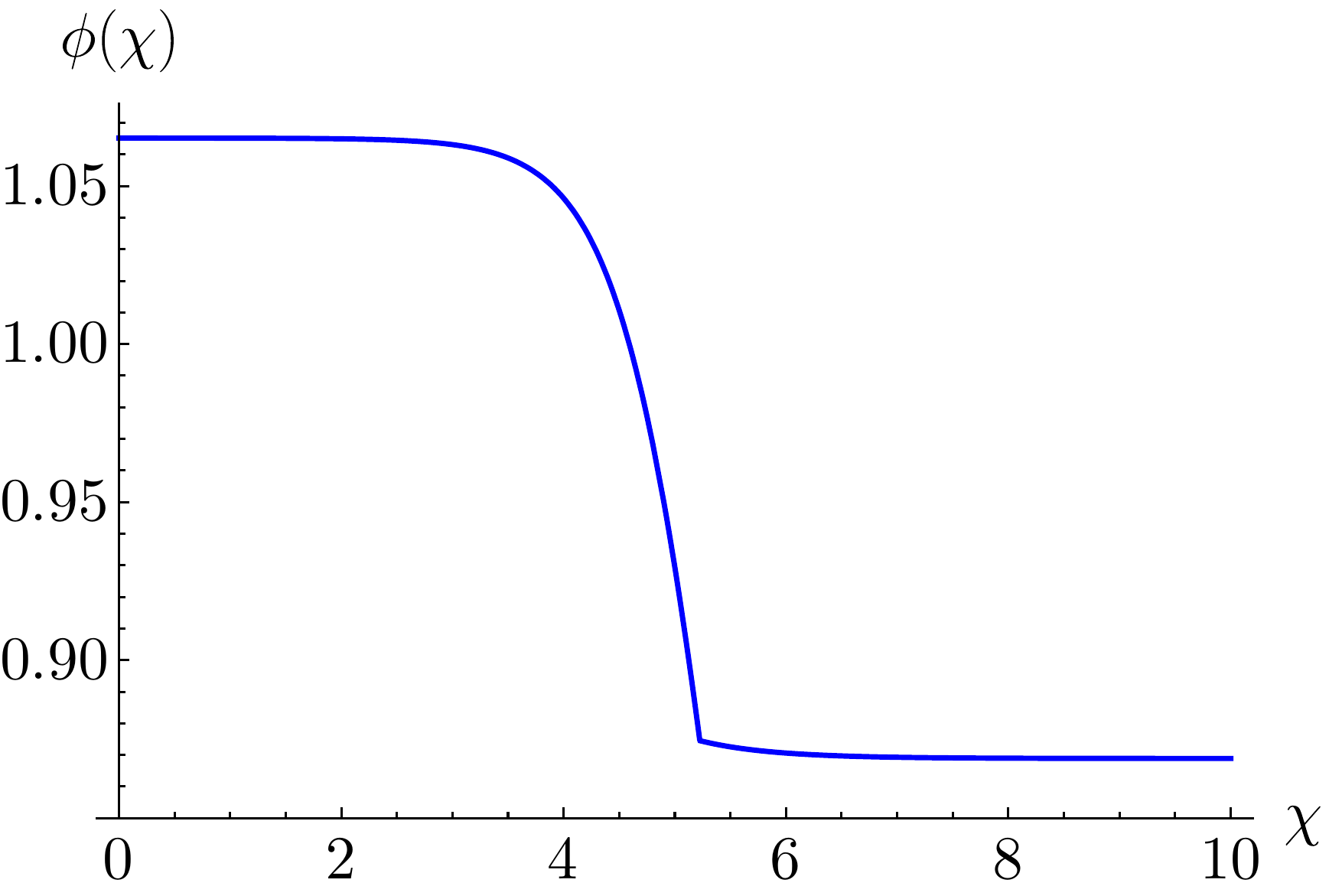}
}
\hfill
\subfloat[]{
\includegraphics[width=0.48\textwidth]{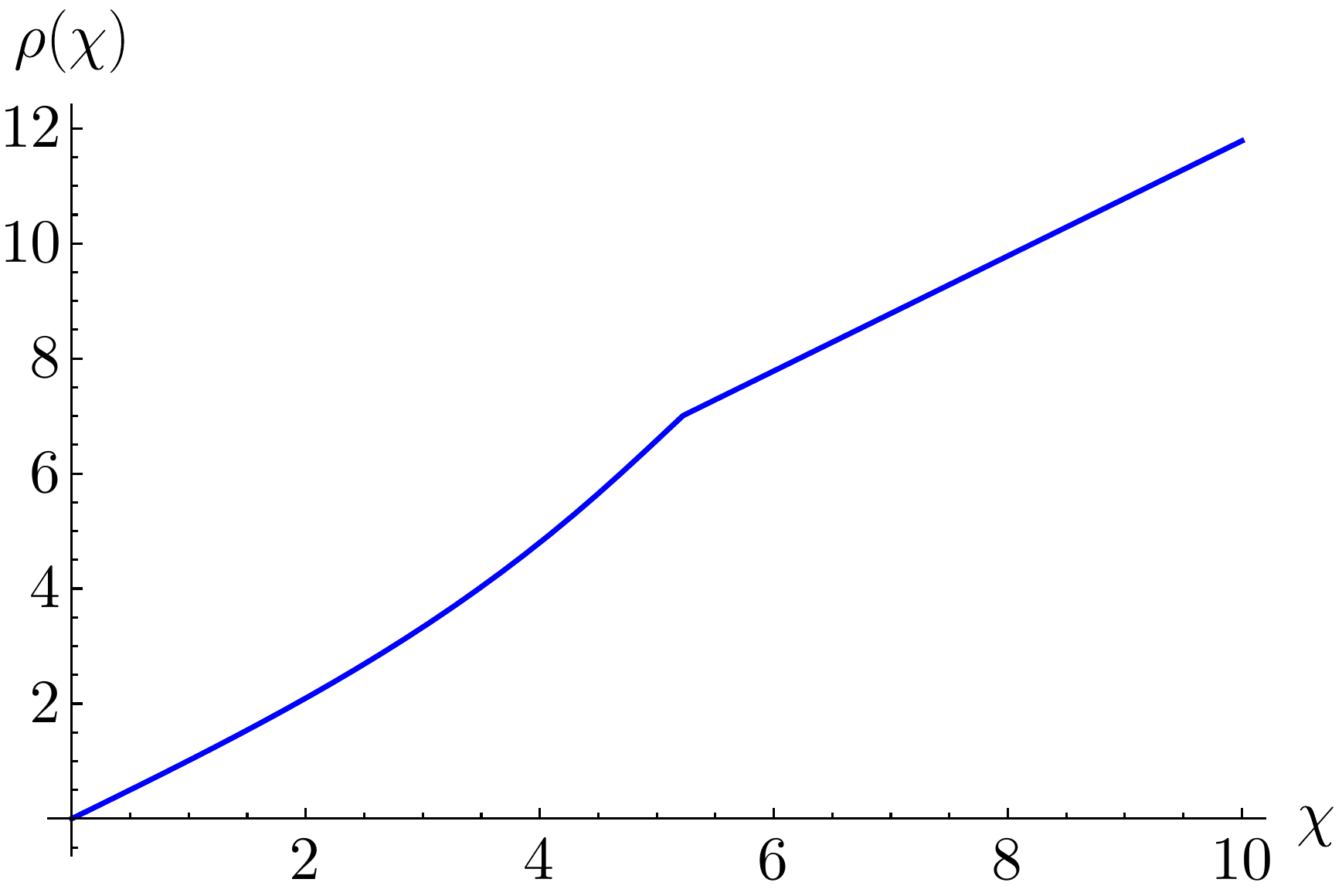}
}
\caption{(a) Scalar field profile and (b) scale factor, for the potential shown in figure \ref{fig:pots}(a), corresponding to the membrane radius which maximized the Euclidean action difference.}
\label{fig:mink_profiles}
\end{figure}
	
	Tuning the supersymmetry-breaking parameter $\mu$, we can also analyze an almost Minkowskian false vacuum ($\mu = 0.33$), see figure \ref{fig:pots}(a). Note that Minkowski false vacua are still non-compact spaces, and thus they should be analyzed in exactly the same fashion as AdS vacua.
	
	After computing the Euclidean action difference for several radii, we found that the radius with maximum Euclidean action was $R = 5.2$ in the Minkowskian case. The profiles corresponding to a setting with such a membrane are shown in figure \ref{fig:mink_profiles}.
	
	Finally, as explicitly shown in \cite{Brown:1988kg,Masoumi:2016pqb}, the requirement of energy conservation can be generalized to tunnelings where gravitational effects are considered and the false vacuum is Minkowskian, by requiring the ADM mass vanishes. After switching to Lorentzian signature, and integrating once the $0-0$ component of Einstein's equation (the Hamiltonian contraint), it is possible to obtain the following expression for the ADM mass (see \cite{Masoumi:2016pqb}):
\begin{align}
M_{\text{ADM}} = 4\pi \int_0^\infty d\chi \ \rho' \rho^2 \left[ \left| \phi' \right|^2 + \tilde{V} + 2 |q\phi| e^{K/2} \ \delta (\chi - R) \right] = 0~.
\end{align}
Here we have identified the Euclidean $\rho$ coordinate with the Lorentzian radial one, $r$, and the integral is taken at the $\tau=0$ spacelike surface\footnote{Note that $\rho' (R)$ should be evaluated as the mean between the derivative of the scale factor right before and after the membrane radius.}. This means that the right hand side of this expression represents the spatial integral of the energy density, that is, the $T_{00}$ component of the energy-momentum tensor. 

We have explicitly checked that this condition is satisfied for the solution depicted in figure \ref{fig:mink_profiles}.
	
\subsubsection{dS to AdS transitions}
	
	\begin{figure}[t]
		\centering
		\subfloat[]{
			\includegraphics[width=0.48\textwidth]{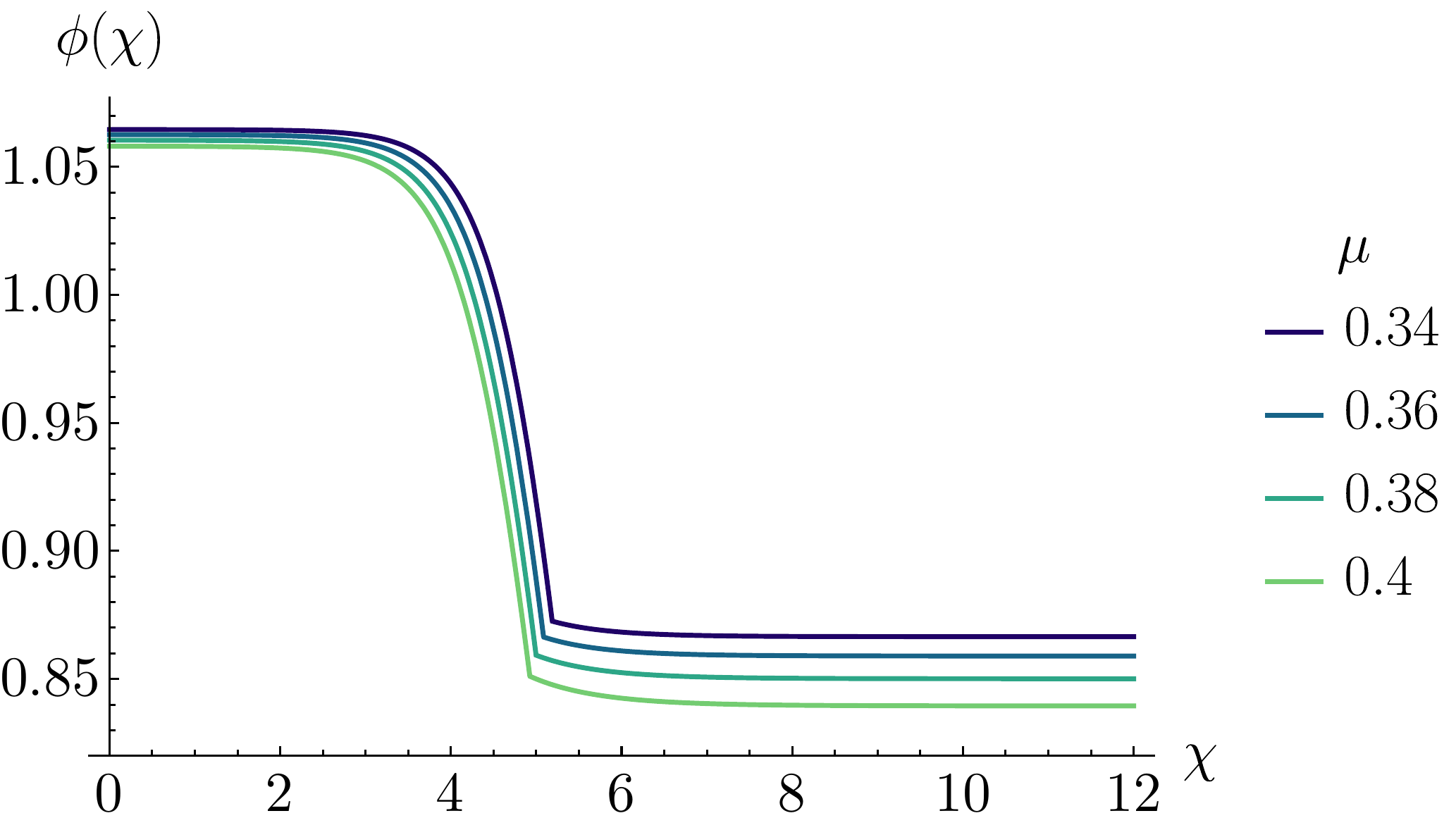}
		}
		\hfill
		\subfloat[]{
			\includegraphics[width=0.48\textwidth]{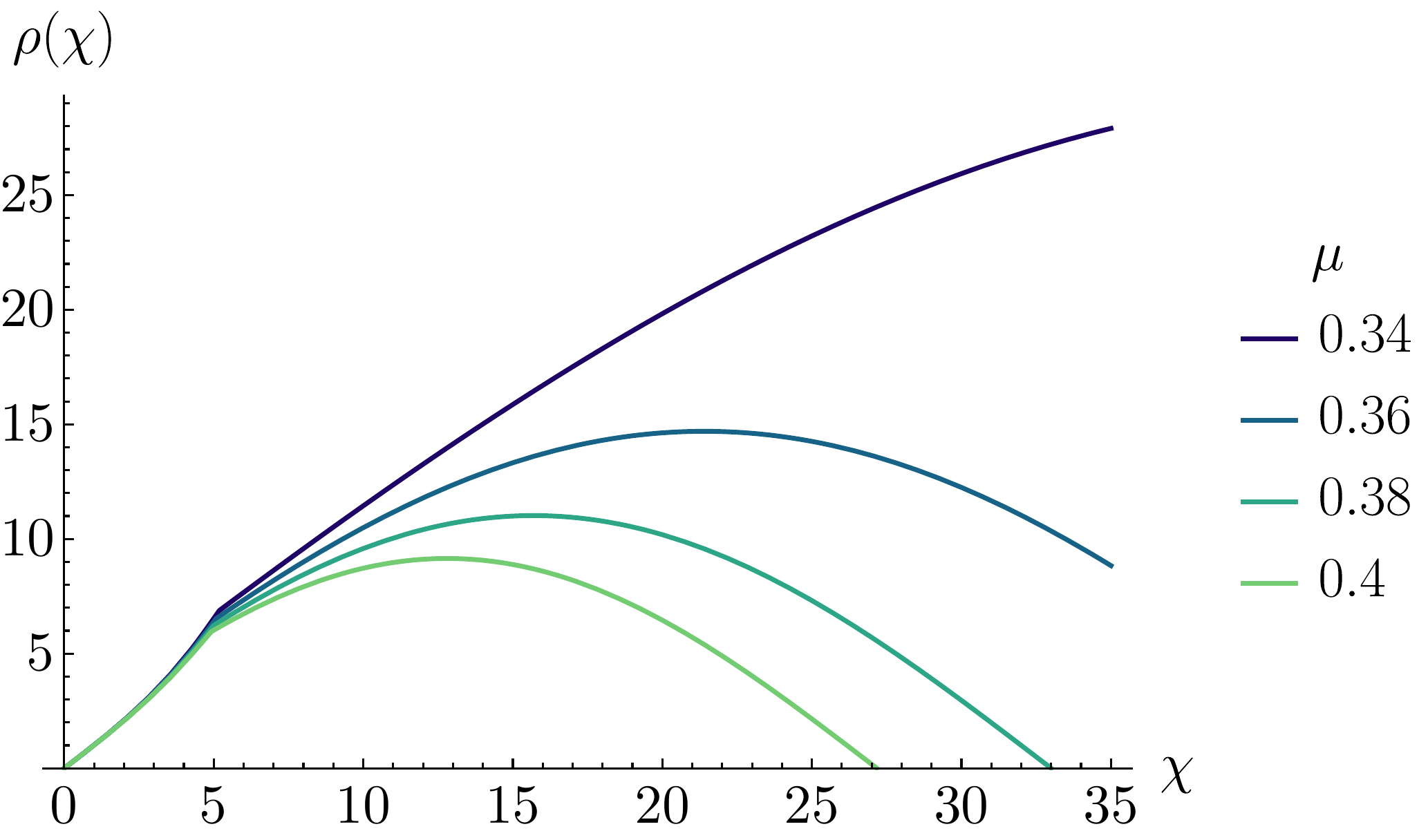}
		}
		\caption{(a) Scalar field profile and (b) scale factor, for the dS $\to$ AdS decays between the potentials depicted in fig.~\ref{fig:pots}(b). All profiles plotted here satisfy $\rho (\chi_{\text{max}}) = 0$ and $\rho ' (\chi_{\text{max}}) = -1$ simultaneously.}
		\label{fig:ds_profiles}
	\end{figure}
	
	By making the supersymmetry-breaking parameter $\mu$ sufficiently big, we can actually make the false vacuum lie in a positive potential value, while the true vacuum rests at $V<0$.
	As opposed to the previous cases, the compact geometry of this kind of instantons imposes boundary conditions at both the origin and the maximum value of the radial coordinate (see, e.g., \cite{Weinberg:2012pjx}), where $\rho(0)=\rho (\chi_{\text{max}})=0$. Of course, the Euclidean action \eqref{eq:SE_grav} will be integrated up to $\chi_{\text{max}}$. Indeed, the background consisting of a constant scalar potential $V_{\text{fv}}>0$ is described, in terms of our ansatz \eqref{eq:metric_sugra}, by the following scale factor:
	\begin{align}
	\rho_{\text{fv}} = H^{-1}\sin \left( H \chi \right) \ , \quad \chi \in [0,\chi_{\text{max}}]
	\end{align}
	where $\chi_{\text{max}}$ represents the first positive and non-zero root of $\rho_{\text{fv}}$. Furthermore, the integral of the background action can be found analytically, and is given by
	\begin{align}
	S_{\text{bg}} = - \frac{24 \pi^2}{V_{\text{fv}}}.
	\end{align}
	
	Since the whole instanton is compact in dS $\to$ AdS decays, the only boundary conditions we can impose on the scalar field are\footnote{Note that, in our case where a membrane is present, each of these boundaries is defined with respect to a different scalar potential (since the flux is different at each side of the membrane). Regardless, both $\phi(0)$ and $\phi(\chi_{\text{max}})$ should lie close enough to the minimum of their respective potentials.} $\phi'(0) = \phi'(\chi_{\text{max}})$=0. On the other hand, in order to find a non-singular instanton, for $\chi \to \chi_{\text{max}}$ one must also require $\rho'(\chi)=-1+\mathcal{O}((\chi-\chi_{\text{max}})^3)$. We have found these requirements to be so restrictive, that for each potential in terms of $\mu$ shown in figure \ref{fig:pots}(b), only a \emph{single} membrane radius solved the equations of motion correctly with respect to all of these boundary conditions.
	
	The profiles of the scalar field and scale factor are shown in figure \ref{fig:ds_profiles}.
	The geometries shown in fig.~\ref{fig:ds_profiles}(b) have been numerically shown to satisfy $\rho'(\chi_{\text{max}}) = -1$. In all cases, the scale factor clearly shows an Euclidean AdS-like space starting at $\chi = 0$, which evolves up to the membrane radius, from where a compact Euclidean dS-like evolution follows.\footnote{Note that these profiles are full numerical solutions to the instanton equations; i.e., no thin-wall approximation has been used in order to compute either the scalar field or the scale factor.}
	
	\section{Conclusions}
	\label{sec:conclusions}
	
	In this paper, we have studied a combination of the Coleman-de Luccia \cite{Coleman:1977py,Coleman:1980aw} and Brown-Teitelboim \cite{Brown:1987dd,Brown:1988kg} formalisms for vacuum decay using both scalar fields \emph{and} membranes, motivated from our study of flux compactifications in supersymmetric theories. We have reviewed the existing theory on single three-form multiplets   \cite{Gates:1980ay,Binetruy:1996xw,Ovrut:1997ur,Kuzenko:2010am,Kuzenko:2010ni,Bandos:2011fw,Farakos:2017jme} which combines scalar fields and real three-form fields in a supersymmetric fashion, in the context of $N=1, D=4$ supersymmetry and supergravity.
	Such a system has been extensively studied in \cite{Bandos:2018gjp}, with special focus on domain wall solutions with the inclusion of the sources given by fundamental supersymmetric membranes. These types of models provide a basis to quantitatively describe transitions between vacua defined by different potentials related by changes of integer flux contributions, such as the ones appearing in flux compactifications of String Theory \cite{Bousso:2000xa}.
	
	We have shown that adding soft supersymmetry-breaking terms to such a theory allows for rich phenomenological applications. Indeed, these terms enable false vacuum decays to occur through the nucleation of membranes. We have studied such systems using Euclidean methods and found the instanton solutions involving the form fields, scalar fields and membranes, both without and with gravity included. Furthermore, we have checked that we retrieve the correct supersymmetric limit as the supersymmetry-breaking terms are made smaller, so the flat membrane case can be taken as the limiting instanton solution corresponding to a membrane with infinite radius.
	
	These processes have very interesting applications from a cosmological viewpoint, as they can incorporate an emerging universe inside the created bubble with an inflationary period. This type of models of open inflation have been extensively studied in the context of the Landscape of String Theory  \cite{Freivogel:2005vv,Blanco-Pillado:2012uao,Masoumi:2016eag,Masoumi:2017gmh,Masoumi:2017xbe,Blanco-Pillado:2017nin,Blanco-Pillado:2019mnq}. In this work we have only investigated transitions between very simple potentials, however, more phenomenologically interesting results may be obtained by considering more realistic models. The transitions studied in this work essentially generalize the false vacuum decays previously discussed in the literature. These new ones naturally incorporate the contributions of the three-forms to the flux potential and the membranes which allow for transitions between vacua. Our formalism therefore includes any backreaction of these objects to the instanton solution, selecting an initial point for the open inflation period inside the bubble. This opens up the possibility of studying the observational consequences of these models more accurately.
	
	On another note, recent proposals \cite{Banerjee:2018qey} have investigated the possibility of embedding the Universe within a positively curved membrane created due to a false vacuum decay in a higher-dimensional theory. While we have studied these models in a four-dimensional environment, it could be interesting to generalize our results to higher dimensions to check the compatibility of these processes with the ones discussed in \cite{Banerjee:2018qey}.
	
	From a numerical perspective, we should note that the models we have solved here were quite simple, in that they only involved a single (effectively real) scalar field. Such models are quite easy to solve using an undershoot/overshoot algorithm, even when membranes and gravity are present in the setup.  Solving non-perturbative decays with two or more scalar fields is no simple matter. Thus, it might be useful to construct algorithms to compute false vacuum decays which effectively incorporate the first-derivative jumps on both the scalar fields and the scale factor. For this purpose it will also be interesting to explore these kind of models within the new formalism to study vacuum decay recently discussed in \cite{Espinosa:2018hue,Espinosa:2018voj}.
	
	Of course, the examples we have explored here can be generalized in several ways, aside from using more involved superpotentials. For once, we have broken supersymmetry by including explicit soft supersymmetry-breaking terms, motivated from phenomenological considerations \cite{Chung:2003fi}. However, recall that the tension of the membrane was determined by the requirement that the action of whole system consisting of supergravity, matter supermultiplets (which contain scalar fields and form fields) and the bosonic membrane preserves half of the supersymmetry. Therefore, another interesting direction to explore would consist on breaking supersymmetry by somehow deturning the coefficients present in the Nambu-Goto and Wess-Zumino terms of the membrane.
	
	Finally, in the models explored here we have considered single three-form multiplets, which allow for one real three-form field for each complex scalar field in the model (see appendix \ref{ch:3f_multiplets} and \cite{Farakos:2017jme}). However, a general type IIB compactification requires $2h^{1,2}+2$ real (and integer) fluxes, where $h^{1,2}$ is the number of complex structure moduli. In order to build such a model in our context, we can use the so-called double three-form supermultiplet, which allows for the inclusion of \emph{complex} three-forms (i.e., two real fluxes for each complex scalar field). Such a system has been considered in \cite{Bandos:2018gjp}, and in \cite{Lanza:2019xxg} where some restrictions have been pointed out on the possible EFTs. It may be interesting to re-check these restrictions when considering false vacuum decay mediated by supersymmetry-breaking spherical membranes.
	
	Furthermore, extended objects such as effective strings and membranes have recently made their way into the Swampland program \cite{Lanza:2020qmt, Lanza:2021udy,Casas:2022mnz}. In particular, our results may be useful to test the AdS Instability Conjecture \cite{Ooguri:2016pdq} in completely backreacted systems.
	
	Finally, the tunnelling processes we have discussed in the present paper describe the transitions between different flux vacua. One may wonder whether we can use this formalism to discuss the transitions where all the flux is absorbed by the charged membrane. In this case the interior of the bubble will not have any flux and therefore the geometry of the internal dimensions can collapse. However, one can show that this collapse can be smooth in the higher dimensional theory and the solution will resemble the bubble of nothing (BON) geometry discussed by Witten in the simple 5-dimensional Kaluza-Klein model \cite{Witten:1981gj}. These type of instabilities were first discussed in the context of a higher dimensional field theory models of flux compactifications in a series of papers in \cite{Blanco-Pillado:2010xww,Blanco-Pillado:2010vdp,Blanco-Pillado:2011fcm} (See also more recent discussions of these BONs in related models in \cite{Dine:2004uw,Horowitz:2007pr,Brown:2010mf,Ooguri:2017njy,GarciaEtxebarria:2020xsr,Draper:2021ujg,Draper:2021qtc}). It would therefore be interesting to study these solutions in our supersymmetric setup. These models would be a perfect
ground to study the conjecture put forward in \cite{Blanco-Pillado:2016xvf} where it was stated that the supersymmetric limit of these BON solutions would in fact be some kind of static {\it end of the world} brane that preserves some supersymmetry. We have initiated some work in this direction and we hope to be able to report on these results in a future publication.

\section*{Acknowledgments}

We are grateful to Thibaut Coudarchet for interesting discussions.  This study
is supported in part by the PID2021-123703NB-C21 and PID2021-125700NB-C21 (IB) grants funded by MCIN/ AEI /10.13039/501100011033/ and by ERDF;  ''A way of making Europe'', the Basque Government grant (IT-1628-22) and the Basque Foundation for Science (IKERBASQUE).

\appendix
\renewcommand{\theequation}{A.\arabic{equation}}
\section{Three-form multiplets in supersymmetry}
\label{ch:3f_multiplets}
	
In this appendix, we will briefly review the simplest way to construct the supersymmetric Lagrangians used throughout the main text, which include 3-forms among their components. More specifically, we will work out the expressions for \emph{single} 3-form multiplets, which are written in terms of chiral superfields in which the real part of the F-component (the auxiliary field) is substituted by a 3-form (as opposed to \emph{double} 3-form multiplets, which are described by other special chiral superfields with both real and imaginary parts of the F-component of which are replaced by 3-form fields). Most of the following formalism may be found in \cite{Bandos:2010yy,Farakos:2017jme,Lanza:2019nfa}, and we will try to stick to the conventions of \cite{Wess:1992cp} unless otherwise stated.
	
In the superspace formalism, supersymmetric covariant derivatives are given by
\begin{align}
	D_\alpha = \partial_\alpha + i \sigma_{\alpha \dot\alpha}^\mu \bar{\theta}^{\dot\alpha} \partial_\mu, \qquad \bar{D}_{\dot\alpha} = - \partial_{\dot\alpha} - i \theta^\alpha \sigma_{\alpha \dot\alpha}^\mu \partial_\mu.
\end{align}
Their algebra, which is given by
\begin{equation}
    {}\{D_\alpha, D_\beta\} =0\; , \qquad \{D_\alpha, \bar{D}_{\dot\alpha}\}=-2i \sigma_{\alpha \dot\alpha}^\mu \partial_\mu\; , \qquad   \{ \bar{D}_{\dot\alpha}, \bar{D}_{\dot\beta}\}=0
\end{equation}
implies the following expressions:
\begin{equation}
D_\alpha D_\beta =\frac 1 2 \epsilon_{\alpha \beta}D^2\; , \qquad D_\alpha D^2=0\; , \qquad \bar{D}_{\dot\alpha} \bar{D}_{\dot\beta}=-\frac 1 2  \epsilon_{\dot\alpha\dot\beta} \bar{D}^2\; , \qquad \bar{D}_{\dot\alpha}  \bar{D}^2=0\; . \qquad
\end{equation}

	A \emph{chiral} supefield is defined by the constraint $\bar{D}_{\dot\alpha} \Phi = 0$, and has the following component expansion in terms of the Grassmanian variables $\theta^\alpha$:
	\begin{align}
	\Phi &= \phi + \sqrt{2} \theta \psi + \theta \theta F + i \theta \sigma^\mu \bar{\theta} \partial_\mu \phi - \frac{i}{\sqrt{2}} \theta \theta  \partial_\mu \psi \sigma^\mu \bar{\theta} + \frac{1}{4} \theta \theta \bar{\theta} \bar{\theta} \square \phi \\
    &=  e^{i \theta \sigma^\mu \bar{\theta} \partial_\mu }\left(\phi + \sqrt{2} \theta \psi + \theta \theta F\right)\; ,
	\end{align}
	where $\phi (x) $ and $F (x)$ are complex scalar fields and $\psi  (x)$ is a Weyl spinor.
 It is convenient to use their equivalent definition in terms of the leading components of the superfield and of its covariant derivatives:
	\begin{align}
	\left. \Phi \right| &= \phi \\
	\frac{1}{\sqrt{2}} \left. D_\alpha \Phi \right| &=  \psi_\alpha \\
	- \frac{1}{4} \left. D^2 \Phi \right| &= F,
	\end{align}
	where the vertical line means that we are taking the superfield with $\theta = \bar{\theta} = 0$.
	
	Given a set of chiral superfields $\vec{\Phi} = \left\lbrace \Phi^1 , \ldots , \Phi^n \right\rbrace$, the most general interacting supersymmetric theory describing their dynamics is given by the Lagrangian
	\begin{align}
	\mathcal{L} &= \int d^2 \theta d^2 \bar\theta \ K (\vec{\Phi}, \bar{\vec{\Phi}}) + \left[ \int d^2 \theta \ W (\vec{\Phi}) + \text{h.c.} \right] \\[7pt]
	&= - K_{a \bar{b}} \partial_\mu \phi^a \partial^\mu \bar{\phi}^{\bar{b}} + K_{a \bar{b}} F^a \bar{F}^{\bar{b}} + F^a W_a + \bar{F}^a \bar{W}_a + \ldots \label{eq:L_off_shell}
	\end{align}
	where $a,b \in \lbrace 1, \ldots, n \rbrace$,  $K (\vec{\Phi}, \bar{\vec{\Phi}})$ is  an arbitrary real function of the chiral  superfields and thier complex conjugate anti-chiral superfields called \emph{Kähler potential}, $W(\vec{\Phi})$ is an arbitrary holomorphic function of the chiral superfields called  \emph{superpotential} and subscripts denote derivatives with respect to the scalar fields as in $K_{a \bar{b}} =\frac {\partial}{ \partial\phi^a} \frac {\partial} {\partial \phi^b} K$. In \eqref{eq:L_off_shell} we have omitted the terms involving the fermionic fields for simplicity; unless otherwise stated, we will not consider them below or also in the main text.
	
	It can be easily shown that, when the auxiliary fields are integrated out applying their algebraic equations of motion, the above Lagrangian reads
	\begin{align}
	\left.\mathcal{L} \right|_{\text{bos., on-sh.}} = - K_{a \bar{b}} \partial_\mu \phi^a \partial^\mu \bar{\phi}^{\bar{b}} - K^{a\bar{b}} W_a \bar{W}_{\bar{b}}.
	\end{align}
	where $K^{a\bar{b}}$ is the inverse of $K_{a\bar{b}}$.

\subsection{Including 3-forms into the chiral superfields: special chiral superfields }	

The material of this subsection is based on the constructions of \cite{Gates:1980ay,Binetruy:1996xw,Kuzenko:2010am,Bandos:2010yy,Bandos:2018gjp}.

The chirality constraint $\bar{D}_{\dot\alpha} \Phi = 0$ can be easily solved in terms of an unconstrained complex scalar superfield
$P=(\bar{P})^*$ which is called \emph{prepotential}: $\Phi = \bar{D}\bar{D}P$ ($\bar{\Phi} = DD\bar{P}$). The simplest special chiral superfield one may use which  includes a  3--form as  the real part of its highest F-component is obtained by expressing this complex prepotential by an uncostrained real superfield  $V= \bar{V}$.
It is convenient to define its $\theta$-expansion as follows:
	\begin{align}
	V(x,\theta,\bar\theta) = &C + i \theta \chi  -i \bar\theta \bar{\chi} + i \theta \theta \bar\phi   - i \bar\theta \bar\theta \phi  - \theta \sigma^\mu \bar{\theta} v_\mu
	\nonumber \\[7pt]
	&+ i \theta \theta \bar{\theta} \left[ \bar{\lambda}  + \frac{i}{2} \bar{\sigma}^\mu \partial_\mu \chi  \right] - i \bar\theta \bar\theta \theta \left[ \lambda  + \frac{i}{2} \sigma^\mu \partial_\mu \bar\chi  \right]
	+ \frac{1}{2} \theta \theta \bar{\theta} \bar\theta \left[ D - \frac{1}{2} \square C \right] .
	\label{eq:real_superfield_1}
	\end{align}
	Here $u (\vec{x})$ and $D  (\vec{x})$ are real scalar fields, $\phi  (\vec{x})$ is a complex scalar field, $v_\mu (\vec{x})$ is a real vector field and $\chi (\vec{x})$ and $\lambda (\vec{x})$ are Weyl spinors.
	
A special chiral superfield $Y$ can be obtained from the generic chiral superfield $\Phi=\bar{D}\bar{D}P$ by expressing the complex prepotential by $P=-\frac i 4 V$ the real superfield $V$, so that\footnote{Note that this definition allows for some freedom in choosing $V$, since \eqref{Y=bDbDV} is invariant under the gauge superspace symmetry
$V \to V + L$, where  $L$ is the so-called \emph{real linear superfield} which satisfies $\bar{D}^2 L = 0=D^2L$
. Furthermore,
the definition of $Y$ may be generalized to the case where the real superfield $V$ is not independent but composed from a so-called complex linear superfield $\Sigma$, which obeys $D^2\Sigma=0$, and its complex conjugate
 $\bar{\Sigma}$, which obys
 $\bar{D}^2\bar{\Sigma}=0$. This last case may be used to construct special chiral superfields which describe a double 3-form supermultiplet.
See \cite{Farakos:2017jme, Bandos:2018gjp, Lanza:2019nfa} for further detail.}
	\begin{align}\label{Y=bDbDV}
	Y:= - \frac{i}{4} \bar{D}^2 V
	\end{align}
	(the prefactor is chosen for later convenience). It is easy to check that it fulfills the condition $\bar{D}_{\dot\alpha} Y = 0$ from its definition. Its components can be projected using
	\begin{align}
	\left. Y \right| &= -\frac{i}{4} \left. \bar{D}^2 V \right| = \phi \label{superfield-components-1}\\
	\left. D_\alpha Y \right| &=- \frac{i}{4} \left. D_\alpha \bar{D}^2 V \right| = \lambda_\alpha \label{superfield-components-2}\\
	- \frac{1}{4} \left. D^2 Y \right| &= \frac{i}{16} \left. D^2 \bar{D}^2 V \right| = \frac{1}{2} \left( \partial^\mu v_\mu + i D \right) \label{superfield-components-3}
	\end{align}
Note that, component-wise, $Y$ is almost identical to the original chiral field $\Phi$ in that it contains a complex scalar field, a complex Weyl fermion and a complex auxiliary field, albeit in this case we are mostly interested in the real part of the latter, which is given by the devergence of a 4-vector field $v^\mu$.

	For our purposes, it will be convenient to consider $v_\mu$ as the one-form associated through Hodge duality to a three-form. Indeed, the Hodge dual of the three-form is given by the following vector field\footnote{In our conventions, $\epsilon^{0123}=-\epsilon_{0123} = 1$ for Lorentzian signature (which we will use throughout this appendix), while $\epsilon^{0123}=\epsilon_{0123} = 1$ for Euclidean one.}
	\begin{align}
	(\ast A_3)_{\mu} \equiv A_\mu = \frac{1}{3!} \epsilon_{\mu \nu \rho \sigma} A^{\nu \rho \sigma}
	\label{eq:hodge_dual}
	\end{align}
	where the indices have been raised using the flat spacetime metric. The divergence of this vector field is related to the Hodge dual of the 4-form field strength $F_4$, associated to $A_{\nu \rho \sigma}$ through\footnote{We have omitted the contribution of the metric determinant in these expressions to clarify the definitions. In the case of a curved spacetime, these expressions are $(\ast A_3)_{\mu} \equiv A_\mu = \frac{1}{3!} \sqrt{-g} \epsilon_{\mu \nu \rho \sigma} A^{\nu \rho \sigma} $ and $\ast F_4 = \frac{1}{\sqrt{-g}} \partial_\mu (\sqrt{-g} A^\mu)$.}
	\begin{align}
	\ast F_4 = \frac{1}{4!} \epsilon_{\mu \nu \rho \sigma} F^{\mu \nu \rho \sigma} = \partial_\mu A^\mu
	\label{eq:F4_flat}
	\end{align}
	With these expressions in hand, we find that if $v_\mu$ is  identified with $A_\mu$, i.e, with the Hodge dual of the 3-form field, the composite auxiliary field of the special chiral superfield $Y$ reads
	\begin{align}
	F_Y = - \frac{1}{4} \left. D^2 Y \right|  = \frac{1}{2} \left( \ast F_4 + i D \right).
	\end{align}

The advantage of treating the vector $v_\mu$ as dual to 3-form is that a membrane can be  coupled
'electrically' (or minially) to the 3-form field (see below) and thus to the composite auxiliary field in the special chiral superfield $Y$.
 Since $D$ does not enter the membrane part of the full action, we will be able to remove it from the action by solving its algebraic equations early on, leaving $\ast F_4$ untouched for its interplay with the scalar field and the membrane.

We can easily apply these expressions to a single three-form multiplet $Y$, with a certain Kähler potential $K(Y,\bar{Y})$ and superpotential $W(Y)$. Plugging the component fields of $Y$ into the bosonic Lagrangian \eqref{eq:L_off_shell}, we find
	\begin{align}
	\left.\mathcal{L} \right|_{\text{bos.}} &=  - K_{\phi \bar{\phi}} \partial_\mu \phi \partial^\mu \bar{\phi} + \frac{1}{4} K_{\phi \bar{\phi}}  \left( \ast F_4 + i D \right)  \left( \ast F_4 - i D \right) +  \frac{1}{2}\left( \ast F_4 + i D \right) W_\phi +  \frac{1}{2} \left( \ast F_4 - i D \right) \bar{W}_{\bar{\phi}} \nonumber \\[7pt]
	&= - K_{\phi \bar{\phi}} \partial_\mu \phi \partial^\mu \bar{\phi} + \frac{1}{4} K_{\phi \bar{\phi}} \left( \ast F_4 \right)^2 + \frac{1}{2} \left( \ast F_4 \right) \left( W_\phi + \bar{W}_{\bar{\phi}} \right)  + \frac{1}{4} K_{\phi \bar{\phi}} D^2 + \frac{i}{2} D \left( W_\phi - \bar{W}_{\bar{\phi}} \right) \label{eq:L_off_shell_1d}
	\end{align}
	The equation of motion for the auxiliary field $D$ reads
	\begin{align}
	D=- i K^{\phi \bar{\phi}} \left( W_\phi - \bar{W}_{\bar{\phi}} \right)
	\end{align}
	which is completely algebraic, as expected. Plugging this into \eqref{eq:L_off_shell_1d} yields
	\small
	\begin{align}
	\left.\mathcal{L} \right|_{\text{bos.}} &=  - K_{\phi \bar{\phi}} \partial_\mu \phi \partial^\mu \bar{\phi} +  \frac{1}{4} K_{\phi \bar{\phi}} \left( \ast F_4 \right)^2 + \frac{1}{2} \left( \ast F_4 \right) \left( W_\phi + \bar{W}_{\bar{\phi}} \right) + \frac{1}{4} K^{\phi \bar{\phi}} \left( W_\phi - \bar{W}_{\bar{\phi}} \right)^2 \\[7pt]
	&=  - K_{\phi \bar \phi} ~\partial_{\mu} \phi \partial^{\mu} \bar \phi - \frac{1}{4\cdot 4!} K_{\phi \bar \phi} ~F^{\mu \nu \sigma \rho} F_{\mu \nu \sigma \rho} + \frac{1}{2\cdot 4!} \left(  W_{\phi} + \bar{W}_{\bar \phi}\right) ~\epsilon^{\mu \nu \sigma \rho} F_{\mu \nu \sigma \rho} + \frac{1}{4}  K^{\phi \bar{\phi}} \left(  W_{\phi} - \bar{W}_{\bar \phi}\right)^2
	\label{eq:L_3}
	\end{align}
	\normalsize
	where, in the second step, we have rewritten the Hodge duals in terms of their original tensor fields for clarity.
	
In order to proceed, recall that the physical field we wish to extremise in the action is not the field strength, but rather its antisymmetric tensor potential $A_{\mu \nu \rho}$. In the following, it will be more convenient to work with the Hodge-dual $A^\mu$, which is related to $\ast F_4$ by eq.~\eqref{eq:F4_flat}. The variation of \eqref{eq:L_3} with respect to $A_\mu$ is
	\begin{align}
	& \frac{1}{2} K_{\phi \bar{\phi}} \left( \ast F_4 \right) \left( \partial^\mu \delta A_\mu \right) +  \frac{1}{2} \left( \partial^\mu \delta A_\mu \right) \left( W_\phi + \bar{W}_{\bar{\phi}} \right) \nonumber \\[7pt]
	&\quad = \delta A_\mu  \left[ \frac{1}{2} \partial^\mu \left( - K_{\phi \bar{\phi}} \left( \ast F_4 \right) + W_\phi + \bar{W}_{\bar{\phi}} \right) \right] + \partial^\mu \left[ \delta A_\mu  \frac{1}{2} \left( K_{\phi \bar{\phi}} \left( \ast F_4 \right) - W_\phi - \bar{W}_{\bar{\phi}} \right)\right] \; .
	\label{eq:A3_variation}
	\end{align}
	We can clearly see that the vanishing of the first term in the r.h.s. will give us an equation of motion for the 3-form gauge field, while the second one will produce a boundary term. As originally discussed in \cite{Brown:1988kg}, in order to deal with the second term in \eqref{eq:A3_variation} we will need to add a boundary term to our original Lagrangian  to convert this contribution into an expression containing the variations of gauge invariant quantities
 ($F_4$ and scalar fields). Indeed, otherwise the variational problem with respect to gauge potential would not be well posed (see \cite{Farakos:2017jme} for recent discussion).
 This boundary term not only ensures the consistency of the variational problem but, as we will see shortly, it will have a noticeable effect on the final, on-shell result.

 The required boundary term is given by
	\begin{align}
	\mathcal{L}_{bd} &= - \frac{1}{2} \partial^\mu \left[  A_\mu   \left( K_{\phi \bar{\phi}} \left( \ast F_4 \right) - W_\phi - \bar{W}_{\bar{\phi}} \right)\right] \nonumber \\[7pt]
	&= \frac{1}{2 \cdot 3!} \partial^\mu \left[ A^{\nu\rho\sigma} \left( K_{\phi\bar{\phi}} F_{\mu \nu \rho \sigma} + \epsilon_{\mu \nu \rho \sigma} \left( W_{\phi} + \bar{ W}_{\bar{ \phi}} \right) \right) \right]
	\end{align}
	while the equation of motion for the form field is
	\begin{align}
	\partial_\mu \left( \frac{1}{2} K_{\phi \bar{\phi}} (\ast F_4) + \Re W_{\phi} \right) = 0 \quad \Rightarrow  \quad   \ast F_4 = -2 K^{\phi \bar{\phi}}\left( \text{Re} W_\phi - n \right)
	\label{eq:on_shell_F4}
	\end{align}
	where $n \in \mathbb{R}$ is an arbitrary real (integration) constant. As this constant appears in the expression for the 4-form field strength, it can be identified with the flux of the 3-form gauge field.
	
	Plugging this result into \eqref{eq:L_3} and taking into account the non-vanishing contribution of the boundary term yields
	\begin{align}
	\left. \mathcal{L} \right|_{\text{bos.,on-sh.}} = - K_{\phi \bar{\phi}} \partial_\mu \phi \partial^\mu \bar{\phi} - K^{\phi \bar{\phi}}\left( W_\phi - n  \right)  \left( \bar{W}_{\bar{\phi}} - n  \right) .
	\end{align}
	From this final form of the Lagrangian we can conclude that the contribution of 3-forms in a supersymmetric setup results in a linear contribution to our original superpotential  in which the field is multiplied by a  constant  associated with the flux of the 3-form gauge field. Thus, setting the 3-forms on shell reduces the original theory to a model of a scalar field described by the original Kähler potential and an effective superpotential given by
	\begin{align}
	\hat{W} (\phi) \equiv W(\phi) - n \phi.
	\end{align}

	\renewcommand{\theequation}{B.\arabic{equation}}
	\section{Rigid supersymmetry, bosonic membranes, and BPS equations}
	\label{sec:BPS_eqs}

	As it was shown in \cite{Bandos:2001jx,Bandos:2005ww}, the action of the interacting system composed of the supergravity multiplet and a bosonic membrane is invariant under a half of local supersymmetry, provided the membrane term of the interacting action is given by the bosonic `limit' of a supermembrane. Furthermore, the preserved part of the local supersymmetry reflects the local fermionic $\kappa$--symmetry of the original supermembrane action.

	In the case of an interacting system composed of supersymmetric matter and a supermembrane, there is no rigorous way to find rigid supersymmetry invariance of the interacting system including the matter supermultiplet and a bosonic membrane. However, as we will show below, there exists a trick allowing us to see the $\kappa$-symmetry of the original supermembrane part of the action for the interacting system. It implies the manipulation of the supersymmetry transformation of the bosonic membrane action with the use of some ansatze both for the fields and for membrane configuration. 	Actually, such a possibility, when it exists, reflects the existence of purely bosonic supersymmetric solutions of the complete supersymmetric system.\footnote{The relation of the $\kappa$-symmetry of a super-p-brane  with supersymmetry preserved by solutions of equations was described for the first time in \cite{Bergshoeff:1997kr}. } In the following we will show how using this arguments we obtained the simple form of the BPS equations for our brane in our model.
	
Let us start by looking at the supersymmetry transformation in our model. Although we did not write those transformations of the 3-form supermultiplet in the main text,
they can be easily restored from the association of the spacetime fields with the superfield components in eqs. (\ref{superfield-components-1}, \ref{superfield-components-2}, \ref{superfield-components-3}) and the identification of supersymmetry as fermionic supertranslations in superspace. This leads to the following supersymmetric transformations for the fields,
	\begin{eqnarray}
		\delta_\epsilon \phi &=&\epsilon^\alpha  D_\alpha Y\vert = \epsilon^\alpha \lambda_\alpha\; , \qquad \\ \label{susy=la}
		\delta_\epsilon \lambda_\alpha &=& \epsilon^\beta  D_\beta D_\alpha Y\vert + \bar{\epsilon}_{\dot\beta} \bar{D}^{\dot\beta}  D_\alpha Y\vert =
		2 \epsilon_\alpha F_Y +2i (\sigma^\mu \bar{\epsilon})_\alpha  \partial_\mu \phi \qquad\nonumber \\ && =  \epsilon_\alpha (\partial_\mu A^\mu +i{\rm D}) +2i (\sigma^\mu \bar{\epsilon})_\alpha  \partial_\mu \phi \; , \qquad
	\end{eqnarray}
	and
	\begin{equation} \delta_\epsilon F_Y = \frac 1 2 (\partial_\mu \delta_\epsilon  A^\mu +i\delta_\epsilon{\rm D}) = -\frac 1 4 \bar{\epsilon}_{\dot\beta} \bar{D}^{\dot\beta}  D^2 Y\vert = i(\partial_\mu \lambda\sigma^\mu \bar{\epsilon}) \, .
	\end{equation}
	These transformations also imply
	\begin{equation} \delta_\epsilon{\rm D} = \partial_\mu \lambda\sigma^\mu \bar{\epsilon}+  {\epsilon}\sigma^\mu \partial_\mu\bar{ \lambda}\; , \qquad \delta_\epsilon  A^\mu = i (\partial_\mu \lambda\sigma^\mu \bar{\epsilon}-{\epsilon}\sigma^\mu \partial_\mu\bar{ \lambda})   \; , \qquad
	\end{equation}
	as well as
	\begin{equation} \delta_\epsilon  A_{\mu\nu\rho} = i \epsilon_{\mu\nu\rho\sigma} (\partial_\mu \lambda\sigma^\sigma \bar{\epsilon}-{\epsilon}\sigma^\sigma \partial_\mu\bar{ \lambda})    ~,
	\end{equation}
	for the dual 3-form $A_{\mu\nu\rho} = \epsilon_{\mu\nu\rho\sigma} A^\sigma$.
	
	Let us now perform the above supersymmetry transformation to the scalar and 3-form field in the bosonic membrane action \eqref{eq:s_memb}:
	\begin{equation}\label{susySm=}
		\delta_\epsilon  S_{{\rm membr.}}= - 2|q|\int d^3\xi \sqrt{-h} \, \Re  \left(\epsilon^\alpha \lambda_\alpha \,\frac {\bar{\phi}}{|\bar{\phi}|}\right)
		+2 q \int d^3\xi \, \Im (\lambda \sigma^\mu \bar{\epsilon} )\epsilon_{\mu\nu\rho\sigma} \partial_0x^\nu \partial_1x^\rho \partial_0x^\sigma\; .
	\end{equation}
	Generically, this last expression does not vanish. However, if we consider a flat membrane lying at $x^3=z=0$  in the static gauge, i.e.,
	\begin{equation}\label{static}
		x^a(\xi)=\xi^a\; , \qquad x^3(\xi):=z(\xi)=0\; ,
	\end{equation}
	then $\sqrt{-h}=1$ and  eq.~(\ref{susySm=}) reduces to
	\begin{equation}\label{susySm==}
		\delta_\epsilon  S_{{\rm membr.}}= - 2|q|\int d^3\xi \Re  \left(\epsilon^\alpha \lambda_\alpha \,\frac {\bar{\phi}}{|\bar{\phi}|}- i \frac q {|q|} (\bar{\epsilon}\tilde{\sigma}{}^3\lambda)
		\right)\; .
	\end{equation}
	This formally vanishes if we set
	\begin{equation}\label{susy=kap}
		\epsilon^\alpha= i (\bar{\epsilon}\tilde{\sigma}{}^3)^\alpha\, \left. \frac{q\phi}{|q\phi|}\right|_{z=0} \; .
	\end{equation}
	This equation has nontrivial solutions for a constant fermionic spinor $\epsilon^\alpha$ if $\left. \frac {{q\phi}}{|{q\phi}|}\right|_{z=0} $ is independent of $\xi^a$ (i.e., as long as it represents a constant phase). This condition is clearly satisfied if we assume the scalar field to depend on the $z$ coordinate only, that is,
	\begin{equation}\label{phi=z}
		\phi (x^a, z)=\phi (z)\; .
	\end{equation}
	This and the restriction to a flat membrane given by the last equation in \eqref{static} define the ansatz for the supersymmetric bosonic solution
	of the interacting system of the 3-form multiplet and supermembrane.
	
	Let us repeat that these formal calculations reflect the $\kappa$--symmetry of the complete supermembrane action, which guarantees the existence of the purely bosonic  supersymmetric solution of the equations for the interacting system of a supermembrane and the single 3-form matter supermultiplet.

	Let us look at the transformation of the fields in the bulk. For a purely bosonic solution obtained with the ansatz \eqref{phi=z}, the preservation of supersymmetry implies $\delta_\epsilon\lambda=0$ which, after using (\ref{susy=la}) and the auxiliary field's equation
	\begin{align}
	2F_Y=(*F_4+i{\rm D})=-K^{\phi\phi}\hat{\bar{W}}_{\bar{\phi}} \,
	\end{align}
	 reduces to
	\begin{equation} i(\sigma^3\bar{\epsilon})_\alpha \partial_z\phi =\epsilon_\alpha K^{\phi\phi}\hat{\bar{W}}_{\bar{\phi}}\; .
	\end{equation}
	This equation has a nontrivial solution with the constant fermionic spinor obeying projecting condition
	\begin{equation}\label{susy=ft}
		\epsilon_\alpha = e^{i\eta} i(\sigma^3\bar{\epsilon})_\alpha \qquad \Leftrightarrow \qquad \epsilon^\alpha = - e^{i\eta} i(\bar{\epsilon}\tilde{\sigma}^3)^\alpha
	\end{equation}
	if the scalar field obeys the BPS equation
	\begin{equation}
		\partial_z \phi(z) = e^{i \eta} K^{\phi \bar \phi} \,  \hat{\bar{W}}_{\bar \phi}\; .
		\label{eq:BPS-A}
	\end{equation}
	Furthermore, the supersymmetry preserved by the bosonic configuration will also preserve the supersymmetry of the static flat bosonic membrane confirguration
	given by (\ref{static}) if (\ref{susy=ft}) coincides with (\ref{susy=kap}), i.e. when
	\begin{align}
			e^{i \eta}=-\left. \frac {q\phi}{|q\phi|}\right|_{z=0} \, .
	\end{align}
	
	Given the BPS equation \eqref{eq:BPS-A}, it is easy to check that the phase of $\partial_z \hat{W}$ remains constant \emph{even across the membrane}. Indeed, if we multiply both sides of the equation by $\hat{W}_\phi$, we find
	\begin{align}
		(\partial_z \phi) \hat{W}_\phi = e^{i\eta} \hat{W}_\phi  K^{\phi \bar \phi} \,  \hat{\bar{W}}_{\bar \phi} \qquad \Rightarrow \qquad \partial_z \hat{W} = e^{i\eta} \left[ V(\phi,\bar{\phi}) + |q \phi | \delta (z) \right]
	\end{align}
	Therefore, one finds that $e^{i\eta}$ may be written as
	\begin{align}
		e^{i\eta} = - \left. \frac{q\phi}{|q\phi|}\right|_{z=0} = \frac{\Delta \hat{W}}{|\Delta \hat{W}|}
		\label{eq:BPS_phase}
	\end{align}
	where $\Delta \hat{W} \equiv \hat{W}_{z=+\infty} - \hat{W}_{z=-\infty}$. Note that this last result exactly reproduces the phase obtained in \cite{Cvetic:1992bf}, albeit in our case it includes the effect of the membrane localized at $z=0$, which effectively changes the superpotential when crossing it.
	
	One may also use this result in order to find a closed expression for the tension of the  domain wall solution in the presence of a supermembrane. As shown in \cite{Lanza:2019nfa}, one may rewrite the action \eqref{eq:on_shell_L} as
	\begin{align}
		S = &- \int d^3x  \int dz K_{\phi\bar{\phi}} \left[ \partial_z \phi - e^{i\beta} K^{\phi\bar{\phi}} \hat{\bar{W}}_{\bar{\phi}} \right]  \left[ \partial_z \bar{\phi} - e^{-i\beta} K^{\phi\bar{\phi}} \hat{W}_{\phi} \right]  \nonumber \\[5pt]
		&- \int d^3x \left(2 |q\phi|_{z=0} + 2 \Re \left[ e^{-i\beta} (\Delta \hat{W} + q\phi|_{z=0} ) \right] \right) .
	\end{align}
	where the phase $e^{i\beta}$ is, at this stage, arbitrary. However, upon choosing $e^{i\beta}=e^{i\eta}$ one can see that the expression for the tension of any field configuration is maximized by the solution of the BPS equation \eqref{eq:BPS-A}. Taking this into account the tension becomes
		\begin{align}
		\mathcal{T}_{\text{DW+memb.}} = |2 \Delta\hat{W}| \ .
	\end{align}
Note that even though this expression is formally similar to the one in the scalar field model in \cite{Cvetic:1992bf} the result is now
written in terms of the effective superpotential $\hat{W}$ which includes the jump due to the different fluxes on both sides of the membrane.
	
	As described in \cite{Bandos:2018gjp} and references therein, similar arguments can be applied to obtain the counterpart of \eqref{eq:BPS-A} for dynamical systems including supergravity, which lead to  eqs.~\eqref{dzphi}-\eqref{dzD}.

	\renewcommand{\theequation}{C.\arabic{equation}}
	\section{Three-form multiplets in supergravity}
	\label{sec:appendix_sugra}
	
	In this section we will follow a similar reasoning as the one above, with gravity included. We will first present the action for scalar multiplets described by generic chiral superfields, which we will later on generalize to include special chiral superfields which have vector or 3-form components among their bosonic ingredients. All of the results we present here have also been derived in \cite{Farakos:2017jme,Lanza:2019nfa} using a super-Weyl invariant approach to matter-coupled supergravity, reaching the same conclusions.
	
	We recall that supergravity can be described in terms of the superspace supervielbein $E_M^A (z)$ and the spin connection $\omega_M^{ab}(z)=-\omega_M^{ba}(z) $ subject to a set of torsion constraints. After fixing the so-called Wess-Zumino gauge, the field content reduces to
	\begin{itemize}
		\item $e_a^\mu$, the vielbein,
		\item $\psi_\alpha^\mu$, the gravitino,
		\item $b^\mu$, a real vector auxiliary field,
		\item $M$, a complex scalar auxiliary field.
	\end{itemize}
 The most general superspace action of interacting supergravity  and  scalar supermultiplets in terms of $E_M^A (z)$ and generic chiral superfields (defined in curved supergravity superspace) can be found e.g. in \cite{Wess:1992cp}. The spacetime action for the component fields is then obtained upon fixing the Wess-Zumino gauge and integrating over the fermionic coordinates of superspace. In the conventions of \cite{Wess:1992cp}, the spacetime Lagrangian of the bosonic sector of such an action reads
	\begin{align}\label{Lx=SG+SC=Om}
	\frac{1}{\sqrt{-g}}{\cal L}=\frac 1 2 \, R e^{-\frac 1 3 K}+  \Omega_{a\bar{b}}\partial_\mu \phi^a\partial^\mu\bar{\phi}^{\bar{b}} - \frac 1 3 e^{-\frac 1 3 K} \tilde{M}\tilde{\bar{M}} -  \tilde{M}\bar{W}- \tilde{\bar{M}} W \nonumber \\ +e^{-\frac 1 3 K} K_{a\bar{b}}F^a\bar{F}^{\bar{b}} +F^a(W_a+K_a W)+\bar{F}^{\bar{b}}(\bar{W}_{\bar{b}}+K_{\bar{b}}\bar{W}) \nonumber \\
	- \frac 19 \Omega b_\mu b^\mu - \frac i 3 b^\mu (\partial_\mu \phi^a\, \Omega_a-\partial_\mu \bar{\phi}^{\bar{b}}\, \Omega_{\bar{b}} )\; , \qquad
	\end{align}
 where
	\begin{align}
	\Omega (\Phi,\bar{\Phi}) =-3 e^{-\frac 1 3 K (\Phi,\bar{\Phi})}, \qquad
	\tilde{M}=M+  K_{\bar{a}}\bar{F}^{\bar{a}}\; , \qquad \tilde{\bar{M}}= {\bar{M}}+  K_{a}F^a\; .
	\end{align}
	This action is not written in Einstein frame. Therefore, it is customary to rescale the vielbein as follows:
	\begin{align}\label{W=ea=K/6}
	e_\mu^a\; \mapsto\; e_\mu^a\,e^{\frac 1 6 K}\; , \qquad
	\end{align}
	and to supplement this with a suitable transformation of the spin connection. Furthermore, rescaling the auxiliary fields as
	\begin{align}\label{sW=FiM}
	F^i\,\mapsto \, F^i \,e^{-\frac 1 6 K}\; , \qquad M\,\mapsto \, M \,e^{-\frac 1 6 K}\; , \qquad
	\end{align}
	we arrive at the following action in Einstein frame
	\begin{align}\label{cL=auxEF}
	\frac{1}{\sqrt{-g}}{\cal L}= &\frac 1 2 \,R- K_{a \bar{b}}\partial_\mu \phi^a \partial^\mu\bar{\phi}{}^{\bar{b}} - \frac 1 3 \tilde{M}\tilde{\bar{M}} -  e^{\frac 1 2 K}  \tilde{M}\bar{W}-  e^{\frac 1 2 K}   \tilde{\bar{M}} W \qquad  \nonumber \\ &+  K_{a \bar{b}}F^a \bar{F}{}^{\bar{b}} + e^{\frac 1 2 K} F^a(W_a+K_a W)+ e^{\frac 1 2 K} \bar{F}{}^{\bar{b}}(\bar{W}_{\bar{b}}+K_{\bar{b}}\bar{W})\; . \qquad
	\end{align}
	Notice that in this last expression the auxiliary fields  $b^\mu$ have been integrated out using their equations of motion. If we are dealing with minimal supergravity and scalar multiplets, $\tilde{M}$ and all $F^i$ are independent, and the auxiliary field equations read
	\begin{align}\label{Fi=}
	\tilde{M}=-3e^{\frac 1 2 K} W\; , \qquad F^a K_{a \bar{b}}= - e^{\frac 1 2 K} (\bar{W}_{\bar{b}}+K_{\bar{b}}\bar{W})\; .
	\end{align}
	Substituting this into \eqref{cL=auxEF} we find the well known matter-coupled $\text{N=1}$, $D=4$ supergravity Lagrangian
	\begin{align}
	\frac{1}{\sqrt{-g}}{\cal L}= \frac 1 2 \, {\cal R}- K_{a\bar{b}}\partial_\mu \phi^a \partial^\mu\bar{\phi}{}^{\bar{b}} - V(\phi,\bar{\phi})
	\end{align}
	where the potential is given by
	\begin{align}
	V(\phi,\bar{\phi})=
	e^{K}\left( D_a W  K^{a \bar{b}} D_{\bar{b}} \bar{W}- 3|W|^2\right),
	\end{align}
	and $D_a = \partial_a + K_a$ are the so-called  Kähler-covariant derivatives.
	
	\subsection{Supergravity interacting with 3-form multiplets}
	
	Just as in the non-gravitational case, we will implicitly introduce three-forms by passing from generic to special chiral superfields. In this case,  special chiral superfields describing single three-form  multiplets are given by
	\begin{align}\label{S=bD2P}
	S=-\frac i 4 (\bar{D}^2-8 \mathcal{R}) \, {\cal P} \; , \qquad  {\cal P}=({\cal P})^*
	\end{align}
	where $\mathcal{R}$ is the so-called main chiral superfield of minimal supergravity, whose leading component is proportional to the complex scalar auxiliary field, $\mathcal{R}\vert =-\frac 1{6} M$, and ${\cal P}$ is an unconstrained real superfield. The latter is defined up to shift by a real linear superfield \eqref{S=bD2P}
	\begin{align}\label{P->P+L}
	{\cal P}\mapsto {\cal P} + L, \qquad  (\bar{D}^2-8 \mathcal{R})L=0= (D^2-8\bar{\mathcal{R}})L .
	\end{align}
	This freedom is the manifestation of a gauge symmetry which can be used to fix the Wess-Zumino gauge, where
	\begin{align}\label{P0=0}
	{\cal P} \vert =0\; , \qquad {D}_{\alpha} {\cal P} \vert =0\; , \qquad \bar{D}_{\dot\alpha} {\cal P} \vert =0\; . \qquad
	\end{align}
	The remaining part of the $L$ symmetry, preserving this gauge, coincides with the 2-form gauge symmetry for the
	3-form dual to vector component of the prepotential superfield,
	\begin{align}\label{Cmu=}
	\sigma^a_{\alpha\dot\alpha}[D^{\alpha}, \bar{D}{}^{\dot\alpha}] {\cal P} \vert = 4 A^{a}\; , \qquad   A^{a}=*(A_3 )^a\; .
	\end{align}
    In the gauge \eqref{P0=0}, the highest component of the special chiral superfield of $S$ simplifies to (in the notation of \cite{Wess:1992cp})
	\begin{align}
	\label{eq:FS}
	\mathcal{F} \equiv F_S  = -\frac 1 4 \, D^2S \vert =\frac i {16} \, D^2\bar{D}{}^2 {\cal P} \vert - \frac i {2} \, \mathcal{R}  {D}{}^2 {\cal P} \vert
	= \frac 1 2 \left(D_\mu A^\mu + i{\rm d} \right)+\frac 1 3 (s \bar{M}+2\bar{s}{}  {M}) \; .
	\end{align}
where $s=\left. S \right|$, $M$ and $\bar{M}$ are the scalar auxiliary fields of minimal supergravity, $\rm d$
is a real auxiliary scalar field,  and
\begin{align}
	\ast F_4 = D_{\mu} A^{\mu} = \frac{1}{e} \partial_\mu (e A^\mu) \; , \qquad  e=\det e_\mu^a=\sqrt{-g}\; .
\end{align}
	
Fixing the WZ gauge and integrating over the fermionic coordinates of superspace in the superfield action we arrive at the following Lagrangian in the bosonic limit:
	\begin{align}\label{Lx=SG+s3=Om}
	\frac{1}{\sqrt{-g}}{\cal L}= \frac 1 2 \, R e^{-\frac 1 3 K}+  \Omega_{s\bar{s}}\partial_\mu s \partial^\mu\bar{s} - \frac 1 3 e^{-\frac 1 3 K} (M+  K_{\bar{s}}\bar{\mathcal{F}})\, ({\bar{M}}+  K_s \mathcal{F} ) -  {M}\bar{W}- {\bar{M}} W \nonumber \\ +e^{-\frac 1 3 K} K_{s\bar{s}} \mathcal{F} \bar{\mathcal{F}} + \mathcal{F} W_s+ \bar{\mathcal{F}}\bar{W}_{\bar{s}}
	- \frac 19 \Omega b_\mu b^\mu - \frac i 3 b^\mu (\partial_\mu s \, \Omega_s-\partial_\mu \bar{s} \Omega_{\bar{s}} ) + \frac{1}{\sqrt{-g}}\mathcal{L}_{bd} \,
	\end{align}
 This Lagrangian formally coincides with that of scalar multiplets interacting with supergravity \eqref{cL=auxEF} up to boundary term, and up to the composite nature of the F-component ${\cal F}$ of the special chiral superfields  \eqref{eq:FS}.

 Before substituting the expression for the F-components of the special chiral superfields, \eqref{eq:FS}, it is convenient to perform a Weyl rescaling of the fields with \eqref{W=ea=K/6}, which will bring our action to the Einstein frame. In the follwoing, will carefully consider the case of supergravity interacting with special chiral superfields describing 3-form multiplets, as this case has some peculiarities with respect to the case of generic chiral superfields describing scalar supermultiplets.

\subsubsection{Super-Weyl transformations}
	
 At the beginning of this appendix, when considering the interaction of supergravity an chiral superfields $\Phi$, we assumed that each superfield $\Phi$ and its components $\phi$ and $F $ are inert under Weyl transformations. This is a consistent assumption in the case of a generic chiral superfield.\footnote{This can be checked by writing the chiral field in terms of an unconstrained complex superfield (with a similar definition as \eqref{S=bD2P}). Choosing the Weyl weights of the transformation accordingly, it can be shown that a general chiral superfield $\Phi$ is invariant under these rescalings.} However, since the chiral superfield $S$ has been written in terms of a real superfield $\mathcal{P}$, it is important to check how \emph{super-Weyl transformations}, act on them. These are defined via the following transformations of supervielbein \cite{Wess:1992cp,buchbinder1998ideas}:
	\begin{eqnarray}
	\label{supW-Ea=S}
	E^a &\mapsto & \tilde{E}{}^a= e^{\Upsilon+\bar{\Upsilon}} E^a \; , \qquad  \\
	\label{supW-Ef=S}
	E^{\alpha} &\mapsto & \tilde{E}{}^{\alpha}= e^{2\bar{\Upsilon}-{\Upsilon}} \left(E^{\alpha} - \frac{i}{4} E^a \bar{\cal D}_{\dot{\alpha}} \bar{\Upsilon} \tilde{\sigma}{}_a^{\dot{\alpha}{\alpha}} \right)\; , \qquad \\
	\label{supW-bEf=S}
	\bar{E}{}^{\dot\alpha} &\mapsto & \tilde{\bar{E}}{}^{\dot\alpha}= e^{2{\Upsilon}-\bar{\Upsilon}} \left(\bar{E}{}^{\dot\alpha} + \frac{i}{4} E^a \tilde{\sigma}{}_a^{\dot{\alpha}{\alpha}} {\cal D}_\alpha {\Upsilon} \right)\; , \qquad \end{eqnarray}
 where ${\Upsilon}$ is a chiral superfield
	\begin{eqnarray}
	\label{bDUp=0}
	\bar{D}_{\dot\alpha} {\Upsilon}=0\; , \qquad D_\alpha \bar{\Upsilon}=0  \; .
	\end{eqnarray}
   These transformations must be supplemented by a suitable transformations of the spin connection; however, their explicit form is not needed for our purposes (see \cite{Wess:1992cp,buchbinder1998ideas} for more detail).
	
 It is important to note that the supergravity chiral projector transforms in an inhomogeneous way under super-Weyl transformations:
	\begin{eqnarray}
	\label{sW=bD2-8R}
	(\bar{D}\bar{D}- 8{\mathcal{R}})  \mapsto   e^{-4{\Upsilon}}(\bar{D}\bar{D}- {\mathcal{R}}) e^{2\bar{\Upsilon}} , \qquad (DD-8\bar{\mathcal{R}}) \mapsto e^{-4\bar{\Upsilon}} (DD-8\bar{\mathcal{R}}) e^{2\Upsilon} .
	\end{eqnarray}
	In the case of a generic chiral superfield $\Phi= (\bar{D}\bar{D}- 8{\mathcal{R}})P$, this projector acts on the generic complex superfield potential $P$.  Choosing the transformation of this superfield to be $P\mapsto e^{+4\bar{\Upsilon}}e^{-2\Upsilon} P$ we can actually make $\Phi$ inert under the super-Weyl tranformations.

On the other hand, this is not possible in the case of a special chiral superfield $S$ \eqref{S=bD2P} constructed from the real superfield prepotential ${\cal P}=({\cal P})^*$. In this case, in the light of \eqref{sW=bD2-8R}, the only way to obtain a covariant super-Weyl transformation of the special chiral superfields \eqref{S=bD2P} is to attribute to its real prepotential the transformation rule
	\begin{align}
	{\cal P} \,\mapsto {\cal P} \, e^{-2\Upsilon-2\bar{\Upsilon}}
	\end{align}
	which results in
	\begin{align}
	S \,\mapsto S \, e^{-6\Upsilon}
	\end{align}
and in the following transformations of its leading component\footnote{We do not write transformation of $P\vert$ explicitly because it vanishes in  the Wess-Zumino gauge \eqref{P0=0}.}
	\begin{align}\label{s->s}
	s \,\mapsto s \, e^{-6\Upsilon \vert }\; .
	\end{align}
	
	We will be interested in the purely bosonic part of the super-Weyl transformations with
	\begin{align}\label{sW->W}
	\Upsilon \vert= \frac{1}{12} K =\bar{\Upsilon} \vert\; , \qquad D_\alpha \Upsilon \vert=0\; ,\qquad D^2\Upsilon \vert=0\; ,
	\end{align}
	since, in that case,
	\begin{align}\label{W2=ea=K/6}
	e_\mu^a\; \mapsto\; e_\mu^a\,e^{\frac 1 6 K}\;
	\end{align}
	as needed to write the Lagrangian in Einstein frame, just as in the case of supergravity interacting with chiral multiplets. However, in this scenario, both  $M$ and  $\mathcal{F}$ will be affected by this transformation.  This can be seen from
	\begin{align}
	(D^2 - 8 \bar{\mathcal{R}}) S | = - 4 \mathcal{F} - 8 \mathcal{R} S | = -4 \mathcal{F} + \frac{4}{3} s \bar{M}
	\end{align}
whose transformation with the use of \eqref{s->s} and \eqref{sW->W} results in
	\begin{align}
	s \ &\mapsto \ e^{-\frac{1}{2} K} s \\
	\mathcal{F}  \ &\mapsto \ e^{-\frac 2 3 K} \mathcal{F} \\
	M \ &\mapsto \ e^{-\frac 1 6 K} M\; .
	\end{align}
As far as the scalar field is concerned, it is convenient to combine the Weyl rescaling with the field redefinition \begin{align}\label{s->phi}
	\phi := e^{-\frac 1 2 K} s
	\end{align}
so that the kinetic term of $\phi$ field remains in a canonical form.
	
	Taking all of the above into account and integrating out the auxiliary field $b^\mu$ using its algebraic equations of motion, we find that
	\begin{align}\label{Lx=SG+s3}
	\frac{1}{\sqrt{-g}}{\cal L}= &\frac 1 2 \, R -  K_{\phi \bar{\phi}}\partial_\mu \phi \partial^\mu\bar{\phi}{}^{J} - \frac 1 3 (M+  e^{-\frac 1 2  K} K_{\bar{\phi}}\bar{\mathcal{F}})\, ({\bar{M}}+  e^{-\frac 1 2  K} K_{\phi} \mathcal{F} )  \nonumber \\ &-   e^{\frac 1 2 K}{M}\bar{W}-  e^{\frac 1 2 K} {\bar{M}} W+e^{- K} K_{\phi \bar{\phi}} \mathcal{F} \bar{\mathcal{F}} + \mathcal{F} W_\phi+\bar{\mathcal{F}}\bar{W}_{\bar{\phi}} + \frac{1}{\sqrt{-g}} \mathcal{L}_{bd}.
	\end{align}
	It is convenient to further redefine the supergravity auxiliary field $M$ as
	\begin{align}
	\check{M} \equiv M e^{\frac 1 2 K}
	\end{align}
	so that the Einstein frame Lagrangian reads
	\begin{align}
	\frac{1}{\sqrt{-g}}{\cal L} = &\frac 1 2 \, R -  K_{\phi \bar{\phi}}\partial_\mu \phi \partial^\mu\bar{\phi} - \frac 1 3 e^{-K} (\check{M}+  K_{\bar{\phi}}\bar{\mathcal{F}})\, (\check{\bar{M}}+   K_{\phi} \mathcal{F} ) -  \check{M}\bar{W}-  \check{\bar{M}} W \nonumber \\ &+e^{- K} K_{\phi\bar{\phi}}\mathcal{F}\bar{\mathcal{F}} + \mathcal{F} W_\phi +\bar{\mathcal{F}}\bar{W}_{\bar{\phi}}  + \frac{1}{\sqrt{-g}} \mathcal{L}_{bd}
	\label{eq:rescaled_L}
	\end{align}
with the composite $\mathcal{F}$ field of the form
	\begin{align}\label{Fs3==}
	\mathcal{F} = \frac 1 2 \left(D_\mu A^{\mu} + i{\rm d} \right)+\frac 2 3 \bar{\phi}{}  \check{M}
	+\frac 1 3 \phi \check{\bar{M}} .
	\end{align}

	The boundary term can be obtained in the same fashion as in the previous section. Since the 3-form field enters  our action only through $D_\mu A^\mu =\frac 1 e \partial_\mu (eA^\mu)$, varying the action with respect to this field gives
	\begin{align}
	\delta S &= \int d^4 x \ e \  \frac{\partial \mathcal{L}}{\partial (D_\mu A^\mu )} \ \left( \frac{1}{e} \partial_\mu (e  \ \delta A^\mu) \right) \nonumber \\[7pt]
	&=- \int d^4 x \ e  \ \delta A^\mu \ \partial_\mu \left(  \frac{\partial \mathcal{L}}{\partial (D_\mu A^\mu )} \right) + \int d^4 x \ \partial_\mu \left(e \ \delta A^\mu \frac{\partial \mathcal{L}}{\partial (D_\mu A^\mu)} \right).
	\label{eq:dS}
	\end{align}
	Therefore, we conclude that the boundary term, which makes the variational problem well posed, includes, besides the Gibbons-Hawking term \cite{Gibbons:1976ue} and its superpartner (see \cite{Luckock:1989jr,Moss:2003bk}),
	\begin{align}
	\mathcal{L}_{bd} = - \int d^4 x \ \partial_\mu \left(e \  A^\mu \frac{\partial \mathcal{L}}{\partial (D_\mu A^\mu)} \right).
	\end{align}
 The  auxiliary field and 3-form equations of motion are then
	\begin{align}
	\frac{\partial \mathcal{L}}{\partial {\rm d}} = 0, \qquad \frac{\partial \mathcal{L}}{\partial \check{M}} = 0, \quad \text{and} \quad
\partial_\nu \left( \frac{\partial \mathcal{L}}{\partial (D_\mu A^\mu)}\right) = 0 \ \Rightarrow  \ \frac{\partial \mathcal{L}}{\partial (D_\mu A^\mu)} = n,
	\label{eq:eoms}
	\end{align}
	where $n \in \mathbb{R}$. These algebraic equations are solved by
	\begin{align}
	{\rm d} &= - i K^{ \phi \bar{ \phi}} e^K \left[D_{ \phi} (W- n  \phi)  - \text{c.c.} \right]
	\label{eq:d_sugra} \\[7pt]
	\check{M} &= e^K \left[ D_{ \phi} (W - n  \phi) K^{ \phi \bar{ \phi}} K_{\bar \phi} - 3 (W-n \phi) \right] \\[7pt]
	D_\mu A^\mu &= e^K \left[ 3 (W-n \phi)\bar{ \phi} - (1+\bar{ \phi}K_{\bar \phi}) K^{ \phi \bar{ \phi}} D_{ \phi} (W-n \phi)  + \text{c.c.} \right]
	\label{eq:F4_sugra}
	\end{align}
	where $D_{\phi} = \partial_\phi + K_\phi$ is the usual Kähler-covariant derivative. Plugging \eqref{eq:d_sugra}-- \eqref{eq:F4_sugra} into \eqref{eq:rescaled_L} and taking into account the contribution of the boundary term yields
	\begin{align}
	\frac{1}{\sqrt{-g}} \mathcal{L} = \frac{1}{2} R - K_{\phi \bar{ \phi}} \partial_\mu \phi \partial^\mu \bar{ \phi} - e^K \left( D_{\phi} \hat{W} K^{\phi \bar{ \phi}} D_{\bar{ \phi}} \hat{\bar{ W}} - 3 \left| \hat{W} \right| \right)
	\end{align}
	where, exactly as in the non-gravitational case, the superpotential is shifted my a term linear in the scalar field multiplied by the constant flux of the 3-form field:
	\begin{align}
	\hat{W} \equiv W - n \phi.
	\label{eq:new_W}
	\end{align}

 Therefore, in summary, after all the auxiliary fields' equations of motion have been solved and used in the action, the contribution of the 3-form fields amount to adding a simple linear term to the superpotential, which is proportional to the constant flux $n$ of the 3-form field. The effective action for the scalar fields and gravity coincides with the usual $\text{N=1}$, $D=4$ matter-coupled supergravity, albeit with a potential constructed from the new effective superpotential \eqref{eq:new_W}.

\bibliography{references}

\end{document}